\documentclass{aa}  

\usepackage{graphicx}
\usepackage{txfonts}
\usepackage{hyperref}
\usepackage{siunitx}
\usepackage{color}

\def\reff{R_{\mathrm{e}}}
\def\mstar{M_*}
\def\asps{\alpha_{\mathrm{sps}}}

\def\zsource{z_{\mathrm{s}}}

\def\zqso{z_{\mathrm{qso}}}

\def\msource{m_{\mathrm{s}}}
\def\mqso{m_{\mathrm{qso}}}

\def\mobs{M_*^{(\mathrm{obs})}}
\def\gammadm{\gamma_{\mathrm{DM}}}
\def\fdm{f_{\mathrm{DM}}}
\def\mdmfive{M_{\mathrm{DM}, 5}}
\def\mhalo{M_{\mathrm{h}}}

\def\tein{\theta_{\mathrm{Ein}}}
\def\teinmin{\theta_{\mathrm{Ein,min}}}
\def\teff{\theta_{\mathrm{e}}}

\def\psilens{\boldsymbol{\psi}_\mathrm{g}}
\def\psisource{\boldsymbol{\psi}_\mathrm{s}}

\def\psisourcenobeta{\boldsymbol{\psi}_\mathrm{s}^{(-\boldsymbol\beta)}}

\def\prlens{{\rm P}_\mathrm{g}}
\def\prsource{{\rm P}_\mathrm{s}}

\def\prsl{{\rm P}_\mathrm{{SL}}}

\def\pdet{{\rm P}_\mathrm{det}}
\def\psel{{\rm P}_\mathrm{sel}}
\def\pfind{{\rm P}_\mathrm{find}}

\def\crosssect{\sigma_\mathrm{{SL}}}

\def\lmst{\lambda_{\mathrm{mst}}}

\def\Fref#1{Figure~\ref{#1}\xspace}
\def\Tref#1{Table~\ref{#1}\xspace}

\def\pr{{\rm P}}

\defcitealias{S+C21}{Paper~I}
\defcitealias{Son21}{Paper~II}
\defcitealias{Son22}{Paper~III}
\setcitestyle{notesep={}}

\begin{document}

   \title{Strong lensing selection effects}
   \titlerunning{Strong lensing selection effects}
   \authorrunning{Sonnenfeld et al.}

   \author{Alessandro Sonnenfeld\inst{\ref{sjtu},\ref{leiden}}\and
           Shun-Sheng Li\inst{\ref{leiden}}\and
           Giulia Despali\inst{\ref{bologna}}\and
           Raphael Gavazzi\inst{\ref{marseille}}\and
           Anowar J. Shajib\inst{\ref{chicago},\ref{kicp},\ref{einstein}}\and
           Edward N. Taylor\inst{\ref{swin}}
          }

   \institute{
Department of Astronomy, School of Physics and Astronomy, Shanghai Jiao Tong University, Shanghai 200240, China\\
              \email{sonnenfeld@sjtu.edu.cn}\label{sjtu}
\and
Leiden Observatory, Leiden University, P.O. Box 9513, 2300 RA Leiden, The Netherlands\label{leiden}\and
Alma Mater Studiorum - Università di Bologna, Dipartimento di Fisica e Astronomia "Augusto Righi", Via Gobetti 93/2, Bologna, Italy\label{bologna}\and
CNRS and CNES, Laboratoire d'Astrophysique de Marseille, Aix-Marseille Universit\'{e} 38, Rue Fr\'{e}d\'{e}ric Joliot-Curie, F-13388 Marseille, France\label{marseille}\and
Department of Astronomy \& Astrophysics, University of Chicago,
Chicago, IL 60637, USA\label{chicago}\and
Kavli Institute for Cosmological Physics, University of Chicago,
Chicago, IL 60637, USA\label{kicp}\and
NHFP Einstein Fellow\label{einstein}\and
Centre for Astrophysics and Supercomputing, Swinburne University of Technology, Hawthorn 3122, Australia\label{swin}
             }

   \date{}

 
  \abstract
    {
Strong lenses are a biased subset of the general population of galaxies.
}
   {
The goal of this work is to quantify how lens galaxies and lensed sources differ from their parent distribution, namely the strong lensing bias.
} 
   {
We first studied how the strong lensing cross-section varies as a function of lens and source properties. 
Then, we simulated strong lensing surveys with data similar to that expected for \textit{Euclid} and measured the strong lensing bias in different scenarios.
We focused particularly on two quantities: the stellar population synthesis mismatch parameter, $\asps$, defined as the ratio between the true stellar mass of a galaxy and the stellar mass obtained from photometry, and the central dark matter mass at fixed stellar mass and size.
}
   {
Strong lens galaxies are biased towards higher stellar masses, smaller half-mass radii, and higher dark matter masses.
The amplitude of the bias depends on the intrinsic scatter in the mass-related parameters of the galaxy population and on the completeness in Einstein radius of the lens sample.
For values of the scatter that are consistent with observed scaling relations and a minimum detectable Einstein radius of $0.5''$, the strong lensing bias in $\asps$ is $10\%$, while that in the central dark matter mass is $5\%$.
The bias has little dependence on the properties of the source population: samples of galaxy-galaxy lenses and galaxy-quasar lenses that probe the same Einstein radius distribution are biased in a very similar way.
}
   {
Given current uncertainties, strong lensing observations can be used directly to improve our current knowledge of the inner structure of galaxies, without the need to correct for selection effects. Time-delay measurements of $H_0$ from lensed quasars can take advantage of prior information obtained from galaxy-galaxy lenses with similar Einstein radii.
}
   \keywords{
             Gravitational lensing: strong 
               }

   \maketitle
%

\section{Introduction}\label{sect:intro}

Strong gravitational lensing is a powerful tool for studying galaxies and cosmology.
Strong lenses have been used to probe the mass structure of massive galaxies \citep{Aug++10, ORF14, Son++15, Sha++21}, to detect substructure \citep{Veg++12, Hez++16, Nie++20}, to carry out detailed studies of magnified star-forming galaxies \citep{Jon++13}, and to measure the expansion rate of the Universe with time delays \citep[see][ for a review]{T+M16}.

Strong lenses, however, are a biased subset of the general population of galaxies and background sources.
In order for a lens-source pair to be included in a strong lensing survey, the following conditions must be met: (1) the lens must produce multiple images of the source, (2) the multiple images must be detected and resolved, and (3) the system must be recognised as a lens by the survey.

Condition 1 poses constraints on the mass distribution of the lens, as well as on the geometry of the system. 
First of all, not all lenses are capable of producing multiple images, only those whose surface mass density, $\Sigma(\boldsymbol\theta)$, is higher than the critical surface mass density for lensing, $\Sigma_{\mathrm{cr}}$, at at least one position, $\boldsymbol\theta$ \citep{SEF92}.
This condition excludes objects with a very diffuse mass distribution from the population of lenses.
Second, objects with a higher concentration of mass allow for a larger portion of the source plane to be multiply imaged and are therefore over-represented in lens samples. Third, with fixed lens properties, sources that are located farther away along the line of sight are more likely to be lensed.

Condition 2 sets limits on the observable properties of the sources and, again, the lens mass distribution, in a way that depends on the survey characteristics.
On the one hand, brighter sources are more likely to be detected when lensed. On the other hand, lensing magnification allows the detection of fainter objects with respect to the field.
However, if the angular separation between the multiple images is too small, the system may not be classified as a lens.
Since the image separation increases with lens mass, this means that the lens sample is biased against low-mass objects.

Condition 3 can introduce additional biases, depending on the efficiency of the lens finding process on which the survey is based. Most commonly, lenses with a smaller image separation are more difficult to identify even if detected and resolved, because of contamination from the lens galaxy. This introduces an additional bias towards lenses with a higher or more concentrated mass.

The cumulative effect of these conditions can be summarised as follows.
The probability distribution, $\prsl$, of a sample of strong lenses from a given survey with selection criterion $S$ is given by \citep{Son22}
\begin{equation}\label{eq:one}
\prsl(\psilens,\psisource|S) \propto \prlens(\psilens)\prsource(\psisource)\psel(\psilens,\psisource|S).
\end{equation}
In this equation, $\psilens$ is the set of parameters describing the properties of foreground galaxies that are relevant for lensing, such as their redshift and mass distribution; $\psisource$ is the set of parameters describing background sources; $\prlens$ and $\prsource$ describe the parent distribution of foreground galaxies and background sources in the absence of lensing; and $\psel$ describes the probability of selecting a lens-source system with parameters $\psilens$ and $\psisource$ given the criterion, $S$, used to define a lens.
This last factor takes both physical and survey selection effects into account, that is, whether a lens with parameters $\psilens$ produces a strongly lensed image of a source with parameters $\psisource$, and whether such an image can be detected and recognised as a strong lens.

The information needed for the left-hand side of Eq. \ref{eq:one} is directly accessible from strong lensing observations. If the main goal of a lensing survey is to characterise the properties of the strong lens population, then it can be accomplished by directly analysing this term. For many applications of strong lensing, however, the aim is to constrain the properties of the general galaxy or source population, $\prlens$ and $\prsource$, which are coupled in a non-trivial way via the lens selection probability, $\psel$.
In order to obtain an unbiased estimate of either $\prlens$ or $\prsource$, it is necessary to invert Eq. \ref{eq:one}.
In principle, this can be done with a Bayesian hierarchical formalism \citep{Son22}, but knowledge of the lens selection probability, $\psel$, is required.
This factor can be written as the following product:
\begin{equation}\label{eq:two}
\psel(\psilens,\psisource|S) = \pdet(\psilens,\psisource)\pfind(\psilens,\psisource|S),
\end{equation}
where $\pdet$ is the probability of detecting a strong lensing event, and $\pfind(\psilens,\psisource|S,\mathrm{det})$ is the probability of correctly classifying it as such\footnote{\citet{Son22} implicitly assumed $\pfind\equiv1$; therefore, $\pdet$ and $\psel$ can be used interchangeably in the context of that work.}.
The detection probability, $\pdet$, can be obtained via simulations. The main challenge is characterising $\pfind$: in most of the existing strong lensing surveys, the process of determining whether a system is included in a strong lens sample is typically a combination of several cuts, usually involving a non-trivial visual selection step.

For the above reasons, the problem of inverting Eq. \ref{eq:one} is a difficult one to tackle exactly.
A few studies have attempted to explicitly account for strong lensing selection effects, usually by making ad hoc simplifying assumptions \citep{Son++15,O+A17,Son++19}.
Whether it is necessary to invert Eq. \ref{eq:one}, however, depends on the severity of the strong lensing bias that needs to be corrected and on the accuracy requirements for the key quantities of interest.

In this paper we aim to quantify the strength of the strong lensing bias on a series of foreground galaxy and background source parameters. 
In particular, we aim to determine how strong lenses differ from the parent population of foreground galaxies and background sources in terms of (a) the radial mass structure of the lenses (i.e. their stellar and dark matter mass density profiles), (b) the ellipticity of the lenses, and (c) the size-magnitude distribution of the lensed sources.
This determination depends on (1) how the lens detection probability, $\pdet$, varies as a function of galaxy and source properties, (2) the efficiency of a survey in correctly classifying detected strong lenses (i.e. $\pfind$), and (3) the shape of the galaxy and source parameter distribution, $\prlens$ and $\prsource$. To understand this third point we can imagine the limiting case in which both $\prlens$ and $\prsource$ are Dirac delta functions (i.e. all lenses and sources are identical): in this limit, $\prsl$ simply reduces to the product $\prlens\prsource$ up to a scaling constant; this corresponds to a case in which there is no lensing bias.

\citet{MVK09} carried out a thorough study of point (1): they quantified how the properties of a lens galaxy determine its probability of creating a lensing event with a point source.
In this work we revisit the \citet{MVK09} study, expanding it to the extended source case. We simulated individual lenses and examined how the lens detection probability varies with lens and source properties.
In addition, we address points (2) and (3): we simulated large populations of strong lenses using empirical models, we simulated the lens detection and finding phase, and we quantified the lensing bias under various scenarios.
We explored how the results change as a function of the efficiency of a lens survey in discovering small image separation lenses, and as a function of the scatter in mass parameters at fixed light.

Finally, we considered galaxy-quasar lenses. 
The strong lensing bias of lensed quasars was studied by \cite{C+C16} in the context of power-law lens mass models with external convergence.
Here we address the question of how different galaxy-quasar lenses are from galaxy-galaxy lens samples. 
This point is relevant for time-delay cosmography studies, in which measurements of the time delay between the multiple images of a strongly lensed quasar are used to constrain the expansion rate of the Universe. Galaxy-galaxy lenses can in principle be used to help break some of the model degeneracies affecting these measurements \citep{Bir++20,B+T21}, but any difference between the two lens classes can introduce biases if not corrected for.
With this study we aim to quantify this bias.

The structure of this paper is as follows.
In Sect. \ref{sect:basics} we introduce the basics of gravitational lensing.
In Sect. \ref{sect:indlenses} we study individual lens systems. 
In Sect. \ref{sect:lenspop} we describe our simulations of lens surveys.
In Sect. \ref{sect:results} we show the results of our analysis of the simulated lens survey data.
We discuss the results in Sect. \ref{sect:discuss} and draw conclusions in Sect. \ref{sect:concl}.

The Python code used for the simulation and analysis of the lens sample can be found in a dedicated section of a GitHub repository\footnote{\url{https://github.com/astrosonnen/strong_lensing_tools/tree/main/papers/selection_effects}}.


\section{Strong lensing basics}\label{sect:basics}

The lensing properties of an object with respect to a source depend solely on its dimensionless surface mass density distribution, $\kappa(\boldsymbol\theta)$ (also referred to as the convergence). 
This is the ratio between the surface mass density and the critical surface mass density for lensing:
\begin{equation}
\kappa(\boldsymbol\theta) = \dfrac{\Sigma(\boldsymbol\theta)}{\Sigma_{\mathrm{cr}}}.
\end{equation}
The latter quantity is defined as
\begin{equation}
\Sigma_{\mathrm{cr}} = \dfrac{c^2D_\mathrm{s}}{4\pi G D_\mathrm{d} D_{\mathrm{ds}}},
\end{equation}
where $c$ is the speed of light, $G$ the gravitational constant, and $D_\mathrm{d}$, $D_\mathrm{s}$, and $D_{\mathrm{ds}}$ are the angular diameter distances between the observer and the lens, the observer and the source, and the lens and the source, respectively.

Given a source at angular position $\boldsymbol\beta$, images of it form at the positions $\boldsymbol\theta$ that are solutions of the lens equation
\begin{equation}\label{eq:lenseq}
\boldsymbol\beta = \boldsymbol\theta - \boldsymbol\alpha(\boldsymbol\theta),
\end{equation}
where $\boldsymbol\alpha$ is the deflection angle of the lens.
This can be expressed in terms of the dimensionless surface mass density by means of the following integral over the whole sky:
\begin{equation}
\boldsymbol\alpha(\boldsymbol\theta) = \frac{1}{\pi}\int_{\mathbb{R}^2} d^2\boldsymbol\theta' \kappa(\boldsymbol\theta')\dfrac{\boldsymbol\theta - \boldsymbol\theta'}{|\boldsymbol\theta - \boldsymbol\theta'|^2}.
\end{equation}
The images of the background source are in general magnified in total flux and in size, while preserving the original surface brightness.

\subsection{The axisymmetric lens}\label{ssec:axisymm}

In the special case of an axisymmetric lens we can simplify the notation by considering a single coordinate axis with origin at the centre of the lens. We label the components of the image position, source position and deflection angle along this axis as $\theta$, $\beta$, and $\alpha$, respectively.
The lens equation for an axisymmetric lens then becomes
\begin{equation}\label{eq:1dlenseq}
\beta = \theta - \alpha(\theta),
\end{equation}
and the expression for the deflection angle reduces to
\begin{equation}
\alpha(\theta) = \frac{2}{\theta}\int_0^{\theta} d\theta' \kappa(\theta') \theta'.
\end{equation}
This can also be expressed in terms of the total projected mass enclosed within a circle with angular radius equal to $\theta$:
\begin{equation}
\alpha(\theta) = \frac{1}{\pi\theta}\dfrac{M(<\theta)}{\Sigma_{\mathrm{cr}}D_{\mathrm{d}}^2}.
\end{equation}

A very important quantity for describing the strength of a strong lens is the Einstein radius, $\tein$. For an axisymmetric lens, this is the radius corresponding to the solution of Eq. \ref{eq:1dlenseq} for a source placed at the same angular position as the lens centre ($\beta=0$).
The circle with radius equal to $\tein$ is known as the tangential critical curve. Images that form there have infinite magnification in the tangential direction.
It can be shown that $\tein$ satisfies the following condition,
\begin{equation}\label{eq:criteq}
\bar{\kappa}(<\tein) = 1,
\end{equation}
where $\bar{\kappa}(<\theta)$ is the average surface mass density within a radius equal to $\theta$:
\begin{equation}
\bar{\kappa}(<\theta) \equiv \frac{2}{\theta^2}\int_0^{\theta}d\theta' \kappa(\theta')\theta'.
\end{equation}

\begin{figure}
\includegraphics[width=\columnwidth]{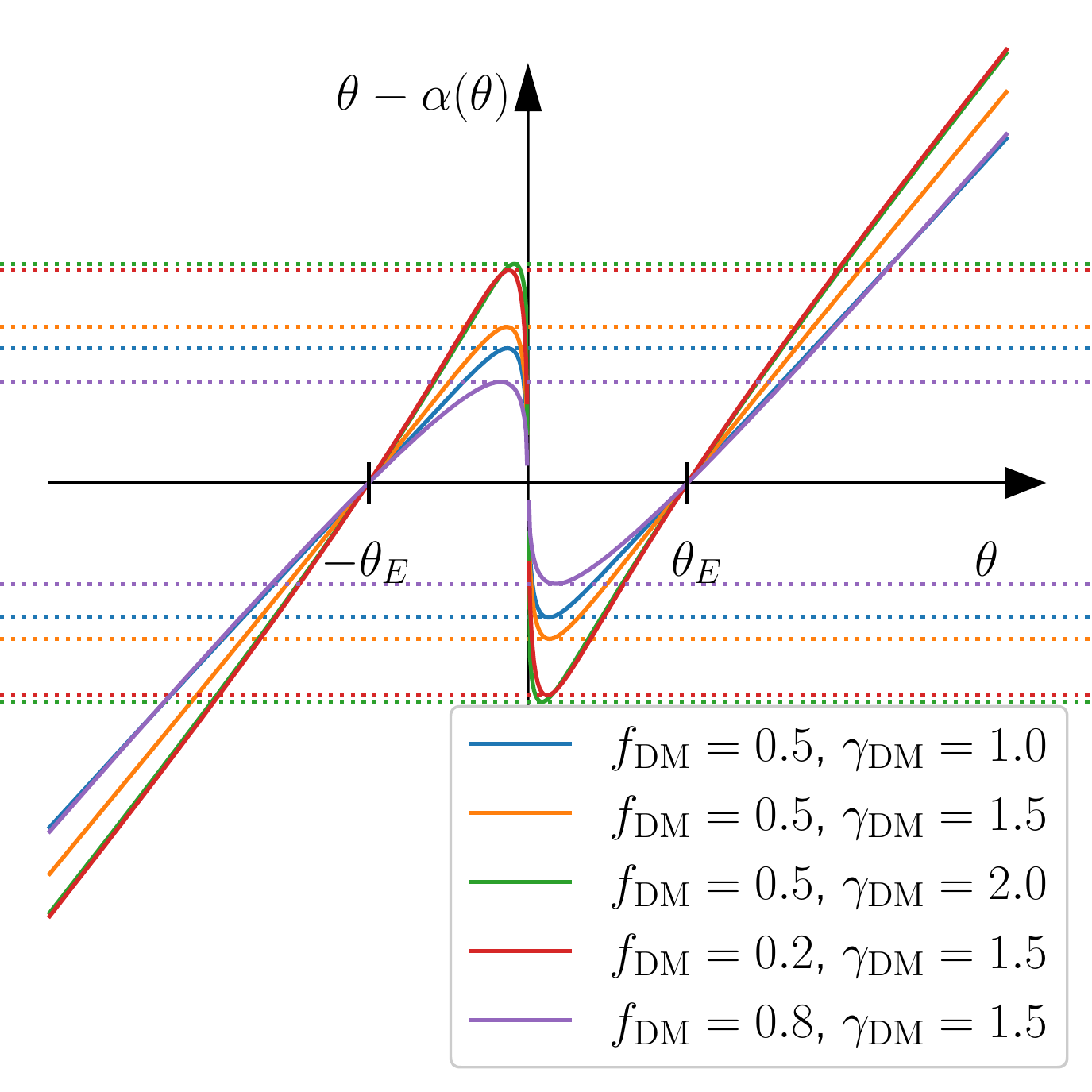}
\caption{
Lens equation of axisymmetric lenses. Solid lines represent the right-hand side of Eq. \ref{eq:1dlenseq} for five lenses with different density profiles. 
Given a source at position $\beta$, its lensed images form at the values of $\theta$ where the solid line intersects a horizontal line located at $\beta$ on the vertical axis. 
Dotted lines represent positions of the radial caustics. Sources located within the radial caustic of a given lens produce three images.
The lens models used in this simulation consist of a stellar component and a dark matter halo, as described in Sect. \ref{ssec:profile}. Their density profile is plotted in Fig. \ref{fig:kappa}.
\label{fig:scheme}
}
\end{figure}

Axisymmetric lenses of the kind considered in this work 
typically produce either one or three images of a point source.
\Fref{fig:scheme} helps visualise this property.
Plotted in \Fref{fig:scheme} is the quantity $\theta - \alpha(\theta)$ as a function of the position in the image plane $\theta$, for a few lens models.
According to Eq. \ref{eq:1dlenseq}, images form at the locations where this quantity equals the position of the source. 
Therefore, given a source position $\beta$, the number of images and their location can be determined by drawing a horizontal line at the value $\beta$ on the vertical axis, and finding the points where this line intersects the $\theta - \alpha(\theta)$ curve.

For small values of $\beta$ (for sources close to the lens centre), the number of images that are produced is three: the source is strongly lensed. For large values of $\beta$, instead, only one image is formed.
The value of $\beta$ where the transition occurs is known as the radial caustic, which is marked by the horizontal dotted line in \Fref{fig:scheme}. 
As can be seen from \Fref{fig:scheme}, a source at this location is mapped to the stationary point of the $\theta - \alpha(\theta)$ curve. 
That location on the image plane is known as the radial critical curve. 
Images that form there have a formally infinite magnification in the radial direction\footnote{The slope of the $\theta - \alpha(\theta)$ curve is the inverse of the radial magnification. This can be understood by taking the derivative of Eq. \ref{eq:1dlenseq} with respect to $\theta$.}.

\subsection{The elliptical lens}

In this paper we focus mostly on lens galaxies with elliptical isodensity contours.
Given a surface mass density profile $\Sigma(R)$, a lens with an elliptical mass distribution can be obtained by replacing the radial coordinate with the circularised radius:
\begin{equation}\label{eq:elltransf}
R \rightarrow \sqrt{qx^2 + \frac{y^2}{q}},
\end{equation}
where $x$ and $y$ are Cartesian axes centred on the lens centre, with $x$ pointing towards the major axis, and where $q$ is the minor-to-major axis ratio.

\begin{figure}
\includegraphics[width=\columnwidth]{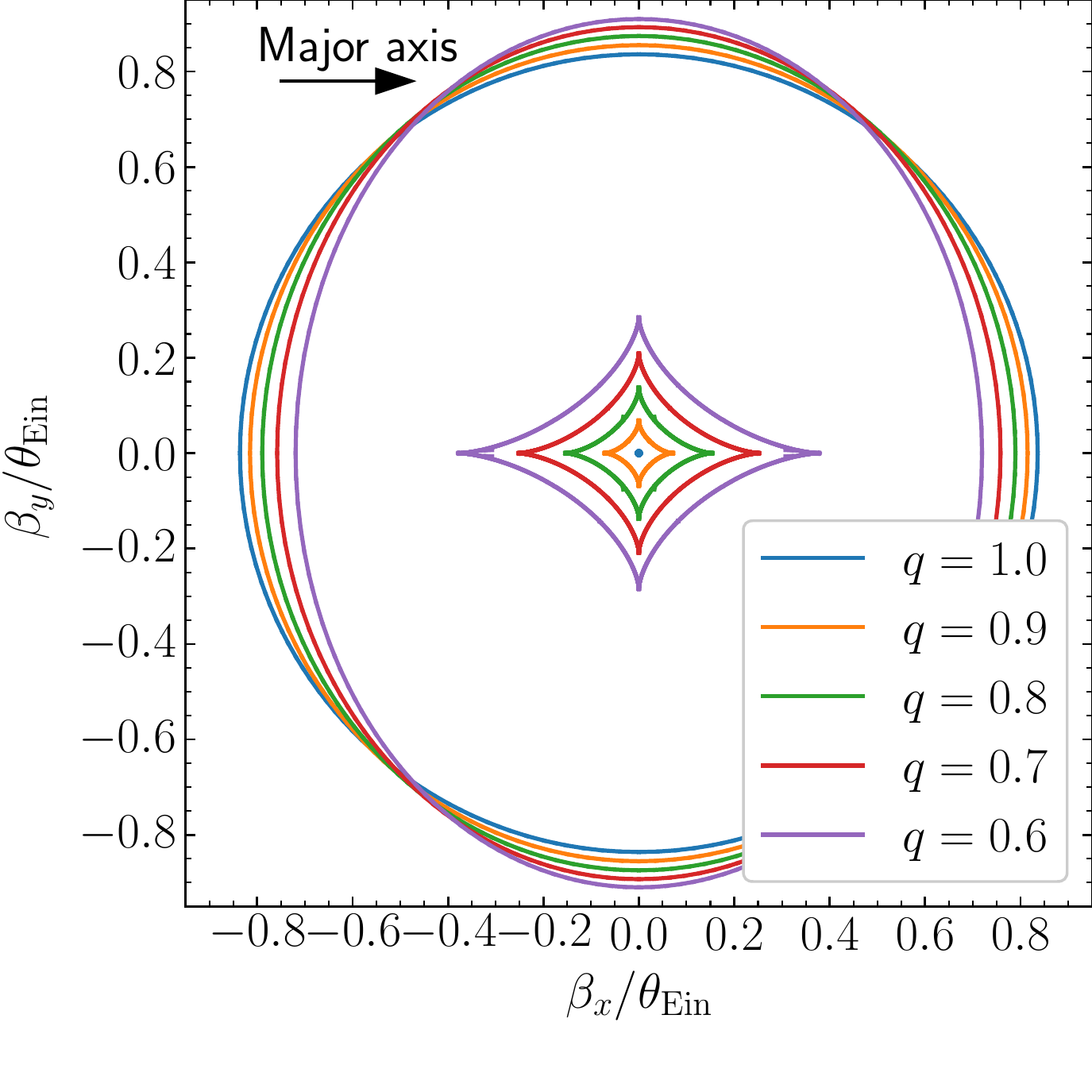}
\caption{
Caustics of lenses with fixed radial structure and different ellipticity.
Source-plane angular coordinates are in units of the Einstein radius.
The outer curve is the radial caustic, while the inner diamond is the tangential caustic. Point sources located outside the caustic are not strongly lensed. Sources that lie in the region enclosed by the two caustics produce three images, while sources inside the tangential caustic produce five images.
The lens model adopted for this experiment consists of a stellar component and a dark matter halo, as described in Sect. \ref{ssec:profile}. The two components have the same ellipticity.
\label{fig:ellcaust}
}
\end{figure}

\Fref{fig:ellcaust} shows the source-plane caustics of elliptical lenses with different values of the axis ratio. We used the software {\sc Glafic} \citep{Ogu21} to obtain the caustic curves.
The outermost curves are radial caustics, while the inner ones are tangential ones.
The most striking difference with respect to the axisymmetric case (blue curves in \Fref{fig:ellcaust}) is the fact that the tangential caustic is transformed from a point into a diamond-like curve.
Sources located within the diamond produce five images (one of which is usually highly de-magnified). Sources that lie in the region enclosed between the two caustics are imaged three times. Sources outside the radial caustic are imaged only once, as in the axisymmetric case.
The fact that the number of images changes by two at a caustic crossing is a general feature of gravitational lenses with no singularities \citep{SEF92}.

\subsection{Lensing event definition: Point sources}\label{ssec:lensdefpoint}

In order to compute the probability of a lensing event we must provide an exact definition for it.
A necessary condition for a lens-source system to qualify as a strong lens is the presence of multiple images of the source.
As we showed above, this requires the source to lie within the region enclosed by the radial caustic.
In order to recognise a strong lens in a real survey, however, it is not sufficient for multiple images to exist: they must be detected.
For this reason, given the detection limit for a point source, $m_{\mathrm{lim}}$, we define as strong lensing event any lens-source system with at least two images brighter than $m_{\mathrm{lim}}$.
This is the same definition used by \citep{VMK09}, upon which the \citet{MVK09} work is based.
Labelling with $m_2$ the magnitude of the second-brightest image, then, in the absence of photometric noise the lens detection probability $\pdet$ becomes
\begin{equation}\label{eq:pointdet}
\pdet(\psilens,\psisource) = \left\{\begin{array}{ll} 1 & \rm{if}\,m_2(\psilens,\psisource) < m_{\mathrm{lim}} \\
0 & \rm{otherwise}\end{array}\right. .
\end{equation}

\subsection{Lensing event definition: Extended sources}\label{ssec:lensdefext}

Defining a strong lensing event in the case of an extended source is less straightforward.
In principle, we should require parts of the source to be multiply imaged.
In practice, it is not always easy to determine whether a lensed image contains multiple images or not.
This is because, when the source is extended, some of the images can be blended together.
In real strong lensing surveys, it is common to find lens candidates in which the lensed source consists of only a single arc. In those cases it is difficult to establish whether the arc is a set of blending images or not, and the decision of including such systems in a strong lens sample is often arbitrary.
Here we adopt the following working definition: an extended source is strongly lensed if the number of surface brightness peaks that are detected is larger than its intrinsic number of peaks in the absence of lensing.

We explain how this definition applies in practice with a few examples.
For simplicity, we focus on the case of a source with a single surface brightness peak, such as a S\'{e}rsic profile. All of the sources considered in this work belong to this family of surface brightness models.
We adopt the following procedure to determine the number of detected peaks.
Given the observed image of a lensed source, we define its footprint as the ensemble of pixels where the source is detected with $S/N>2$. The footprint is in general composed of multiple disconnected regions, corresponding to the different images. In order to only include images that can be clearly identified, we add the condition that the integrated $S/N$ of each separate region must be $S/N > 10$. This condition has the effect of removing from the source footprint any isolated region consisting only of a very small number of pixels. In a real-world application, it would be very hard to classify such marginal detections as images.
If, after applying this cut, the source footprint is spread over multiple separated regions, then the system is classified as a lens.
If the source footprint consists instead of a single region, we iteratively increase the surface brightness threshold used to define the footprint and count the number of isolated regions with $S/N>10$. The maximum number of detected regions defines the number of detected peaks.

\begin{figure}
\includegraphics[width=\columnwidth]{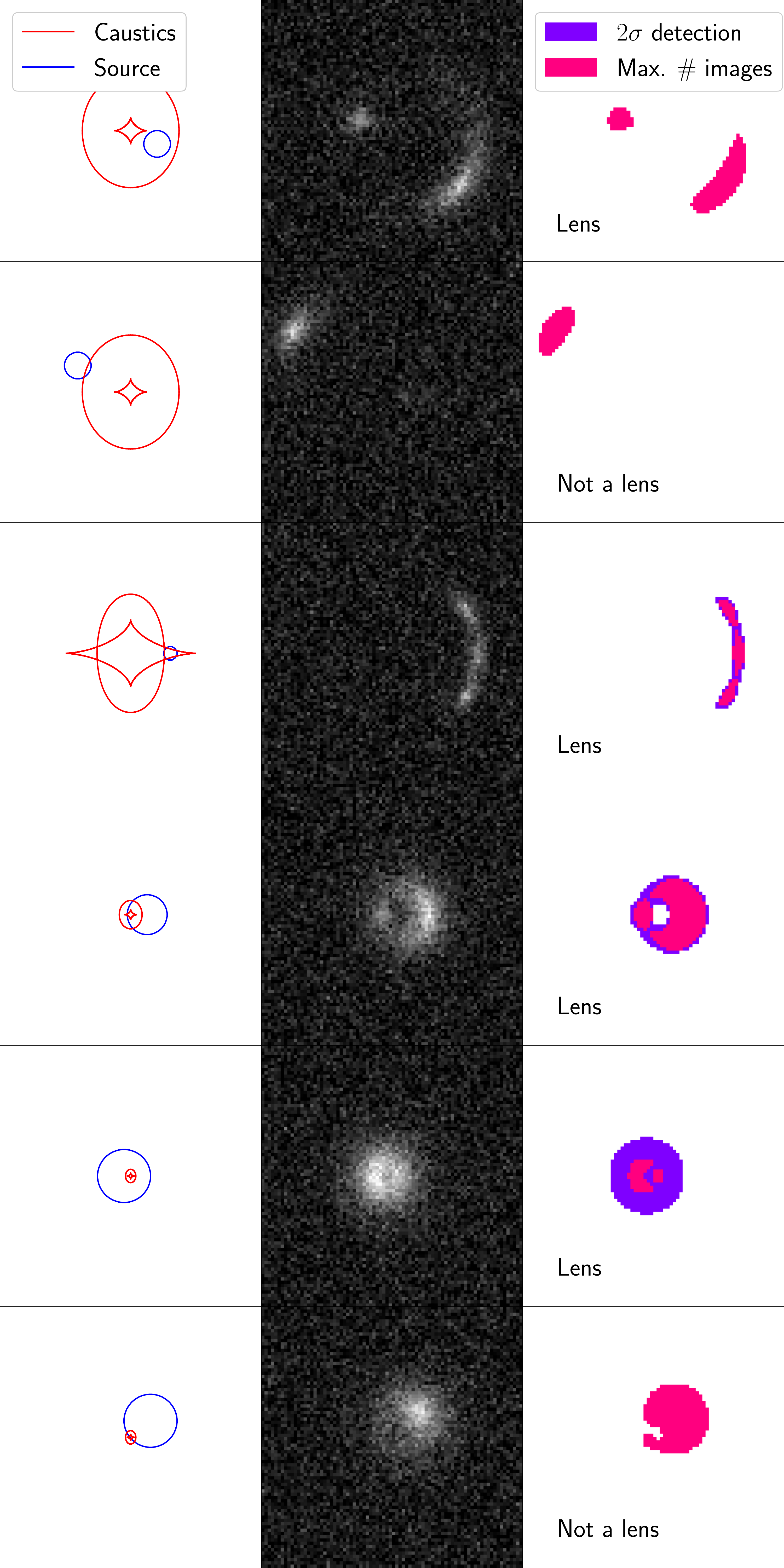}
\caption{
Criterion used to classify lensed images of extended sources. Six examples are shown.
First column: Caustics (red curves) and source position (blue circle). The radius of the circle corresponds to the radius at which the surface brightness is equal to $2\sigma$ the sky background rms fluctuation for a single pixel. In other words, the blue circle delimits the area of the source that can be detected.
Second column: Mock image of the lensed source, with added noise.
Third column: Footprint of the source. The purple footprint is obtained with a $2\sigma$ detection criterion. The coral region is the largest footprint with the highest number of detected images.
\label{fig:lensdef}
}
\end{figure}

\Fref{fig:lensdef} shows a few examples of how this criterion can be used to classify lenses.
The first column shows the caustic structure and source position. The second column shows an image of the lensed source. 
The third column shows the source footprint defined with the procedure described above.
Pink pixels correspond to the largest footprint that maximises the number of detected images, while purple pixels belong to the $2\sigma$ detection footprint. In the first, second and sixth example, these two footprints coincide.
In the third, fourth and fifth example, instead, the $2\sigma$ detection footprint consists of a single region, but an increase of the surface brightness threshold leads to the detection of additional images.

Strictly speaking, our lens definition criterion fails for a perfect Einstein ring. In order to cover such a scenario, we also classify as strongly lensed any sources that produce a footprint with a hole.

We point out that, although our lens definition relies on peak detections, it is not necessary for the peak of the source surface brightness to lie within the caustics in order for the source to be strongly lensed.
This can be observed in the fourth example of Fig. \ref{fig:lensdef}: although the source centroid lies outside of the caustic (as shown in the first column), the outskirts of the source overlap with the lens, and therefore multiple images are produced.

We can identify three different regimes in strong lensing of extended sources, depending on the relative size of the source and of the lens caustics.
In the limit of a very small source, the image configurations that are produced are qualitatively similar to those that can be obtained in the point-source case.
When the source and caustic size are comparable, the multiple images tend to be blended into arcs or rings. 
Most, if not all, of the galaxy-scale strong lenses known belong to these two categories.
However, there is a third regime in strong lensing, corresponding to the case in which the source size is much bigger than the caustics, such as in the fifth example of Fig. \ref{fig:lensdef}.
In this regime, the overall size and total flux of the source are roughly preserved, and the lens produces only a relatively minor perturbation on a localised region of the image.
Strong lenses of this kind can be difficult to detect, especially if the region of the image subject to strong lensing overlaps with the light from the foreground galaxy. 
The ultimate limiting factor to the detection of strong lenses in the large source size regime, however, is the ability to spatially resolve the multiple images. This limit is set by the size of the point spread function (PSF).

All of the examples that we considered were based on sources with a single surface brightness peak.
In general this is not necessarily the case, and the intrinsic number of surface brightness peaks of a strongly lensed source is not known a priori.
When dealing with a real lens candidate, applying our lens definition criterion requires showing that the observed number of surface brightness peaks can be reproduced with a lens model in which the source has a smaller number of peaks.
In samples of lenses defined via visual inspection this process is typically done implicitly, by identifying multiply imaged blobs that belong to the same source element.

To our knowledge, we are the first to propose a surface brightness peak-based definition of a strong lensing event.
A popular alternative definition of a strong lens is one based on magnification: only images that are magnified by more than a given threshold are considered as strongly lensed \citep[see e.g.][]{Hil++07}.
The problem with a definition of this kind is that magnification cannot be determined unambiguously from observations, unless the intrinsic properties of the lensed source are known from other means (e.g. if the source is a standard candle or a standard ruler).
Because of the mass-sheet degeneracy \citep{FGS85}, it is possible to vary the magnification of a lensed image while keeping its observed properties fixed: this could lead to the paradox of two identical-looking lenses that are classified differently on the basis of the underlying magnification.
Although we could still use a magnification-based definition for the sake of carrying out our experiments, it would then be difficult to apply our results to real data.
Our definition of a strong lensing event, instead, is robust with respect to the mass-sheet degeneracy.


\section{Individual lenses}\label{sect:indlenses}

In this section we study how the probability of a strong lensing event varies as a function of lens and source properties.
In order to do so, it is useful to introduce the concept of strong lensing cross-section.
Given a foreground galaxy with parameters $\psilens$, a background source with parameters $\psisource$, and a criterion $S$ to define a strong lensing event, the strong lensing cross-section is defined as \citep{Son22}
\begin{equation}\label{eq:crosssect}
\crosssect = \int_{\mathbb{R}^2} d\boldsymbol\beta \pdet(\psilens,\psisourcenobeta,\boldsymbol\beta|S),
\end{equation}
where $\boldsymbol\beta$ is the position of the background source, $\psisourcenobeta$ is the ensemble of source parameter except for the position, and the integral is carried out over the whole sky.
The definition above is valid for both a point source and an extended source: although there is no unique way of defining the position of an extended source, the integral over the sky ensures that the result is independent of how the source position is defined.
In the limit of low density of background sources, which is satisfied in all practical cases, the probability of a lensing event is proportional to $\crosssect$.
The lensing cross-section defined via Eq. \ref{eq:crosssect} depends solely on the lens detection probability $\pdet$ and does not take into account whether the lens can be correctly classified by a lens finder.
This separate selection step is captured by the term $\pfind$ in Eq. \ref{eq:two}.
In this section we consider exclusively the detectability of a lens, and therefore focus only on $\crosssect$. In Sect. \ref{sect:lenspop}, when considering specific lens survey simulations, we introduce $\pfind$.

We computed $\crosssect$ in a series of different scenarios of increasing complexity.
The model family adopted to describe the radial density profile of the lenses was the same in all of our experiments. We describe this in Sect. \ref{ssec:profile}.
In Sect. \ref{ssec:axisymmpoint} we show calculations of the strong lensing cross-section in the case of axisymmetric lenses and point sources.
In Sect. \ref{ssec:ellpoint} we generalise the lens geometry to elliptical, while in Sect. \ref{ssec:ellext} we replace point sources with extended sources.

\subsection{Lens density profile}\label{ssec:profile}

In this work we focus on massive early-type galaxies as lenses, as these make up the vast majority of known strong lenses.
We describe their mass distribution with a model consisting of two concentric components, one describing the baryons and one for the dark matter.
We assume that the baryonic component consists entirely of stars, thereby neglecting gas, which is known to contribute very little to the mass of early-type galaxies in the inner regions that are probed by strong lensing. 
We then assume that the stars follow a S\'{e}rsic profile, with 
projected surface mass density given by
\begin{equation}
\Sigma(R) = \Sigma_0 \exp{\left\{-b\left(\frac{R}{\reff}\right)^{1/n}\right\}}.
\end{equation}
In the above equation,
\begin{equation}
\Sigma_0 = \frac{\mstar b^{2n}}{2\pi n \reff^2 \Gamma(2n)},
\end{equation}
$\mstar$ is the total mass, $\reff$ is the radius enclosing half of the total mass, $n$ is the S\'{e}rsic index, $\Gamma$ is the incomplete Gamma function, and $b$ is given by \citep{C+B99}
\begin{equation}
b(n) \approx 2n -\frac13 + \frac{4}{405n} + \frac{46}{25515n^2} + O(n^{-3}).
\end{equation}
Throughout this paper we indicate with $\teff$ the angular size of the half-light radius.

We fix the S\'{e}rsic index of the lenses to $n=4$, corresponding to a de Vaucouleurs model.
Although early-type galaxies are often described with a free S\'{e}rsic index, a de Vaucouleurs profile is able to reproduce their surface brightness distribution to a few percent in the radial range $1\rm{kpc} < R < 30\rm{kpc}$ \citep[see e.g.][]{Son20}, which is the region that is most relevant for strong lensing.
Finally, we assume that the light distribution of the stellar component follows its mass distribution exactly. That is, we do not allow for the presence of gradients in the stellar mass-to-light ratio. In Sect. \ref{ssec:limitations} we discuss qualitatively what the implications of relaxing this assumption would be.

We describe the dark matter halo with a generalised Navarro, Frenk, and White (gNFW) profile.
We first define it by its three-dimensional distribution, which for a spherically symmetric profile is
\begin{equation}
\rho(r) = \dfrac{\rho_0}{(r/r_s)^{\gammadm}\left(1 + r/r_s\right)^{3-\gammadm}}.
\end{equation}
The parameter $\gammadm$ is the inner density slope, $\rho_0$ is a normalisation parameter, while $r_s$ is the scale radius. The logarithmic slope of the density profile transitions from $-\gammadm$ to $-3$ at a radius $r\approx r_s$.
The projected surface mass density of a gNFW profile can be expressed in terms of the following integral \citep{WTS01}:
\begin{equation}
\Sigma(R) = 2r_s\rho_0 \left(\frac{R}{r_s}\right)^{1-\gammadm}\int_0^{\pi/2} dx \sin{x}(\sin{x} + R/r_s)^{\gammadm-3}.
\end{equation}

\begin{figure}
\includegraphics[width=\columnwidth]{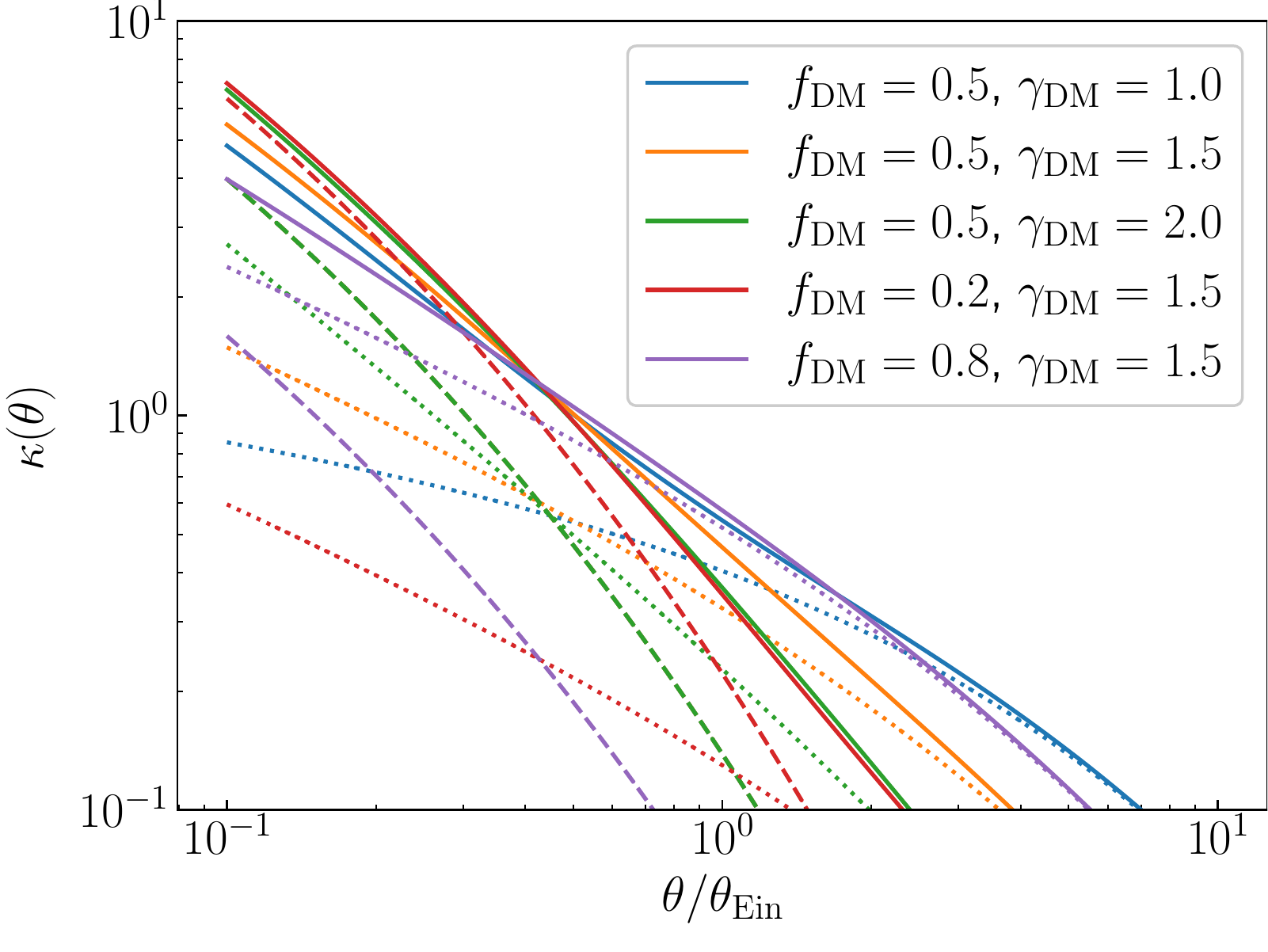}
\caption{
Dimensionless surface mass density profile of S\'{e}rsic + gNFW composite models.
The S\'{e}rsic index of the baryonic component is fixed to $n=4,$ and the scale radius of the dark matter component is fixed to ten times the half-light radius.
All of the profiles are normalised in such a way that the Einstein radius is equal to the half-light radius.
Solid lines show the total density profile, dotted lines the dark matter density profile, and dashed lines the baryonic density profile.
The dashed blue, orange, and green lines are identical, as they correspond to profiles with the same fraction of baryonic mass within the half-light radius.
\label{fig:kappa}
}
\end{figure}

Figure \ref{fig:kappa} shows the dimensionless surface mass density profile of S\'{e}rsic + gNFW models with various values of the inner dark matter density slope $\gammadm$ and of the fraction of projected dark matter mass within the half-light radius, $\fdm$.
All of the profiles have 
a dark matter scale radius equal to ten times $\reff$, and are normalised in such a way that the Einstein radius is equal to the half-mass radius of the stellar component (these assumptions are dropped later).
Two main features emerge from Fig. \ref{fig:kappa}. First, the baryons generally dominate the total density in the inner regions $(\theta < \tein)$, while the dark matter is the main component at large radii.
Second, models with different dark matter fractions and inner dark matter slopes can conspire to produce very similar total density profiles. 
This is the case, for example, of the ($\fdm=0.5,\gammadm=1.5$) and the ($\fdm=0.2,\gammadm=2.0$) models (green and blue lines in Fig. \ref{fig:kappa}).

\subsection{Axisymmetric lenses, point sources}\label{ssec:axisymmpoint}

Axisymmetric lenses with a density profile of the kind introduced in Sect. \ref{ssec:profile} can produce either one or three images of a point source.
This can be seen in Fig. \ref{fig:scheme}, which shows the lens equation for various values of the dark matter fraction and the inner dark matter slope.
All of the lenses shown in this figure have the same Einstein radius, which is equal in size to the half-mass radius of the stellar component.
The radius of the radial caustic, marked by the dotted lines in Fig. \ref{fig:scheme}, is a strong function of the lens properties: it is largest in lenses with a smaller dark matter fraction or steeper dark matter slopes.
As a result, the source plane area that is subject to strong lensing is an even stronger function of these properties, since it scales with the square of the radial caustic radius.
In order to compute the lensing cross-section, however, we must take the magnification into account  because it determines whether multiple images can be detected or not.

\begin{figure}
\includegraphics[width=\columnwidth]{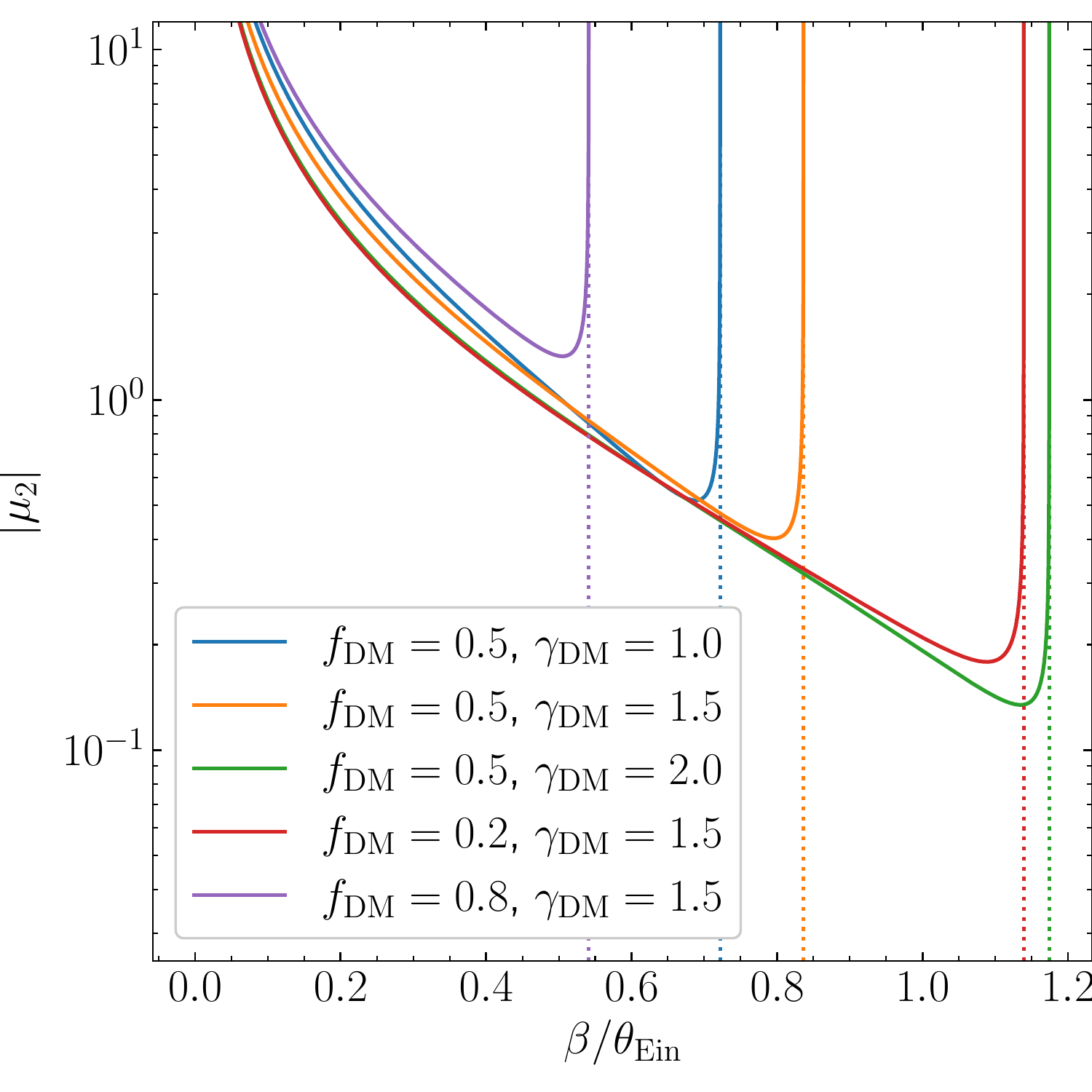}
\caption{
Magnification of the secondary image as a function of the source position.
The lenses are axisymmetric composite models. Their density profile is plotted in Fig. \ref{fig:kappa}.
Vertical dotted lines mark the position of the radial caustic.
\label{fig:1dmag}
}
\end{figure}

Figure \ref{fig:1dmag} shows the magnification of the secondary image as a function of source position. The secondary image is the one located in the region between the radial and tangential critical curves, opposite to the source with respect to the lens centre. In most practical cases this is the second brightest image.
As Fig. \ref{fig:1dmag} shows, the magnification is very large for sources close to the lens centre (small values of $\beta$), decreases with increasing source position and then increases in close proximity to the radial caustic.
While for the model with $f_{\mathrm{DM}}=0.8$ (purple curve) the magnification is above unity everywhere, other lens models can produce highly de-magnified secondary images for large values of $\beta$.
Depending on the intrinsic brightness of the lensed source, these images may or may not be detected.

\begin{figure}
\includegraphics[width=\columnwidth]{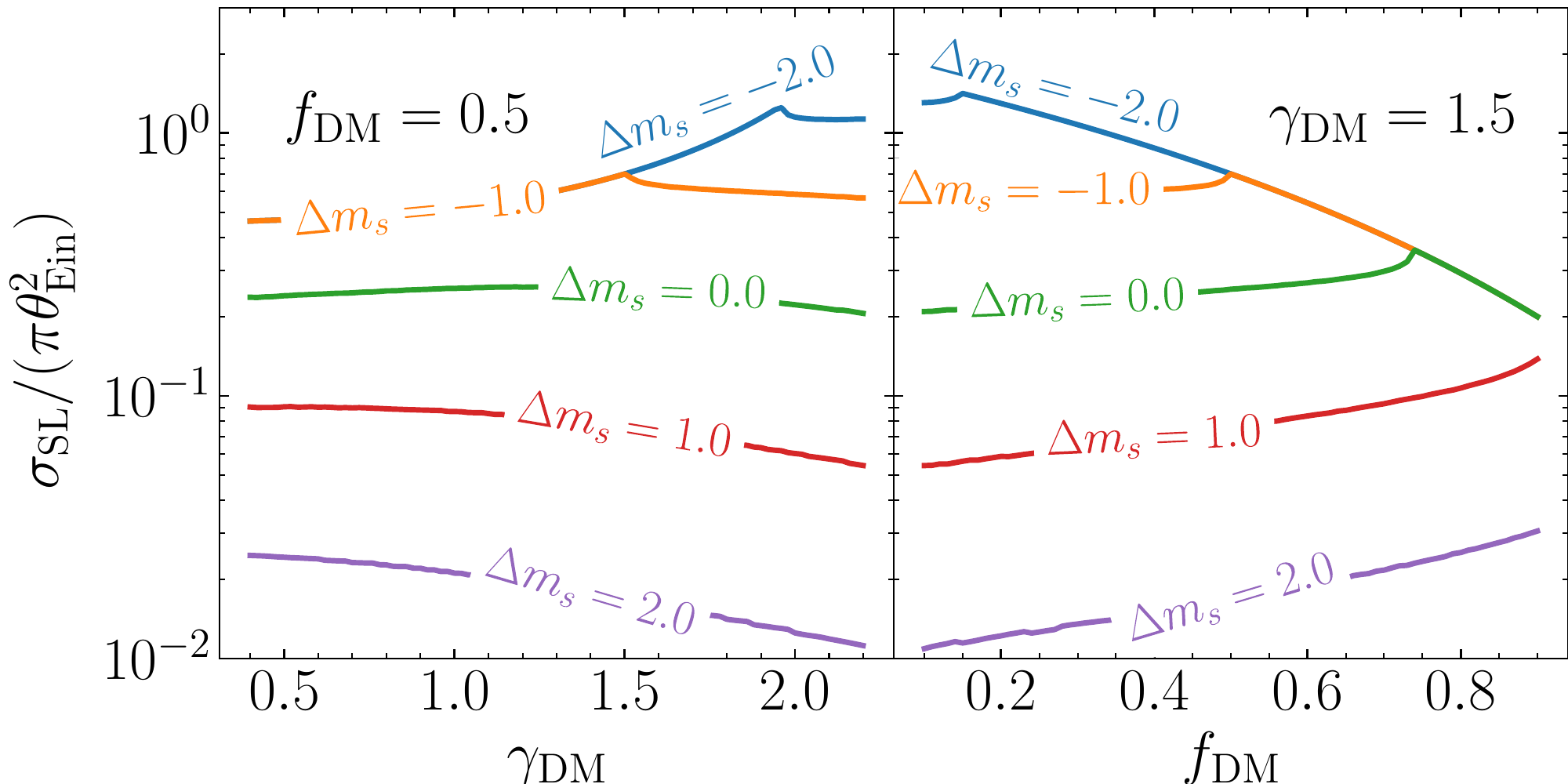}
\caption{
Strong lensing cross-section of an axisymmetric lens and a point source.
The lens galaxy is a composite model, introduced in Sect. \ref{ssec:profile}, with an angular half-light radius equal to the Einstein radius.
The left panel shows the effect of varying the slope $\gammadm$ for fixed $\fdm$; conversely, the right panel shows variations as a function of $\fdm$ with $\gammadm$ fixed.
The system is defined as a strong lens if at least two images are detected.
Different lines correspond to the difference, $\Delta m$, between the source magnitude and the survey detection limit for a point source.
\label{fig:fixedap_cs}
}
\end{figure}

Using the definition of Eq. \ref{eq:crosssect}, we computed the lensing cross-section of a set of axisymmetric lenses, with respect to point sources with different brightnesses.
In particular, we considered model lenses with fixed Einstein radius, with angular half-mass radius $\teff$ fixed to $\tein$, and varying values of the dark matter fraction and dark matter inner slope.
The results are shown in Fig. \ref{fig:fixedap_cs}.
Each line corresponds to a source with a given intrinsic (i.e. unlensed) magnitude, $m_s$. The difference between this magnitude and the limiting magnitude of the survey is indicated as follows:
\begin{equation}
\Delta m \equiv m_s - m_{\mathrm{lim}}.
\end{equation}

We can see a clear trend between $\Delta m$ and the lensing cross-section, in both panels of Fig. \ref{fig:fixedap_cs}: $\crosssect$ is larger for brighter sources.
The trend saturates below a certain $\Delta m$, for sufficiently small values of $\gammadm$ or for sufficiently large values of $\fdm$.
In these cases the lensing cross-section coincides with the full area enclosed within the caustic, and further increasing the source brightness does not result in an increased value of $\crosssect$.

\begin{figure*}
\includegraphics[width=\textwidth]{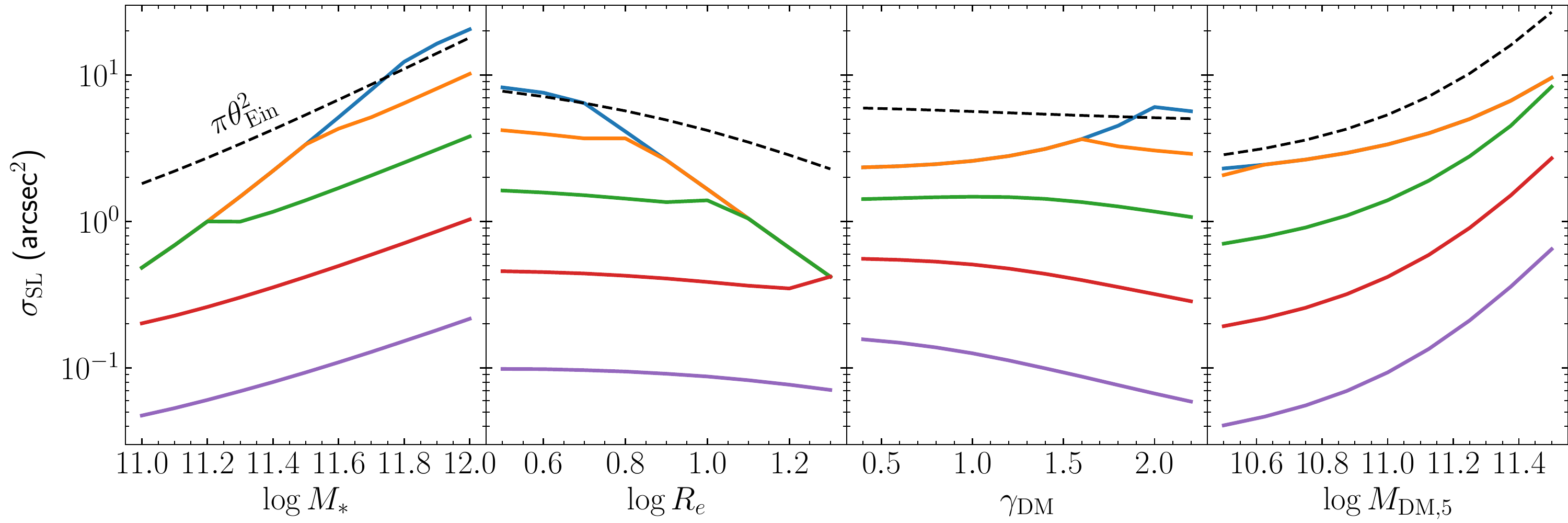}
\caption{
Absolute value of the strong lensing cross-section as a function of various lens properties.
The reference lens is a galaxy at $z=0.3$ with $\log{\mstar}=11.5$, $\reff=7\,{\rm kpc}$, $\gammadm=1.5$, $\log{\mdmfive}=11.0$, $r_s=100$~kpc, and a source at $z=1.5$
In each panel, only one property of the lens is varied, as indicated on the label of the horizontal axis.
Each curve corresponds to a different value of $\Delta m$, in accordance with Fig. \ref{fig:fixedap_cs}.
The dashed line in each panel shows the quantity $\pi\tein^2$.
\label{fig:phys_cs}
}
\end{figure*}

At fixed source brightness, trends with the dark matter inner slope or dark matter fraction are generally weak.
The lines of Fig. \ref{fig:fixedap_cs}, however, have been computed by keeping the Einstein radius fixed while varying $\gammadm$ or $\fdm$.
This is achieved by adjusting other properties of the lens, such as the total mass of the baryonic or the dark matter component (see Fig. \ref{fig:kappa}). 
In practice, when varying one ingredient of the lens density profile, the Einstein radius varies in response.
To get a complete view of how the lensing cross-section depends on different lens properties, we also computed how $\crosssect$ responds in absolute terms by varying one lens parameter at a time.
Figure \ref{fig:phys_cs} shows $\crosssect$ as a function of stellar mass, half-light radius, inner dark matter slope, and projected dark matter mass enclosed within an aperture of $5$~kpc, $\mdmfive$.

The lensing cross-section increases with increasing stellar mass and dark matter mass, decreases with increasing $\reff$ for bright sources, while is only a weak function of $\gammadm$. 
The lack of a clear trend between $\crosssect$ and $\gammadm$ appears to be in contradiction with the result of \citet{MVK09}, who found a strong positive correlation between $\crosssect$ and $\gammadm$.
The origin of this discrepancy lies in the different ways in which $\gammadm$ is varied in the two experiments. While we varied $\gammadm$ at fixed $\mdmfive$, \citet{MVK09} kept fixed the virial mass of the dark matter halo. At fixed virial mass, increasing the inner dark matter slope results in a correspondingly larger dark matter mass in the inner regions, which naturally results in a larger lensing cross-section.

Figures \ref{fig:fixedap_cs} and \ref{fig:phys_cs} show that the trends between lens properties and the strong lensing cross-section can be different for sources with different brightnesses.
The net effect in a strong lensing survey is an average over the source population, weighted by the source luminosity function. 
This implies that surveys that target different families of sources, with different luminosity functions (for instance, galaxies or quasars), can have different strong lensing biases. We explore this possibility in Sect. \ref{sect:results}.

\subsection{Elliptical lenses, point sources}\label{ssec:ellpoint}

We measured $\crosssect$ for lenses with a fixed radial density profile and different ellipticities, with respect to point sources of different brightnesses.
In particular, we set $\fdm=0.5$, $\gammadm=1.5$, $r_s=10\reff$, $\tein=\teff$, and set the ellipticities of both the baryonic and dark matter components to be the same, with the same orientation of the major axis.
This is the lens model used to produce the caustics plot of Fig. \ref{fig:ellcaust}.
Figure \ref{fig:ellcaust} suggests that the size of the source plane area subject to strong lensing, the area enclosed within the outermost caustic, does not vary strongly with the ellipticity of the lens.
Therefore, we expect the strong lensing cross-section to be a weak function of ellipticity.

\begin{figure}
\includegraphics[width=\columnwidth]{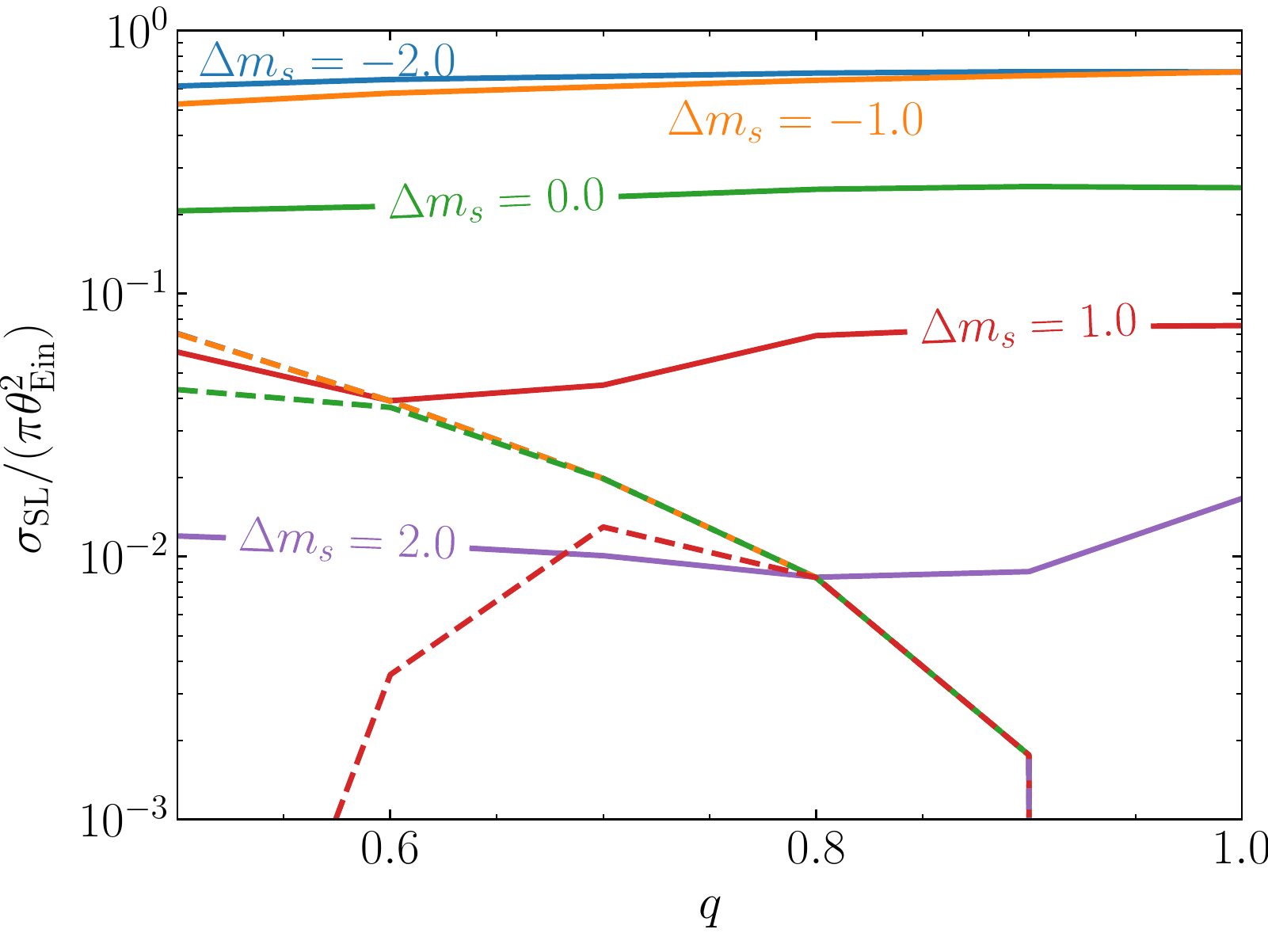}
\caption{
Point source strong lensing cross-section as a function of lens axis ratio.
Solid lines show the cross-section based on the lens event definition of Eq. \ref{eq:pointdet} and dashed lines the cross-section based on the detection of four images (quad cross-section).
Lines of different colours correspond to sources of different intrinsic magnitudes.
The dashed blue line, corresponding to the brightest source magnitude, overlaps completely with the dashed orange line.
The dashed purple line is zero: very faint sources cannot produce any detectable quads.
The parameters of the lens density profile are $\fdm=0.5$, $\gammadm=1.5$, $r_s=10\reff$, and $\teff=\tein$.
The baryonic and dark matter components have the same ellipticity and direction of the major axis.
\label{fig:ellpoint_cs}
}
\end{figure}

We carried out the computation of $\crosssect$ by means of simulation: we generated a large number of point sources randomly distributed over a given area that includes the caustic, then used {\sc Glafic} to solve the lens equation, find the number of images and their magnification. We then measured the fraction of sources that are strongly lensed according to the criterion of Eq. \ref{eq:pointdet} and multiplied this value by the area over which sources are located.
The resulting $\crosssect$ is shown in Fig. \ref{fig:ellpoint_cs} as a function of the minor-to-major axis ratio, $q$.

For bright sources the strong lensing cross-section is approximately constant with axis ratio. This is because, as pointed out earlier, in the bright source regime the cross-section is determined by the area enclosed within the radial caustic, which does not vary much with lens ellipticity.
For faint sources we observe a larger variation with $q$, with a factor of two difference between the largest and smallest value of $\crosssect$ at fixed source brightness.

Most of the sources that result in detectable lenses produce two detectable images. These are sources that are located in the region enclosed between the radial and the tangential caustic. 
If the source is located within the tangential caustic, however, four detectable images are usually created.
Lenses with four visible images, usually referred to as quad lenses, are sometimes given a high priority in certain lensing studies, because they offer more constraints compared to double lenses. For instance, quads make up the majority of the lenses used so far in time-delay studies \citep{Mil++20}.
For this reason we also computed an alternative lensing cross-section, in which the definition of a lensing event requires the detection of four images, instead of two.
This is plotted in Fig. \ref{fig:ellpoint_cs} with dashed lines.
The cross-section for quads is a strong function of lens ellipticity, for bright sources.
This is a consequence of the fact that the area enclosed within the tangential caustic, which is where a source needs to be in order to produce four images, increases with increased lens ellipticity, as Fig. \ref{fig:ellcaust} shows.
For sources that are intrinsically fainter than the detection limit, however, the quad cross-section is extremely small, regardless of ellipticity.

\subsection{Elliptical lenses, extended sources}\label{ssec:ellext}

In the case of an extended source, the complexity of the problem is increased due to the addition of a series of features: the source surface brightness distribution, with its radial profile, shape and orientation, and the PSF. 
Moreover, as we discussed in Sect. \ref{ssec:lensdefext}, there are different regimes in strong lensing of extended sources, depending on the relative size of the source and the caustics of the lens.
For this reason, we split our analysis into two parts. First, we explore the small source size regime, where the source size is comparable to or smaller than the lens caustics. Then, we consider cases in which the source size is bigger than the lens caustics, which we refer to as the large source regime.

\subsubsection{Small source sizes}

\begin{figure}
\includegraphics[width=\columnwidth]{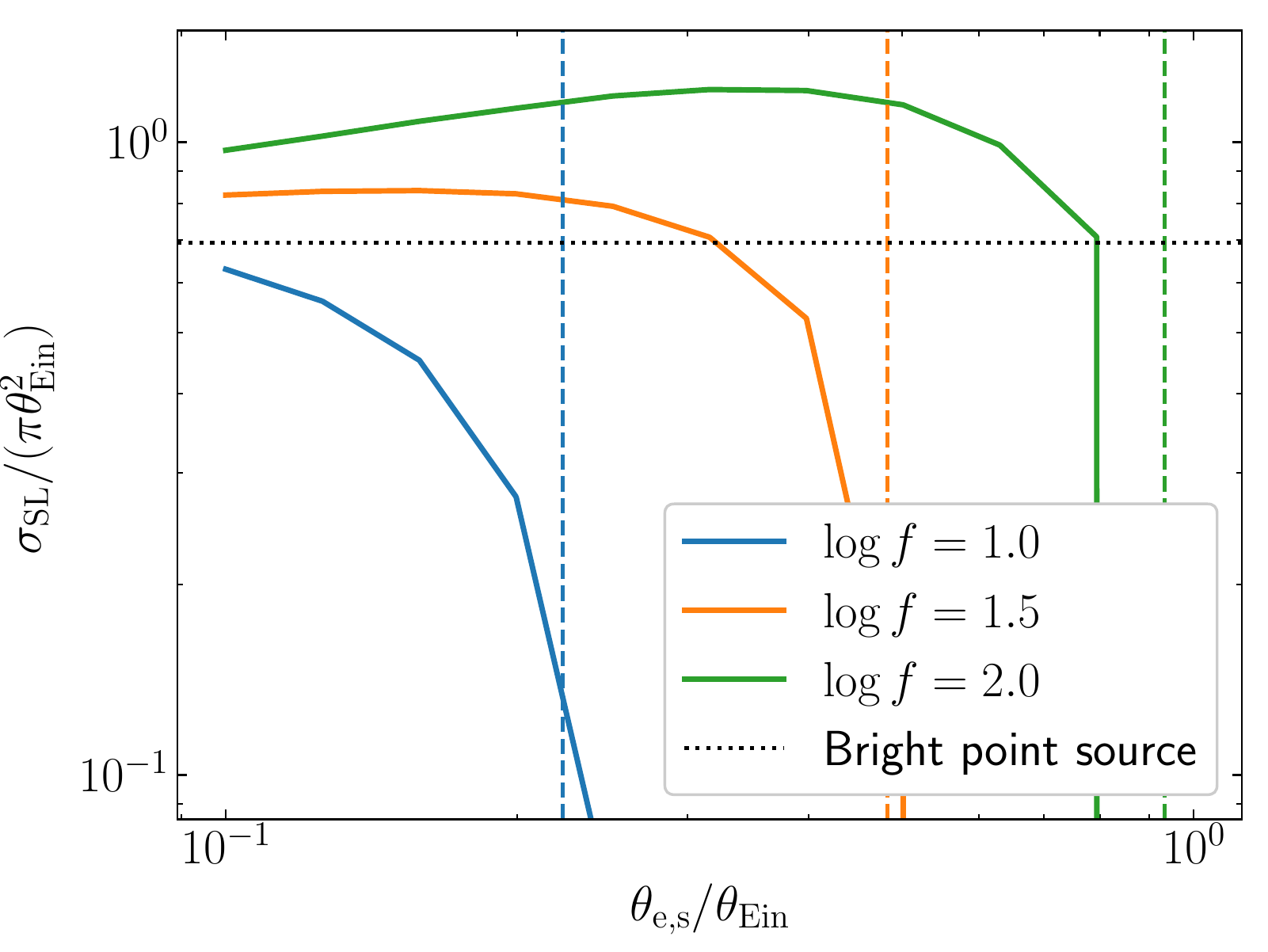}
\caption{
Strong lensing cross-section of an extended source: small source size regime ($\theta_{\mathrm{e,s}} < \tein$).
The cross-section is plotted as a function of the ratio between the half-light radius of the source and the Einstein radius of the lens.
The lens model is fixed to be the same as in Fig. \ref{fig:ellpoint_cs}, with axis ratio $q=0.7$. The source is a circular exponential profile.
Each solid line corresponds to a different value of the total unlensed flux of the source. The flux, $f$, is expressed in terms of the background noise rms fluctuation measured over an area equal to $\tein^2$.
The vertical dashed lines correspond to the maximum size for which a galaxy with a given flux can be detected in the absence of lensing.
The horizontal dotted line indicates the cross-section for a very bright point source (i.e. the area enclosed within the caustics).
\label{fig:extcs}
}
\end{figure}

For the sake of reducing the dimensionality of the analysis, we focused on circularly symmetric sources. We also fixed the surface brightness profile to an exponential disk (i.e. a S\'{e}rsic model with $n=1$).
We then took a lens model with the same parameters used in Sect. \ref{ssec:ellpoint} and a minor-to-major axis ratio of $q=0.7$.
We simulated a large number of images of extended sources with {\sc Glafic} and measured the fraction of them that results in a strong lens, according to the definition of Sect. \ref{ssec:lensdefext}.
For the small source size experiment, we used pixels with a size equal to $1/20\tein$ and convolved the images with a Moffat PSF with a full width at half maximum (FWHM) of two pixels and a $\beta$ parameter of $5.0$.
We also assumed that the background noise is an uncorrelated Gaussian field.
We carried out experiments with sources with different values of the total unlensed flux, $f$, and half-light radius, $\theta_{\mathrm{e,s}}$.
The results are shown in Fig. \ref{fig:extcs}.
The total flux of the source, indicated in the legend, is measured in units of the sky background rms fluctuation within an area equal to the square of the Einstein radius.

As in the point source case, the lensing cross-section increases with increasing total flux, at fixed source size.
At fixed flux, $\crosssect$ stays approximately constant with increasing source size until a given value, then drops rapidly for larger sizes.
From a qualitative point of view, this behaviour can be observed also in the absence of lensing: increasing the size of a galaxy while keeping its flux fixed lowers its average surface brightness. If the surface brightness drops below the sky rms fluctuation level, then it becomes very difficult to detect it.
In order to determine whether there are lensing-specific features in the $\crosssect-\theta_{\mathrm{e,s}}$ relation of Fig. \ref{fig:extcs}, we computed, for each source flux, the maximum half-light radius for which it can be detected in the absence of lensing.
We used the same criterion as that of Sect. \ref{ssec:lensdefext} to define a detection: we defined the source footprint as the ensemble of pixels that are $2\sigma$ above the background and required the total signal-to-noise ratio within the footprint to be larger than ten.
The resulting limiting sizes are shown as vertical lines in Fig. \ref{fig:extcs}. For each given total flux, the non-lensing size limit is similar to the value of $\theta_{\mathrm{e,s}}$ at which the lensing cross-section drops.

This result suggests that, to the first approximation, lensing does not introduce a strong selection in source size.
While perhaps surprising, this follows from the fact that gravitational lensing preserves surface brightness: a source that can be detected in the absence of lensing will produce images with the same surface brightness when lensed, which can be detected as well.
In order to classify a source as strongly lensed, however, we require that multiple images are observed. 
Only sources that lie within a well-defined region give rise to a strong lensing configuration.
If part of the source extends outside of this region, then only a fraction of its flux contributes to creating a set of strongly lensed images.
This lowers the signal-to-noise ratio of the multiple images compared to the point source case.
The result is that $\crosssect$ starts to decrease with increasing source size at values of $\theta_{\mathrm{e,s}}$ that are smaller than the no-lensing detection limit, as observed in Fig. \ref{fig:extcs}.

At the brightest flux explored in the experiment (green line in Fig. \ref{fig:extcs}), the lensing cross-section at small source sizes is larger than the area enclosed within the radial caustic (black dotted line in Fig. \ref{fig:extcs}).
This is because, when the source is very bright, it can give rise to multiple images even while its centroid lies outside of the radial caustic, as long as its surface brightness distribution extends into it.
This also explains why $\crosssect$ increases with increasing source size, before dropping to zero: the more extended the source, the farther away from the lens centre it can be while still producing multiple images.

\subsubsection{Large source sizes}

\begin{figure}
\includegraphics[width=\columnwidth]{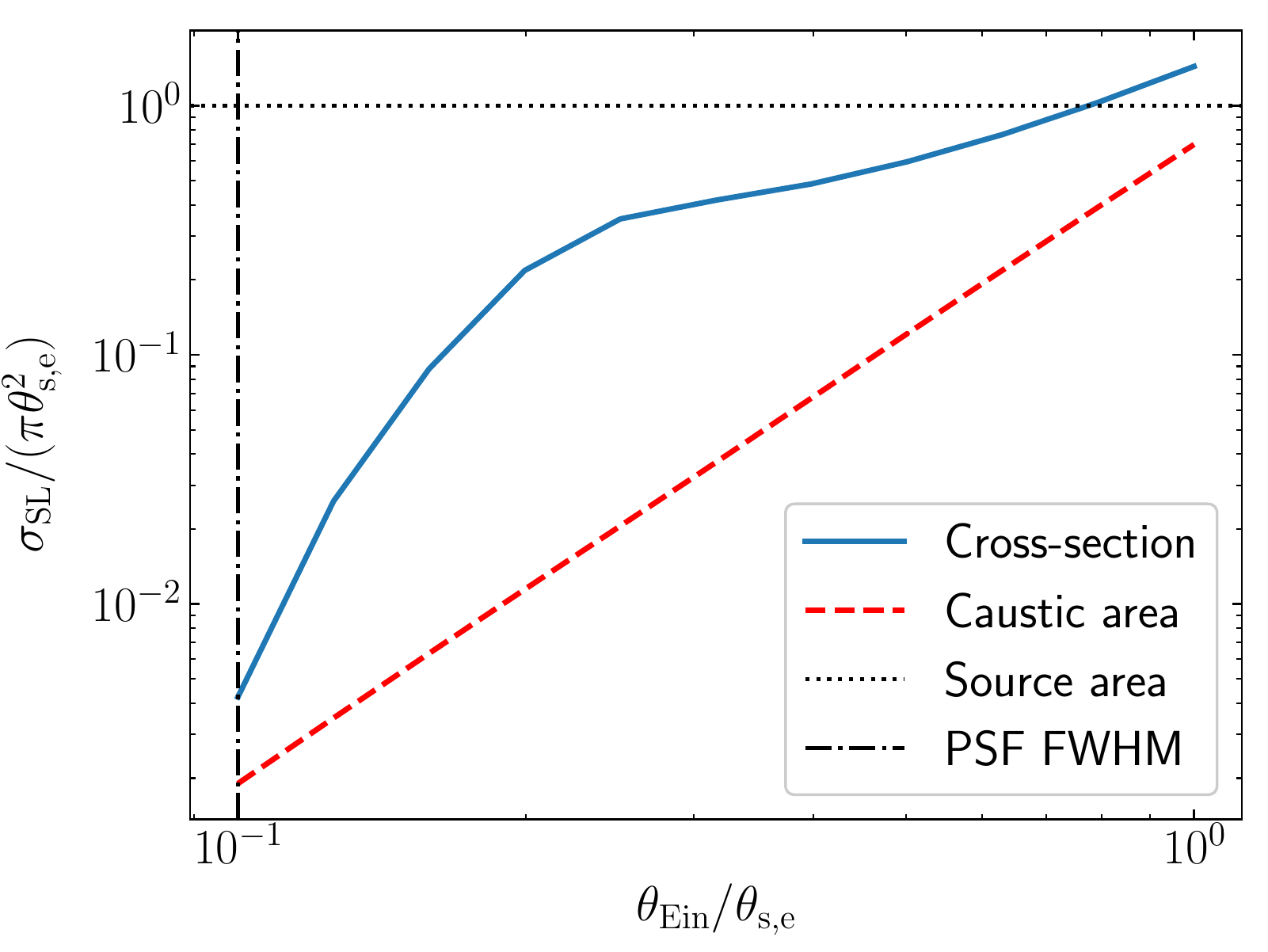}
\caption{
Strong lensing cross-section of an extended source: large source size regime ($\theta_{\mathrm{e,s}} \gtrsim \tein$).
The cross-section is plotted as a function of the ratio between the Einstein radius of the lens and the half-light radius of the source.
The source model is fixed to a circular S\'{e}rsic profile with $n=1$, detected out to the half-light radius.
The source half-light radius measures $20$ pixels, and the PSF has an FWHM of two pixels.
The lens model is varied: starting from the model of Fig. \ref{fig:extcs}, the critical surface mass density is increased to produce progressively smaller values of $\tein$.
The solid line shows the lensing cross-section, in units of the footprint area of the source in the absence of lensing, as a function of the ratio between the lens Einstein radius and the source half-light radius.
The dashed line delimits the area enclosed within the lens caustics, and the dotted line shows the value of the cross-section corresponding to the area of the source.
The dash-dotted line is the FWHM of the PSF.
\label{fig:largesourcecs}
}
\end{figure}

For the large source size case we fixed the surface brightness profile of the source and varied the Einstein radius of the lens.
In particular, we set the source half-light radius to $20$ pixels and adjusted its total flux in such a way that the $2\sigma$ detection footprint in the absence of lensing extends out to the half-light radius.
Then, starting from the lens model used in the previous section, we progressively increased the critical surface mass density to reduce the Einstein radius down to values comparable to the pixel size.
Figure \ref{fig:largesourcecs} shows the resulting strong lensing cross-section as a function of the ratio between lens Einstein radius and source half-light radius.
For values of the Einstein radius close to the size of the PSF, the lensing cross-section (blue solid line) is very small: this is because the perturbations caused by lensing are not well resolved.
For larger values, $\crosssect$ stays approximately constant, around values that are comparable to the source size (horizontal dotted line) and much larger than the area enclosed within the caustics (red dashed line).
We conclude that, in the large source size regime, the main factor that determines the lensing cross-section is the area of the background source, provided that the Einstein radius of the lens is larger than the PSF.

\section{Lens population simulations}\label{sect:lenspop}

In the previous section we showed how the lensing cross-section, which is closely related to the lens detection probability $\pdet$, varies as a function of lens and source properties.
From here on we focus on the effect that those trends have on populations of lenses.
We addressed this question by simulating populations of foreground galaxies and background sources, selecting strong lenses among them, and comparing the properties of the strong lens sample with the parent galaxy population.
We simulated a lens-based search (as opposed to a source-based one), in which strongly lensed images are searched among a stellar mass-selected sample of galaxies.
Our simulations are based on empirical models, in which existing observations of the baryonic component of galaxies are complemented with a set of assumptions on the mass distribution of the lenses.
In Sect. \ref{ssec:lenses} we explain how we built our foreground galaxy sample,
while in Sect. \ref{ssec:sources} we describe the simulation of the background source population.
In Sect. \ref{ssec:obs} we describe how our mock observations of lenses are generated.
In Sect. \ref{ssec:teincuts} we apply a further selection step, based on the angular size of the Einstein radius.

\subsection{Foreground galaxies}\label{ssec:lenses}

Our foreground galaxy population consists of a volume-limited sample of early-type galaxies, complete above a minimum observed stellar mass of $10^{11}M_\odot$.
We chose to focus on early-type galaxies because many strong lensing surveys have preferentially targeted this class of objects in their lens-finding phase \citep[e.g.][]{Bol++06,Gav++12,Son++18a}.
This, in turn, had a dual motivation:
first, early-type galaxies are among the most massive objects in the Universe, and therefore they are more likely to be lenses;
second, their smooth surface brightness distribution and red colour makes it easier to detect arcs from strongly lensed star forming galaxies around them.

We describe lenses with elliptical versions of the two-component model introduced in Sect. \ref{ssec:profile}.
In the following sections we explain how their parameters are generated.

\subsubsection{Stellar mass and redshift distribution}\label{ssub:mstarz}

We generated lens galaxies over a finite redshift range, $0.1 < z < 0.7$.
We chose these lower and upper limits because the value of the critical surface mass density $\Sigma_{\mathrm{cs}}$ becomes very large outside of this range, and the fraction of galaxies that act as strong lenses drops substantially as a result.

We drew stellar masses from the stellar mass function of quiescent galaxies measured by \citet{Muz++13}.
In particular, we used the following comoving number density distribution:
\begin{equation}
\Phi(\mobs) = \Phi^*\left(\frac{\mobs}{\mstar^*}\right)^{\alpha} \exp{\left\{-\frac{\mobs}{\mstar^*}\right\}}.
\end{equation}
We set $\Phi^*=1.009\times10^{-3}\,{\rm Mpc}^{-3}$, $\alpha=-0.92$ and $\log{\mstar^*}=11.21$: these are the best-fit values measured by \citet{Muz++13}.

We refer to the stellar masses drawn from this distribution as the observed stellar masses, $\mobs$, as opposed to the true stellar masses.
The observed stellar mass is meant to represent an estimate of $\mstar$ based on stellar population synthesis modelling, which is the method used by \citet{Muz++13} to measure the galaxy stellar mass function.
Stellar population synthesis measurements, however, are subject to systematic uncertainties, since they have not been calibrated on galaxies whose stellar mass is known by other means.
We quantify the discrepancy between the observed and true stellar mass by means of the stellar population synthesis mismatch parameter $\asps$, defined as follows:
\begin{equation}
\asps \equiv \frac{\mstar}{\mobs}.
\end{equation}
The most important source of systematic uncertainty in the measurement of the stellar mass is the assumption of the stellar initial mass function (IMF). \citet{Muz++13} assumed a Kroupa IMF \citep{Kro01} to derive their measurements. This means that, in the absence of other systematic effects, a galaxy with a Kroupa IMF has a value of $\asps=1$.

For each galaxy in the sample, we randomly drew a value of $\log{\asps}$ from the following distribution:
\begin{equation}
\pr(\log{\asps}) \sim \mathcal{N}(0.1, \sigma_\alpha^2),
\end{equation}
where the notation $\mathcal{N}(\mu,\sigma^2)$ indicates a Gaussian with mean $\mu$ and variance $\sigma^2$.
We set the mean of $\log{\asps}$ to $0.1$, as this is an intermediate value among estimates of $\asps$ from the literature \citep{CvD12, Cap++13, SLC15, Son++15, Son++19}.
We adopted a few different values for the scatter $\sigma_\alpha$. We explain in Sect. \ref{ssub:scat} how these were chosen.

\subsubsection{Stellar mass density distribution}

We described the surface mass density distribution of each galaxy as an elliptical de Vaucouleurs profile (a S\'{e}rsic profile with $n=4$).
Given the observed stellar mass of a galaxy, we assigned a half-mass radius by drawing it from the following distribution in $\log{\reff}$:
\begin{equation}\label{eq:masssize}
\pr(\log{\reff}) \sim \mathcal{N}(1.20 + 0.63(\log{\mobs} - 11.4), 0.14^2).
\end{equation}
Then, we assigned an axis ratio $q$ by drawing it from the following beta distribution:
\begin{equation}\label{eq:qdist}
\pr(q) \propto q^{\alpha-1}(1 - q)^{\beta-1},
\end{equation}
with $\alpha=6.28$ and $\beta=2.05$.
The choice for these distributions was motivated by observations of a sample of early-type galaxies. In Appendix~\ref{sect:appendixa} we explain how this sample was defined and how the coefficients of Eq. \ref{eq:masssize} and Eq. \ref{eq:qdist} were determined. 

\subsubsection{Dark matter distribution}\label{ssub:dmprofile}

We modelled the dark matter distribution of each lens galaxy with an elliptical gNFW halo (see Sect. \ref{ssec:profile} for its definition).
The parameters of the radial density profile were assigned as follows.
Given the stellar mass of a lens, we first determined the value of its halo virial mass, $\mhalo$. We defined this as the mass enclosed within a spherical shell with average density equal to $200$ times the critical density of the Universe.
Using current weak lensing constraints on the halo mass of elliptical galaxies as a reference \citep{Son++22}, we drew halo masses from the following distribution:
\begin{equation}
\pr(\log{\mhalo}) \sim \mathcal{N}(13.0 + 1.0(\log{\mstar} - 11.5), \sigma_h^2),
\end{equation}
which is a Gaussian in $\log{\mhalo}$ with a mean that scales with stellar mass and scatter $\sigma_h$.
The intrinsic scatter in halo mass is not well constrained observationally; therefore, we ran simulations with different values of $\sigma_h$, as for the stellar population synthesis mismatch parameter.

To determine the density profile of the halo we relied on a theoretically motivated model that takes into account the effect of baryons on the dark matter.
We assumed the dark matter distribution to be initially described by an NFW profile with a concentration\footnote{The concentration is the ratio between the halo virial radius and the scale radius, $r_s$.} of five, then used the prescription of \citet{Cau++20} to model the response of the halo to the infall of baryons. 
In the \citet{Cau++20} model the halo response is approximated with an analytical function that depends on the present stellar mass distribution, and typically results in a more concentrated and steeper density profile compared to the original NFW model. 
Finally, we fitted a gNFW profile to the surface mass density of the contracted halo. By doing this, we were able to fully describe the dark matter density profile with three parameters: $\mhalo$, $\gammadm$ and $r_s$.
The dark matter density profile defined in this way is determined uniquely by the halo mass, the stellar mass and the half-light radius (the more concentrated the stellar distribution, the stronger the halo response and the steeper the dark matter density profile).
In principle we could have allowed for additional degrees of freedom, for instance by relaxing the assumption of a fixed initial halo concentration.
In practice, as we explain in Sect. \ref{ssec:limitations}, our main results are not affected by this choice.

Given the radial profile of the dark matter halo, we obtained an elliptical version of it by applying a transformation of the kind of Eq. \ref{eq:elltransf} to its projected surface mass density.
We assumed that the axis ratio and orientation of the halo is the same as that of the stellar component.

\subsubsection{Intrinsic scatter}\label{ssub:scat}

The distribution in stellar mass, halo mass and dark matter inner slope of our sample of simulated foreground galaxies depends on parameters describing the intrinsic scatter in these properties, namely $\sigma_{\mathrm{sps}}$ and $\sigma_h$.
Direct observational constraints on these quantities are poor.
However, we can derive upper limits on them on the basis of observed scaling relations.

Early-type galaxies lie on the stellar mass fundamental plane, a scaling relation between stellar mass, half-light radius and central velocity dispersion \citep{H+B09,deG++21}:
\begin{equation}\label{eq:fpcompact}
\sigma_{\mathrm{e}} \propto \mstar^{(\mathrm{obs})\beta_{\sigma}}\reff^{\xi_{\sigma}}.
\end{equation}
The existence of a fundamental plane relation is a consequence of the virial theorem: the velocity dispersion of a galaxy in dynamical equilibrium is directly related to the three-dimensional mass distribution in its inner regions, which is typically dominated by the stellar component.
At fixed observed stellar mass distribution, however, the central velocity dispersion can vary depending on the stellar population synthesis mismatch parameter, on the dark matter mass and on the dark matter density profile. This gives rise to a spread in the values of the velocity dispersion given $\mobs$ and $\reff$.
Therefore, we can use the observed scatter in velocity dispersion to put an upper limit on the intrinsic scatter in the stellar population synthesis mismatch parameter, halo mass and dark matter slope.

We used Jeans modelling for this purpose.
We generated samples of early-type galaxies with the recipes described above and with different values of $\sigma_{\mathrm{sps}}$ and $\sigma_\mathrm{h}$.
Using the spherical Jeans equation under the assumption of isotropic orbits, we predicted the central velocity dispersion of each galaxy in the sample. Then, we fitted a fundamental plane relation to this mock sample and measured the predicted scatter in velocity dispersion at fixed $\mobs$ and $\reff$.
We then varied $\sigma_{\mathrm{sps}}$ and $\sigma_h$ to match the observed and the predicted scatter.
We did not attempt to match the other parameters of the fundamental plane (i.e. the constant of proportionality of Eq. \ref{eq:fpcompact} and the power-law indices $\beta_{\sigma}$ and $\xi_{\sigma}$), because these are sensitive to the orbital anisotropy of the galaxies, which we are asserting to be zero.
The details of this procedure are given in Appendix~\ref{sect:appendixb}.

We settled on three different sets of intrinsic scatter parameters, as indicated in \Tref{tab:scatter}. We label them the fiducial, the low-scatter and the high-scatter scenarios.
The high-scatter scenario is ruled out both by the fundamental plane and by weak lensing constraints \citep{Son++22}. 
Moreover, our dynamical model is quite simplistic, as it neglects the effects of orbital anisotropy and departures from spherical symmetry, which are additional sources of scatter. 
Nevertheless, we use it in our experiment in order to obtain a more conservative upper limit on the amplitude of the strong lensing bias.
\begin{table}
\caption{Intrinsic scatter scenarios.}
\label{tab:scatter}
\begin{tabular}{lccc}
\hline
\hline
Model name & $\sigma_{\mathrm{sps}}$ & $\sigma_{\mathrm{h}}$ & Predicted FP \\
 & & & scatter \\
\hline
Fiducial & 0.08 & 0.20 & 0.034 \\
Low scatter & 0.05 & 0.10 & 0.021 \\
High scatter & 0.10 & 0.30 & 0.043 \\
\end{tabular}
\tablefoot{
Adopted values of the intrinsic scatter parameters $\sigma_{\mathrm{sps}}$ and $\sigma_{\mathrm{h}}$ in different simulations. For each set of values, the fourth column indicates the scatter around the fundamental plane predicted via the spherical Jeans analysis of Appendix~\ref{sect:appendixb}.
The observed fundamental plane scatter is $0.035$.
}
\end{table}

\subsection{Background sources}\label{ssec:sources}

\subsubsection{Galaxies}\label{ssub:extsources}

Our background galaxy population is taken from the \textsc{surfs}-based KiDS-Legacy-Like Simulation (SKiLLS) input catalogue, a hybrid simulation catalogue integrating cosmological simulation with high-quality imaging observations \citep{Li++23}. The cosmological simulation is obtained from the Synthetic UniveRses For Surveys (\textsc{surfs}) simulations, a set of $N$-body simulations from \citet{Ela++18}. The galaxy properties, including the star formation history and the metallicity history, are from an open-source semi-analytic model named \textsc{Shark}\footnote{\url{https://github.com/ICRAR/shark}}~\citep{Lag++18}. The original photometry is drawn from a stellar population synthesis technique using stellar synthesis libraries with physically motivated dust attenuation and re-emission models~\citep{Rob++20}. \citet{Li++23} further applied an empirical correction to the original synthetic photometry to better agree with the COSMOS2015 observations~\citep{Lai++16}. The galaxy morphology is described by a S\'ersic profile with three parameters: the half-light radius in angular units $\theta_{
\mathrm{e,s}}$, the S\'ersic index $n_\mathrm{s}$, and the axis ratio $q_\mathrm{s}$. These structural parameters are learned from the imaging data obtained with the Advanced Camera for Surveys (ACS) instrument on the \textit{Hubble} Space Telescope~\citep{Gri++12}. We refer to \citet{Li++23} for details on the learning algorithm and validation.

The complete SKiLLS catalogue contains ${\sim}108~{\rm deg}^2$ of galaxies with redshift up to $2.5$ and $r$-band apparent magnitude down to $27$.
We applied a lower limit to the source redshift, by selecting only sources with $z_s > 0.8$.
This ensures that all of the sources lie behind all of the lenses.
A similar cut could be applied to a real survey using photometric redshifts, to reduce the incidence of false positives (e.g. arc-like features physically associated with the lens galaxy) in the lens finding phase.
The resulting number density of sources is $70$~arcmin$^{-2}$.
We approximated their spatial distribution as uniform in the sky, that is, we neglected clustering of the sources.

\subsubsection{Quasars}\label{ssub:quasars}

We described the population of background quasars with the following double power-law luminosity function in the rest-frame UV absolute magnitude, $M$:
\begin{equation}
\Phi(M,\zqso) = \dfrac{\Phi(M^*)}{10^{0.4(\alpha_Q+1)(M - M^*)} + 10^{0.4(\beta_Q+1)(M - M^*)}}.
\end{equation}
Following \citet{Man++17}, we set $\alpha_Q=-1.35$ (faint-end slope), $\beta_Q=-3.23$ (bright-end slope), and adopted a redshift-evolving normalisation
\begin{equation}
\log{\Phi^*} = -6.0991 + 0.0209\zqso + 0.0171\zqso^2
\end{equation}
and characteristic magnitude
\begin{equation}
M^* = -22.5216 - 1.6510\zqso + 0.2869\zqso^2.
\end{equation}
Given the redshift and rest-frame UV luminosity of a quasar, we then computed the apparent magnitude in the observed $i-$band, $\mqso$, using a quasar spectral template\footnote{\url{https://archive.stsci.edu/hlsps/reference-atlases/cdbs/grid/comp_qso/}} built from optical and near-infrared spectra obtained by \citet{vdB++01,Gli++06}.

For the sake of consistency with the population of background extended sources, we limited the redshift distribution of quasars to the range $0.8 < \zqso < 2.5$.
We then truncated the distribution in $\mqso$ at two magnitudes fainter than the detection limit (which is specified in Sect. \ref{ssec:obs}).
Finally, we randomly placed quasars in the source plane with a projected number density of $70$~arcmin$^{-2}$. This is a much larger number density than observed in the real universe, but we are allowed to do so because we are not interested in predicting the absolute number of strong lenses, so this is a legitimate choice. The advantage of boosting the number density of quasars is that it allows us to produce a large number of lenses without the need for generating too big a population of foreground galaxies.

\subsection{Observations}\label{ssec:obs}

For each of the three intrinsic scatter scenarios, we drew a population of foreground galaxies covering \num{1000} square degrees.
The expectation value of the number of galaxies in the corresponding volume, given the redshift and stellar mass cuts described in Sect. \ref{ssub:mstarz}, is around \num{300000}.
We approximated the foreground galaxies as isolated: when determining whether a galaxy acts as a strong lens, we only modelled the contribution to the lensing signal from the galaxy itself, and neglected that of the environment.
We discuss the possible implications of this approximation in Sect. \ref{sect:discuss}.
For each lens, we determined its caustics relative to the highest source redshift, $\zsource=2.5$. We then placed sources randomly behind the lens. If at least one source fell within a circular region enclosing the caustics, we proceeded to compute its lensed images.

For the simulation with extended sources, we produced images with properties similar to those expected for the \textit{Euclid} Wide survey \citep{Sca++22}.
We used a pixel size of $0.1''$ and we applied a Moffat PSF with an FWHM of $0.2''$ and a $\beta$ parameter of $5.0$.
Finally, we assumed a background noise level such that an extended source with half-light radius $0.5''$ and an apparent magnitude in the absence of lensing of $\msource=25$ is detected with $S/N=10$.
We then applied the peak detection-based lens selection criterion introduced in Sect. \ref{ssec:lensdefext} to find the strong lenses.

The fiducial scatter simulation with extended background sources produced a sample of $2113$ lenses, corresponding to a number density of $2.1$~deg$^{-2}$. This is about a factor of five smaller than the number density predicted by \citet{Col15} for the \textit{Euclid} survey. This is a result of differences in the description of the source population, in the criteria used to define a strong lensing event, and in the redshift and stellar mass cuts that we applied to define the foreground galaxy population.

For the lensed quasars we did not simulate pixel-level data, but simply computed the observed magnitudes of the multiple images.
Following the definition of Sect. \ref{ssec:lensdefpoint}, we included in the sample of strong lenses only systems with at least two images brighter than a limiting magnitude $m_{\mathrm{lim}}$.
We set $m_{\mathrm{lim}}=23.3$, which corresponds to the $10\sigma$ detection limit of the Legacy Survey of Space and Time (LSST) in a single visit \citep{O+M10}. This is motivated by the fact that the quasar lenses are meant to simulate a sample assembled for the purpose of carrying out time-delay measurements, which in turn require combining single-visit detections over many epochs.
The scenario that we are simulating, then, is that of a lens search in a \textit{Euclid}-like survey, followed-up with LSST time-domain observations.

The resulting number of quasar lenses in our simulation with the fiducial scatter is $1621$. This number is meaningless, given that the simulation was created with an unrealistically large number density of quasars.
More interesting is the relative number of quad lenses with respect to the total, which is about $9\%$. This is a slightly smaller value than the fraction of quads predicted by \citet{O+M10}.
The reason for this discrepancy lies in the differences between the lens mass models in the two simulations.

\subsection{Lens finding probability}\label{ssec:teincuts}

\begin{figure}
\includegraphics[width=\columnwidth]{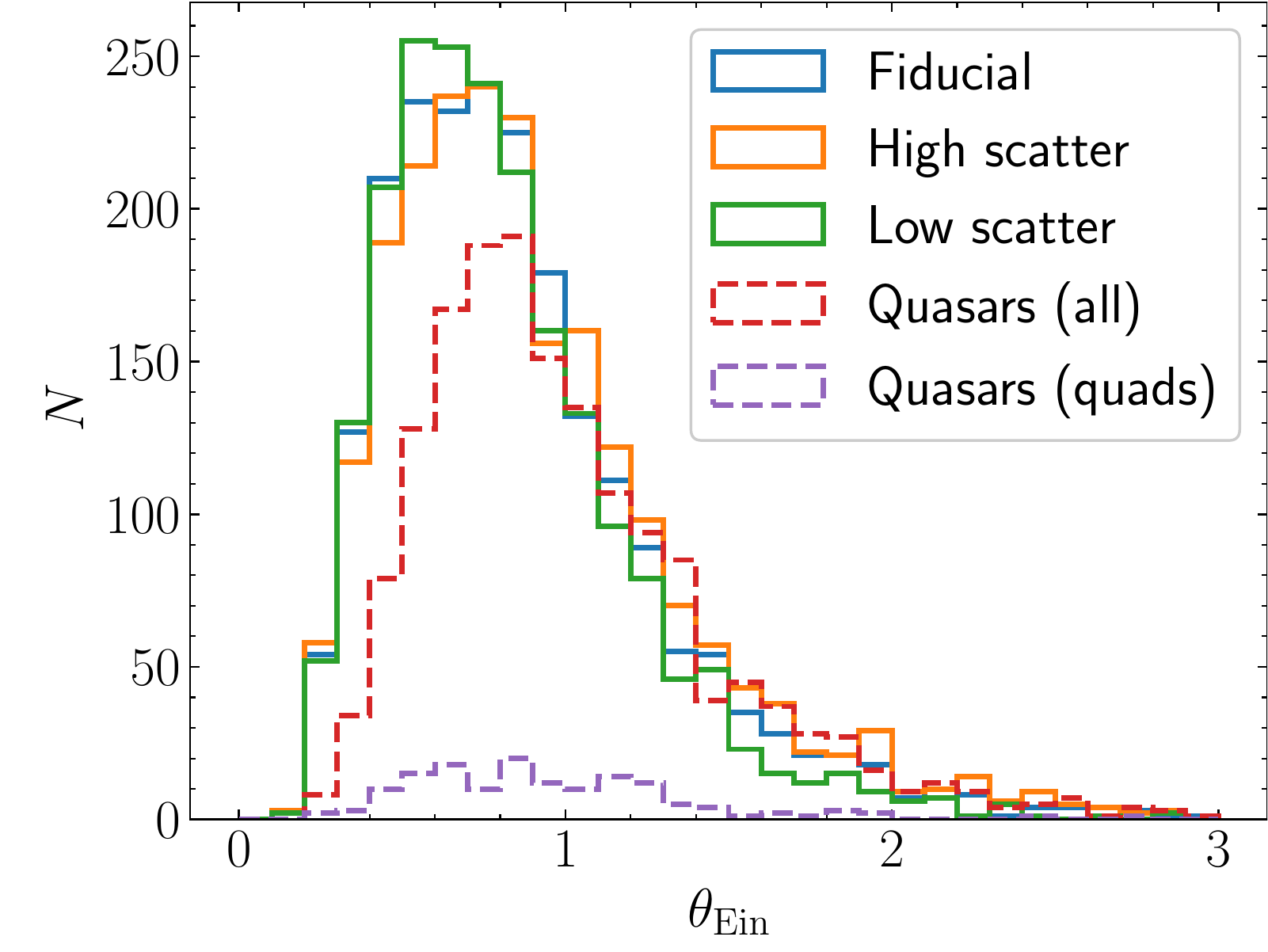}
\caption{
Einstein radius distribution of the simulated lens samples.
These are: galaxy-galaxy lenses in the fiducial, high scatter, and low scatter scenarios and galaxy-quasar lenses with fiducial scatter, considering all lenses or only lenses that produce four images.
\label{fig:teinhist}
}
\end{figure}

Figure \ref{fig:teinhist} shows the distribution in Einstein radius of the simulated lens samples.
In all cases, the distribution peaks at $\tein\approx0.7''$.
Although all of the lenses in these samples are detected, it does not necessarily follow that they would all be included in a strong lensing study.
There can be a few reasons for excluding certain lenses from a sample. 
One is the low accuracy of lens finders: current automated lens finding algorithms tend to produce lens candidates samples with low purity \citep[see e.g.][]{Son++18a,Pet++19,Sav++22}. Such samples are then visually inspected, and only those candidates that can be clearly distinguished from false-positives are kept. This visual inspection step tends to disfavour lenses with a small image separation, because of the contamination from the light of the lens galaxy.

Another possible reason for refining a sample of lens candidates is the availability of redshift measurements. Redshifts of both the lens and the source are needed in order to convert a lens model into a measurement of mass. When working with large samples of lenses, obtaining spectroscopic measurements is not a viable option, and photometric redshifts are a necessity. Measuring the photometric redshift of a strongly lensed source, however, is challenging, especially when the Einstein radius is small and the source light is blended with the light from the lens \citep{Lan++23}.

Both of these scenarios can result in samples that are incomplete below a certain value of the Einstein radius. 
We simulate this situation via the following Einstein radius-dependent lens finding probability:
\begin{equation}
\pfind(\tein|S,\mathrm{det}) = \left\{\begin{array}{ll} 1 & \rm{if}\,\tein > \teinmin \\
0 & \rm{otherwise}\end{array}\right. .
\end{equation}
In words, all lenses with Einstein radii larger than $\theta_{\mathrm{Ein,min}}$ are included in the sample, while all those with smaller Einstein radius are excluded.
We refer to $\teinmin$ as the completeness limit: our simulated lens samples are complete down to $\tein=\teinmin$.
We explored scenarios with different values of $\teinmin$.
As the next section shows, the larger the minimum Einstein radius, the higher the strong lensing bias.


\section{Results}\label{sect:results}

In this section we present the results of the lens population simulations. 
Section \ref{ssec:galbias} shows the results of the galaxy-galaxy lens experiment, while Sect. \ref{ssec:quasarbias} focuses on the population of galaxy-quasar lenses.
Given the number of parameters that are needed to describe our model, providing a complete characterisation of the strong lensing bias is a problem with relatively high dimensionality, and is beyond the scope of this paper.
For the sake of conciseness, we focus instead on the quantities that we consider most important.
Nevertheless, the output of our simulations is available online\footnote{\url{https://github.com/astrosonnen/strong_lensing_tools/tree/main/papers/selection_effects}}. We encourage readers who are interested in studying aspects of the strong lensing bias that are not covered in this section to download our data and analyse them directly.

\subsection{Galaxy-galaxy lenses}\label{ssec:galbias}

In this section we show the results of the experiments with populations of galaxy-galaxy lenses.
We first present the results in a qualitative way, then proceed to quantify the amplitude of the strong lensing bias in various quantities of interest.

\begin{figure*}
\includegraphics[width=\textwidth]{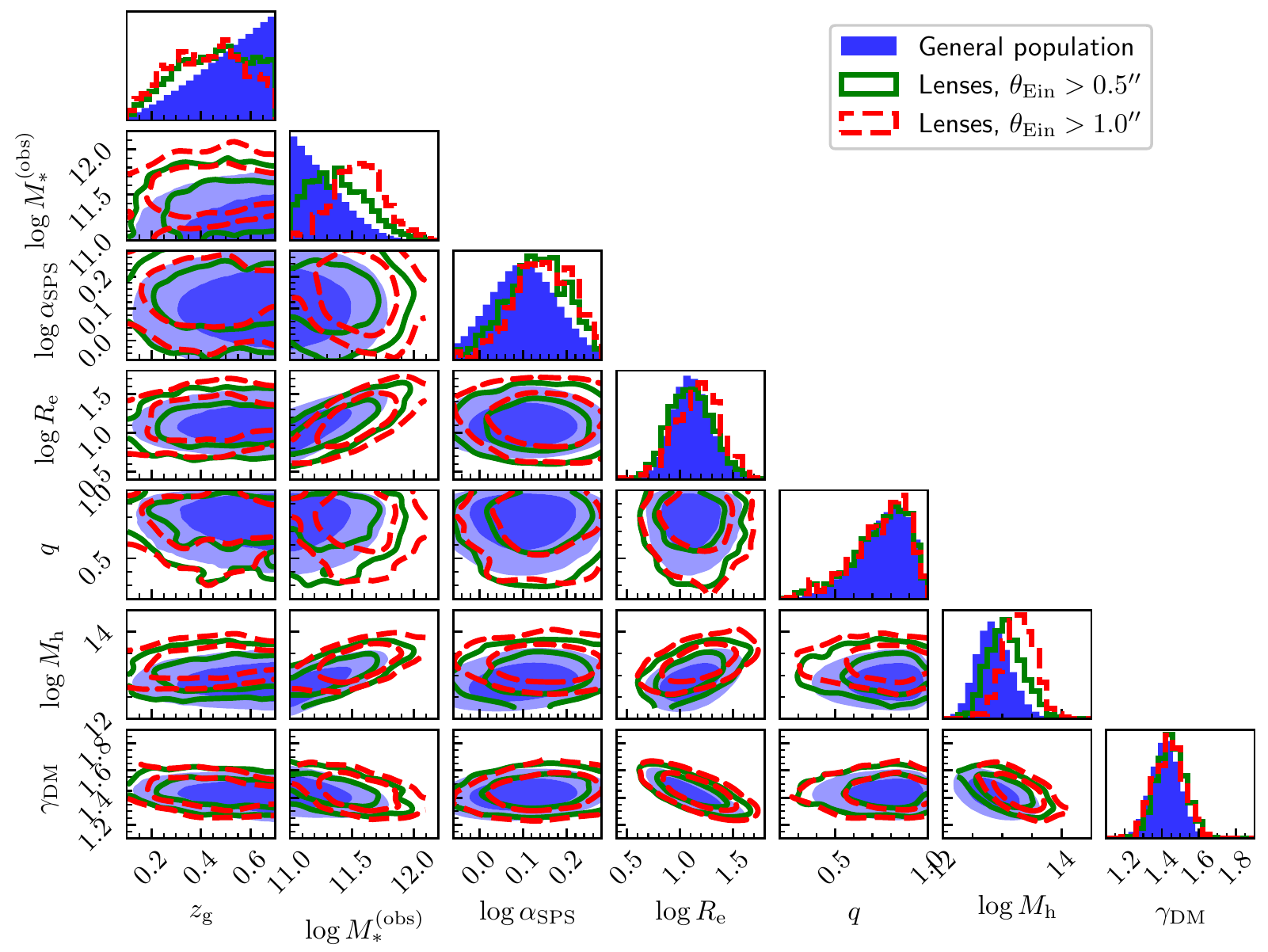}
\caption{
Comparison between the properties of lens samples and the parent population: distribution in foreground galaxy parameters.
Filled contours represent the distribution of the parent sample, solid green lines the distribution of the lenses with Einstein radii larger than $0.5''$, and solid red lines the distribution of the lenses with Einstein radii larger than $1.0''$.
\label{fig:lenspars}
}
\end{figure*}

Figure \ref{fig:lenspars} shows the distribution in the parameters of the foreground galaxies of the lens systems, compared to those of the parent population, for the fiducial scatter scenario with two different values of the minimum Einstein radius: $0.5''$ and $1.0''$.
A completeness limit of $0.5''$ is close to what can currently be achieved via visual inspection of high-resolution space-based images \citep{Gar++22}, while the $1.0''$ limit can be seen as a more conservative case.

The most striking difference between the samples is in the stellar mass: strong lensing selects preferentially galaxies with larger values of $\mobs$.
Lenses tend to also have a larger halo mass, a smaller half-light radius at fixed stellar mass, and a larger stellar population synthesis mismatch parameter. The distribution in ellipticity and inner dark matter slope of the lenses instead look very similar to that of the parent population.
Additionally, we can see that the amplitude of the strong lensing bias appears to be always larger in the lens sample with the more restrictive selection on Einstein radius.
We quantify the amplitude of these biases later in this section.

\begin{figure*}
\includegraphics[width=\textwidth]{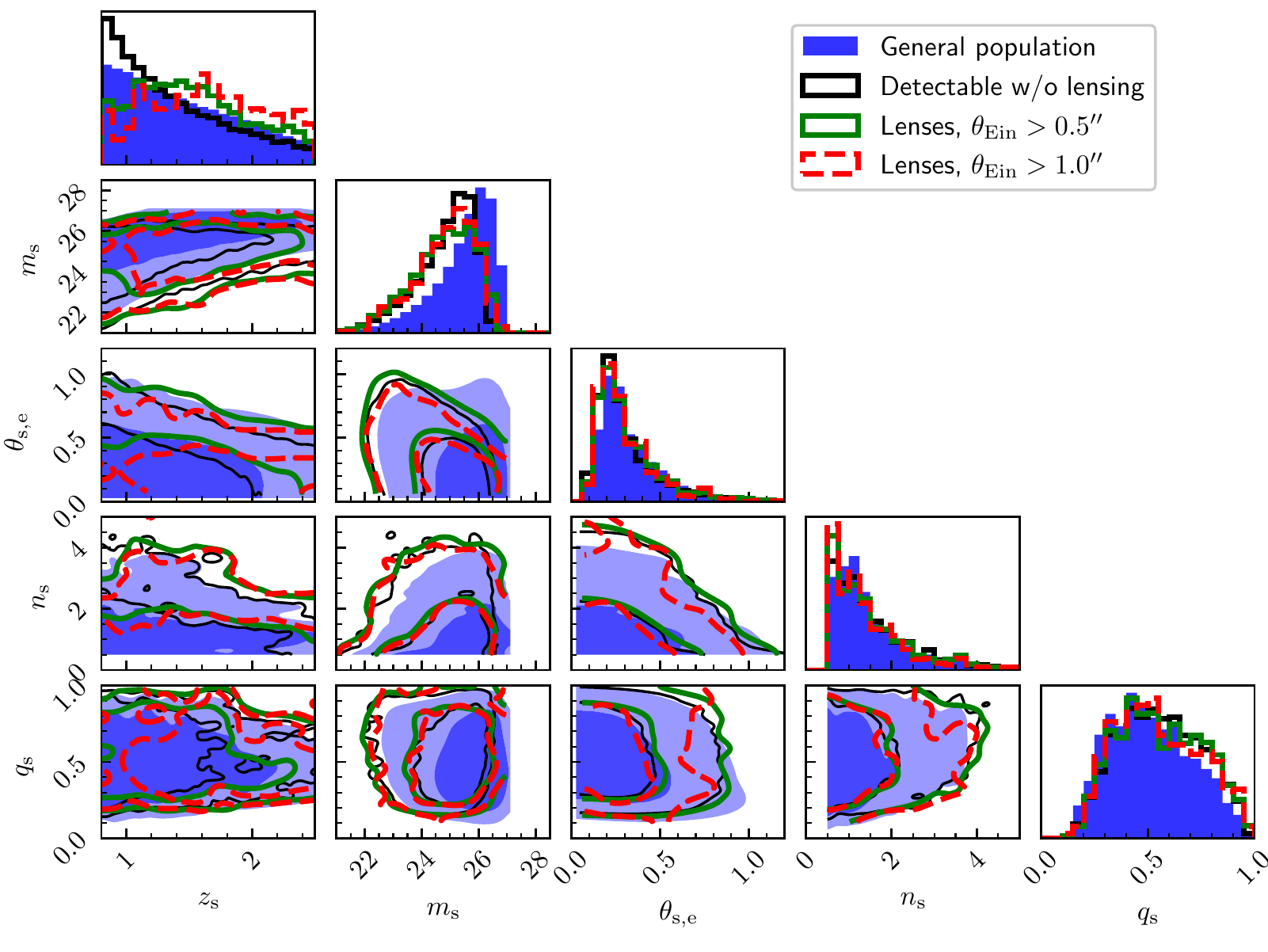}
\caption{
Comparison between the properties of lens samples and the parent population: distribution in background source parameters.
Filled contours show the distribution of the parent sample, solid black lines the distribution of the detectable sources, solid green lines the distribution of the lenses with Einstein radii larger than $0.5''$, and solid red lines the distribution of the lenses with Einstein radii larger than $1.0''$.
\label{fig:sourcepars}
}
\end{figure*}

Figure \ref{fig:sourcepars} shows the distribution in parameters describing the background source population, for the same simulations of Fig. \ref{fig:lenspars}. 
Additionally, Fig. \ref{fig:sourcepars} shows the subset of the parent population that consists of detectable sources.
These are background galaxies that, in the absence of lensing, can be detected according to the same criterion used for the lensed sources (i.e. the $S/N$ over their $2\sigma$ footprint is larger than ten).
The detection limit of the survey is at $\msource\approx25$ (the actual limit varies depending on the surface brightness distribution parameters).
Because our simulated background source population extends to much fainter magnitudes, the distribution of detectable sources differs substantially from that of the parent population.
We can then consider two different strong lensing bias definitions: one that quantifies the difference in lensed source properties with respect to the parent population, and one that describes the difference with respect to the detectable source population. We are mostly interested in the second definition.

Strong lensing tends to preferentially select sources at higher redshift, especially in the more restrictive case with $\tein > 1.0''$.
This is because, at fixed lens properties, increasing the source redshift lowers the critical surface mass density and, consequently, increases the size of the caustics and the Einstein radius.
The distribution in the magnitude of the lensed sources is also very different from that of the parent distribution, as it drops rapidly for values larger than $\msource\approx25$.
Interestingly, however, there does not seem to be a large difference with respect to the distribution of detectable sources in the absence of lensing.
This result can appear to be somewhat counter-intuitive: lensing magnification should allow the detection of sources that are intrinsically fainter than the detection limit. To some extent, this is the case: the distribution of lensed sources shows a slight excess of fainter galaxies compared to the unlensed case. However, as we quantify later in this section, the difference is far from large.
The reason for this behaviour lies in the fact that, in the detection of both lensed and unlensed sources, the most important quantity is surface brightness, which is preserved by gravitational lensing when the source is larger than the PSF size.
As we showed in Sect. \ref{ssec:ellext}, the lensing cross-section drops to zero once the surface brightness of the source reaches a value that would make it undetectable in the absence of lensing.
For this reason, also the distribution in half-light radius is very similar between the lensed sources and the detectable source population.

Qualitatively, the results shown in Figs. \ref{fig:lenspars} and \ref{fig:sourcepars} match our expectations from Sect. \ref{sect:indlenses}.
In the rest of this section we quantify the lensing bias.
We present the results in three different parts.
First, we focus on the properties of the lens galaxies that can be observed directly. These are quantities that can be derived from photometry and spectroscopy with minimal assumptions: the lens redshift, the observed stellar mass, the half-light radius and the axis ratio\footnote{Strictly speaking, the axis ratio in mass is not necessarily observable, but in the context of our simulations it is, since light and mass have the same ellipticity.}.
Any bias in these quantities can be determined relatively easily in a real strong lens survey.
In the second part, we consider the parameters related to the mass distribution: the stellar population synthesis mismatch parameter and the dark matter distribution parameters. Determining the bias on these parameters is much more difficult, but these are quantities of great interest from a galaxy science point of view.
In the third part, we focus on source distribution parameters. 

\subsubsection{Bias in observable lens parameters}

\begin{figure}
\includegraphics[width=\columnwidth]{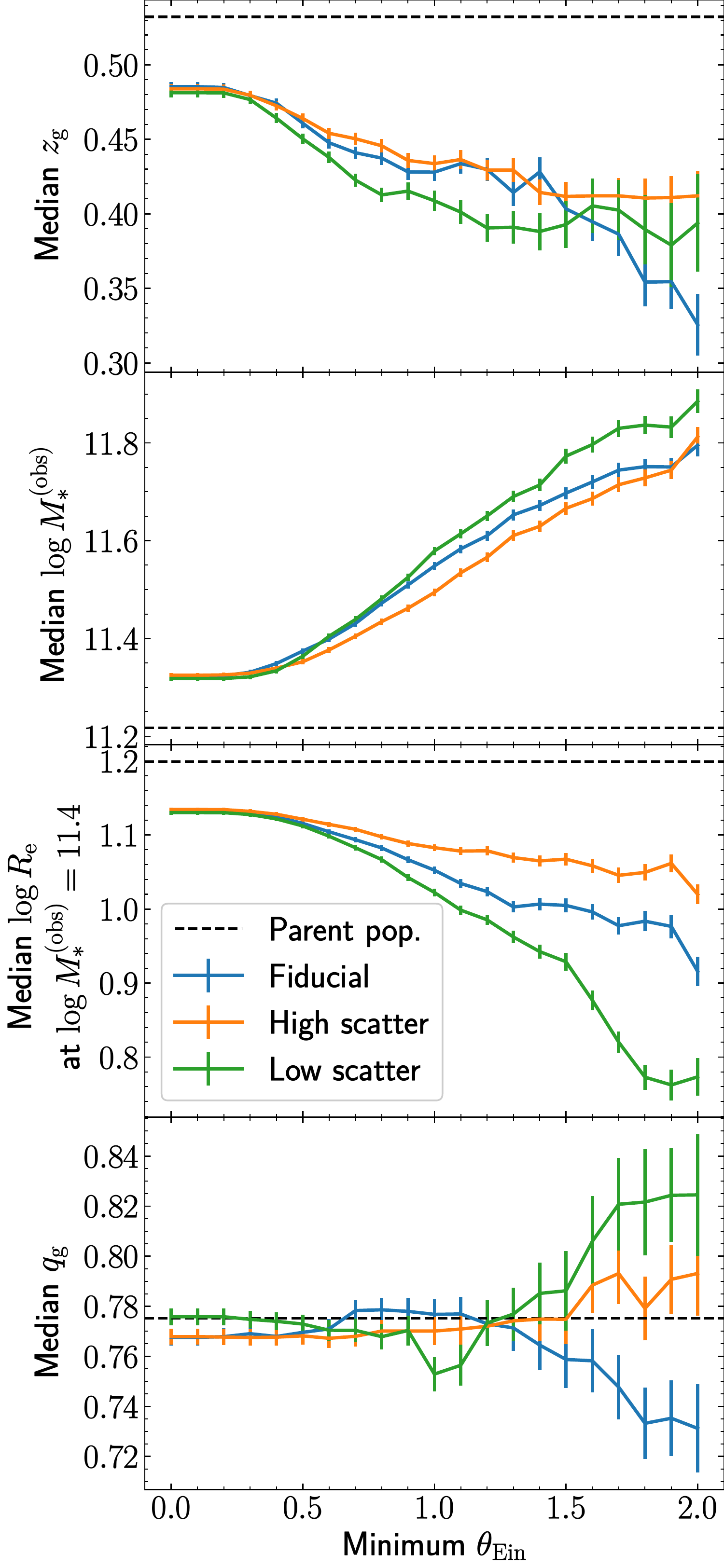}
\caption{
Bias on lens observable properties as a function of the minimum Einstein radius.
First panel: Median redshift.
Second panel: Median observed stellar mass.
Third panel: Median half-light radius at an observed mass of $\log{\mobs}=11.4$.
Fourth panel: Median axis ratio.
In each panel, the dashed black line indicates the value of the parent population.
Error bars indicate the standard deviation of the mean of the lens sample.
\label{fig:lensobsbias}
}
\end{figure}

Figure \ref{fig:lensobsbias} shows the median redshift, median $\mobs$, median size for a given $\mobs$, and median axis ratio of various lens population simulations, as a function of the minimum Einstein radius.
We defined the median size at fixed $\mobs$ by fitting the following mass-size relation to the $\reff-\mobs$ distribution:
\begin{equation}
\log{\reff} \sim \mu_{R,0} + \beta_R(\log{\mobs} - 11.4).
\end{equation}
The quantity shown in the third panel of Fig. \ref{fig:lensobsbias} is the parameter $\mu_{R,0}$, which is the average $\log{\reff}$ at an observed stellar mass of $\log{\mobs}=11.4$.
The horizontal dashed line in each panel shows the value of the parent population: the larger the distance between the curve of a simulation and this line, the higher the strong lensing bias.

As was already visible in Fig. \ref{fig:lenspars}, there is a clear bias towards lower redshift, higher stellar mass and smaller sizes, with the bias becoming stronger for more restrictive cuts in Einstein radius.
Biases in stellar mass and size are stronger for the simulation with low scatter in $\asps$ and $\mhalo$.
This can be explained as follows.
For a given value of the observed stellar mass, strong lensing selection favours galaxies with a larger $\asps$ or $\mhalo$, or with a smaller half-light radius.
Since the galaxy stellar mass function of the parent population is steep, lenses tend to have a relatively small $\mobs$ and large values of $\asps$ or $\mhalo$ for their observed stellar mass (this is shown more clearly in the next section).
If the intrinsic scatter in $\asps$ and $\mhalo$ is low, however, the number of galaxies with a small $\mobs$ and a large stellar or halo mass is greatly reduced.
Only galaxies with a large $\mobs$ or a small size can therefore act as strong lenses.
This is an interesting result, because it suggests that it is in principle possible to use the strong lensing bias on $\mobs$ or $\reff$, which is observable, as a way to constrain the amplitude of the intrinsic scatter in the mass parameters, which is poorly known.

\subsubsection{Bias in lens mass parameters}

\begin{figure}
\includegraphics[width=\columnwidth]{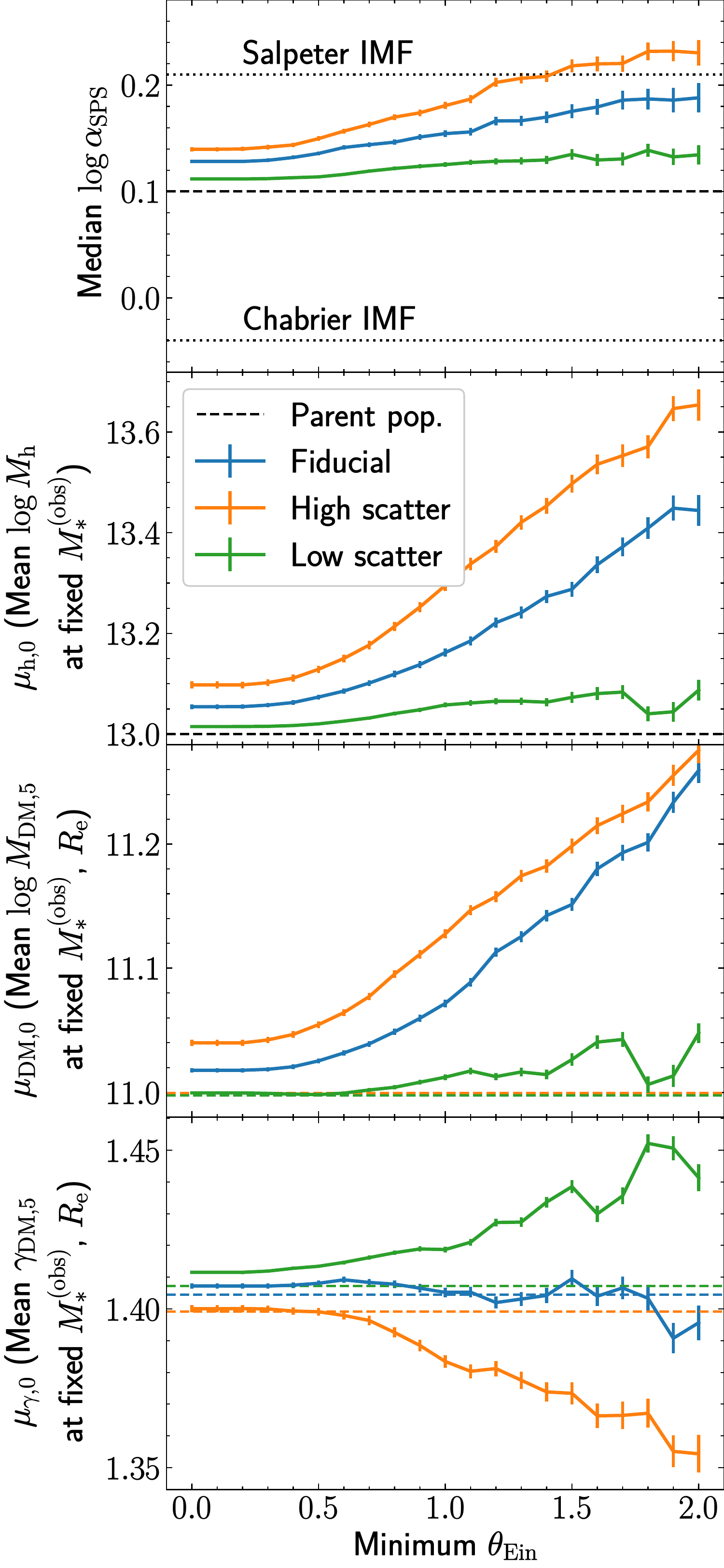}
\caption{
Bias on lens mass properties as a function of the minimum Einstein radius.
First panel: Median $\log{\asps}$. Dotted lines indicate values of $\asps$ corresponding to a Chabrier and a Salpeter IMF ($\asps=1$ corresponds to a Kroupa IMF).
Second panel: Median $\mhalo$ at $\log{\mobs}=11.4$.
Third panel: Median $\mdmfive$ at $\log{\mobs}=11.4$ and $\log{\reff}=1.2$.
Fourth panel: Median $\gammadm$ at $\log{\mobs}=11.4$ and $\log{\reff}=1.2$.
In each panel, the dashed line indicates the value of the parent population.
Error bars indicate the standard deviation of the mean of the lens sample.
\label{fig:lensmassbias}
}
\end{figure}

Figure \ref{fig:lensmassbias} shows the median of the distribution in various mass-related quantities, as a function of the minimum Einstein radius.
The top panel is the median stellar population synthesis mismatch parameter.
In all simulations, strong lenses are biased towards larger values than the parent population. The bias is larger the higher the intrinsic scatter in $\asps$ and $\mhalo$, and increases with increasing $\theta_{\mathrm{Ein,min}}$.
For the fiducial model, the bias on $\asps$ can be as small as $0.03$~dex (7\%), if no cut on Einstein radius is applied.
However, it can rise up to $0.09$~dex (23\%) when considering only lenses with $\tein > 2''$.
For comparison, we also show two reference values of $\asps$, corresponding to a Chabrier IMF \citep{Cha03} and a Salpeter IMF \citep{Sal55}.
These values roughly bracket the current systematic uncertainty on $\asps$.

The second panel of Fig. \ref{fig:lensmassbias} shows the median halo mass of galaxies with an observed stellar mass of $\log{\mobs}=11.4$.
We measured this quantity by fitting the following relation to the $\mhalo-\mobs$ distribution,
\begin{equation}
\log{\mhalo} \sim \mu_{\mathrm{h},0} + \beta_{\mathrm{h}}(\log{\mobs} - 11.4),
\end{equation}
and taking the resulting value of the parameter $\mu_{\mathrm{h},0}$.
Similarly to the $\asps$ case, the strong lenses are biased towards larger values of the halo mass, with the bias being larger for higher-scatter simulations and more restrictive cuts on $\tein$.
In the fiducial scatter scenario with a completeness limit of $\theta_{\mathrm{Ein,min}}=1.0''$, the halo masses of lenses are on average $0.16$~dex larger than those of their parent population.

The third panel of Fig. \ref{fig:lensmassbias} shows the median projected dark matter mass enclosed within an aperture of $5$~kpc, $\mdmfive$, at fixed observed stellar mass and half-light radius.
We obtained this quantity by fitting the following relation to the $\mdmfive-\mobs-\reff$ distribution:
\begin{equation}\label{eq:dmscaling}
\log{\mdmfive} \sim \mu_{\mathrm{DM},0} + \beta_{\mathrm{DM}}(\log{\mobs} - 11.4) + \xi_{\mathrm{DM}}(\log{\reff} - 1.2).
\end{equation}
Figure \ref{fig:lensmassbias} shows the value of $\mu_{\mathrm{DM},0}$.
The bias on this quantity is qualitatively similar to that on the total halo mass, but much smaller in amplitude.
There are two reasons for this.
First, the dependence of $\mdmfive$ on $\mhalo$ is shallower than linear. This is because, as the virial mass increases, the virial radius increases as well: the extra mass is spread over a larger volume, and therefore the mass within the inner region does not increase proportionally.
Second, while $\mu_{\mathrm{h},0}$ is the halo mass at fixed stellar mass, $\mu_{\mathrm{DM},0}$ is measured at fixed half-light radius as well.
The density profile of the dark matter halos in our simulation have a dependence on galaxy size: the response of dark matter to baryons is stronger for more concentrated stellar distributions (see Sect. \ref{ssub:dmprofile}).
By capturing this dependence in the model of Eq. \ref{eq:dmscaling}, the residual scatter in $\mdmfive$ around the mean is reduced, and so is the strong lensing bias.
This, however, is a minor effect: we repeated the analysis while setting the dependence on size to zero and found minimal differences on the derived values of $\mu_{\mathrm{DM},0}$.
In the fiducial scenario, the bias on $\mdmfive$ is as small as $0.02$~dex for $\theta_{\mathrm{Ein,min}} < 0.5''$, and is negligible in the low-scatter scenario.

In the fourth panel of Fig. \ref{fig:lensmassbias} we show the average inner dark matter slope at fixed observed stellar mass and half-light radius, which we measured by fitting the following model
\begin{equation}\label{eq:dmscaling}
\gammadm \sim \mu_{\gamma,0} + \beta_{\gamma}(\log{\mobs} - 11.4) + \xi_{\gamma}(\log{\reff} - 1.2),
\end{equation}
and taking the resulting value of $\mu_{\gamma,0}$.
Different simulations show different trends in the average $\gammadm$.
These are the results of underlying correlations between the dark matter density profile and the stellar and halo mass.
In our model, galaxies of a given size with a larger stellar mass have a steeper dark matter slope, because the response of the dark matter to the infall of baryons is stronger. Vice versa, galaxies with a larger halo mass have a shallower density slope.
At fixed $\mobs$ and $\reff$, strong lenses have both a larger stellar mass (because their $\asps$ is larger) and a larger halo mass.
In the simulation with low scatter, the bias in stellar mass is more important; therefore, $\gammadm$ has a positive bias. In the simulation with large scatter, the bias in halo mass dominates, and therefore $\gammadm$ is negatively biased.

\subsubsection{Bias in source parameters}

\begin{figure}
\includegraphics[width=\columnwidth]{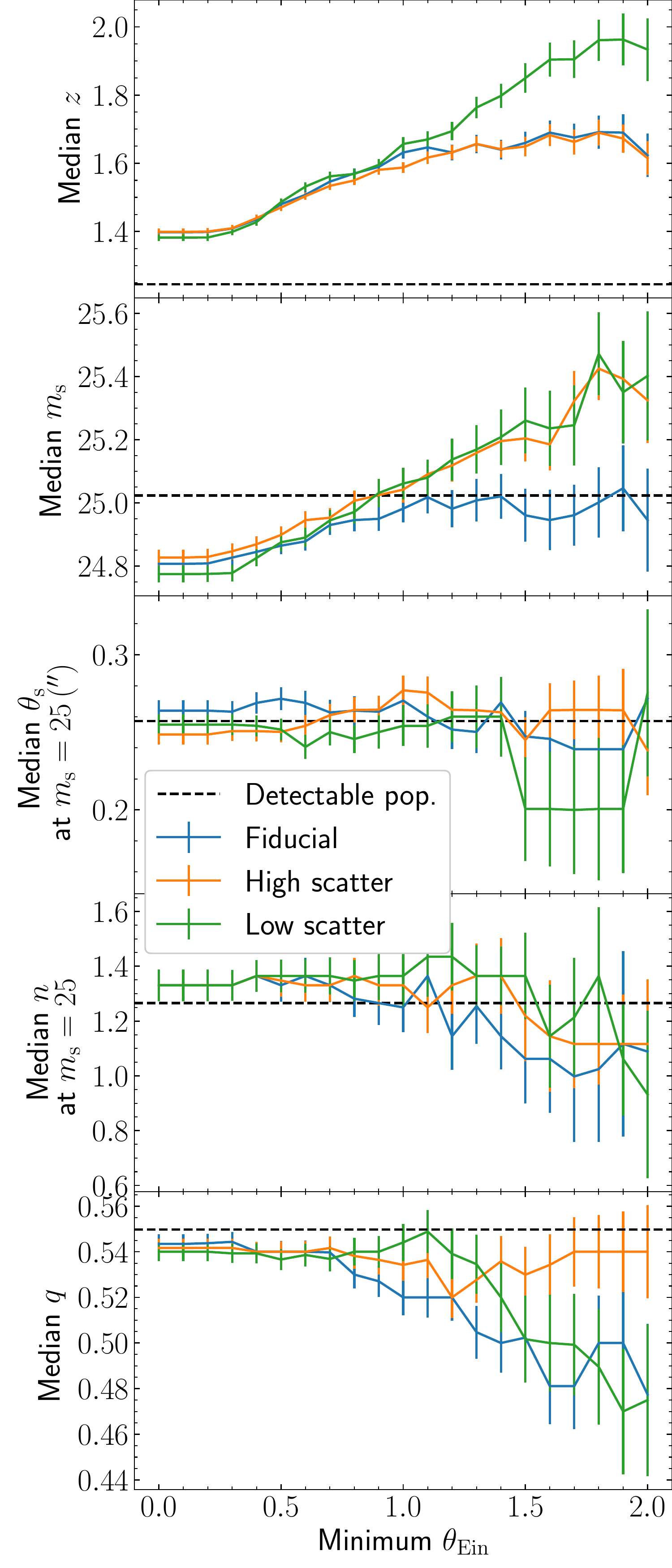}
\caption{
Bias on source properties as a function of the minimum Einstein radius.
First panel: Median source redshift.
Second panel: Median source magnitude.
Third panel: Median half-light radius in the magnitude bin $24.8 < \msource < 25.2$.
Fourth panel: Median S\'{e}rsic index in the magnitude bin $24.8 < \msource < 25.2$.
Fifth panel: Median axis ratio.
In each panel, the dashed line indicates the value of the population of detectable sources.
Error bars indicate the standard deviation of the mean of either the full sample (first, second, and fifth panel) or the bin (third and fourth panel).
\label{fig:sourcebias}
}
\end{figure}

Figure \ref{fig:sourcebias} shows the lensing bias in source-related parameters, with respect to the population of sources that are detectable without lensing.
As previously seen in Fig. \ref{fig:sourcepars}, lensed sources are biased towards higher redshift (top panel of Fig. \ref{fig:sourcebias}), with the trend being larger for more restrictive cuts on the Einstein radius.
The second panel shows the bias in the source magnitude. This bias is negative for small values of $\theta_{\mathrm{Ein,min}}$, that is, lensed sources tend to be brighter than their field counterparts.
Naively one might have expected the opposite trend, as strong lensing magnification allows the detection of sources that are intrinsically fainter than the detection limit.
However, as the analysis of Sect. \ref{sect:indlenses} clearly shows, the lensing cross-section is always larger for brighter sources, and that explains the sign of the bias.
At the same time, the median does not capture the whole picture of the bias in $\msource$. For instance, in the fiducial simulation with no cut on the Einstein radius, the $90$th percentile of the $\msource$ distribution is $26.05$, while that of the population of detectable sources is $0.14$~mag brighter: indeed, strong lensing allows the detection of fainter sources.

The third and fourth panels of Fig. \ref{fig:sourcepars} show the bias in half-light radius and S\'{e}rsic index, respectively, for sources within a magnitude bin centred on $\msource=25$ and with a $0.4$~mag width.
We detected no clear sign of bias.
Finally, the fifth panel shows the bias in source axis ratio. Also in this, case no obvious sign of bias was detected, except in the largest values of $\theta_{\mathrm{Ein,min}}$.

\subsection{Galaxy-quasar lenses}\label{ssec:quasarbias}

\subsubsection{Bias in lens parameters}

The main goal of the experiment with lensed quasars is to check whether there are any differences in the strong lensing bias with respect to the extended source case, at fixed properties of the foreground galaxy population.
For this reason, we only ran simulations with lensed quasars in the fiducial scatter scenario and compared the results with those from the extended source simulation.
We are also interested in understanding how the subset of quad lenses differs from the entire population of quasar lenses; therefore, we also analysed that subsample on its own.

\begin{figure}
\includegraphics[width=\columnwidth]{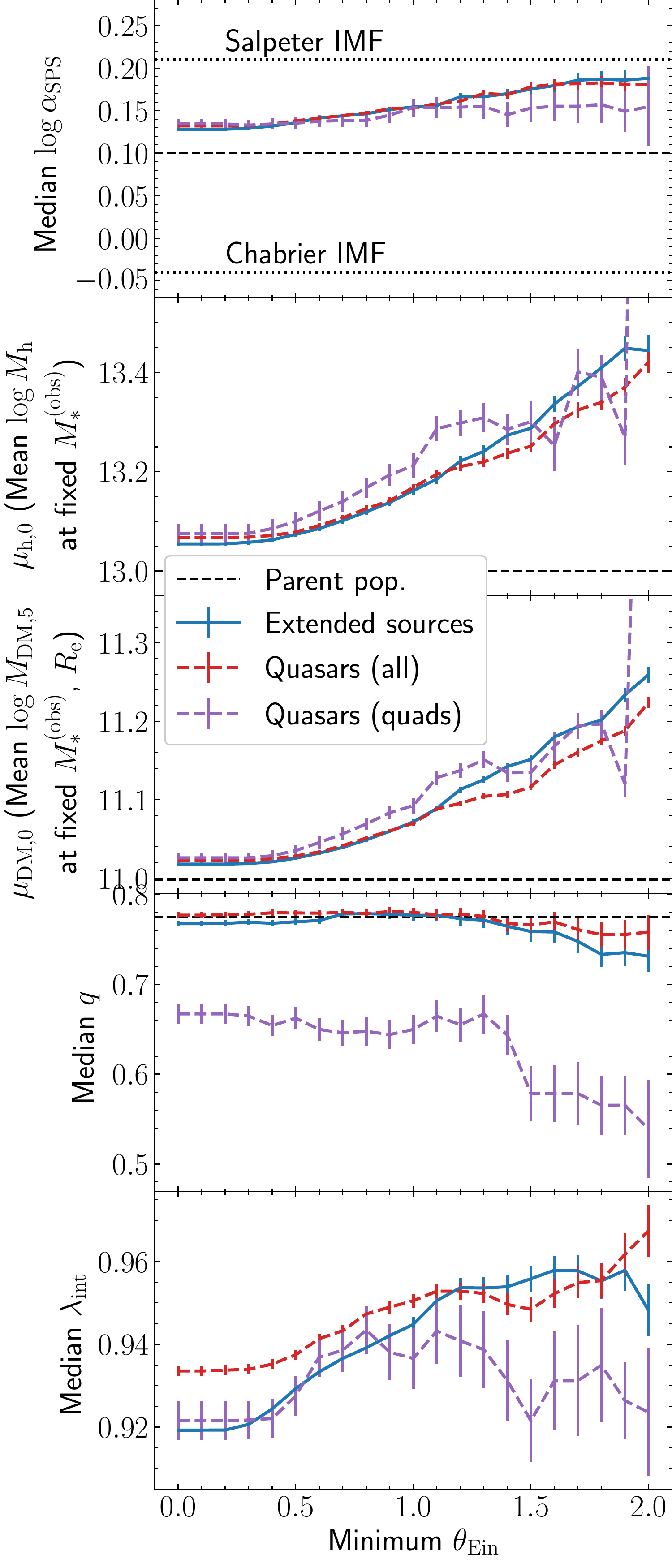}
\caption{
Bias in the population of quasar lenses (all and quads only), compared to the extended source simulation.
First panel: Median $\log{\asps}$.
Second panel: Median $\mhalo$ at $\log{\mstar}=11.5$.
Third panel: Median $\mdmfive$ at $\log{\mstar}=11.5$ and $\log{\reff}=1.2$.
Fourth panel: Median lens axis ratio.
Fifth panel: Median mass-sheet transformation parameter, defined in Appendix~\ref{sect:appendixc}.
In each panel, the dashed line indicates the value of the parent population.
Error bars indicate the standard deviation of the mean of the lens sample.
\label{fig:quasarbias}
}
\end{figure}

The first four rows of Fig. \ref{fig:quasarbias} show the strong lensing bias as a function of minimum Einstein radius in the following quantities: stellar population synthesis mismatch parameter, halo mass at fixed stellar mass, enclosed dark matter mass at fixed stellar mass and half-light radius, and axis ratio.
The biases of the population of lensed quasars (red curves) are very similar to that of lensed galaxies (blue curves), especially for values of $\theta_{\mathrm{Ein},min}<1''$, 
When considering only quad lenses, however, there are some differences, the most remarkable of which is the bias in the axis ratio: quad lenses tend to be on average galaxies with a higher ellipticity.
This is a well-known effect \citep[see e.g.][]{KKS97} that can be explained in the context of the results of Sect. \ref{ssec:axisymmpoint}: the higher the ellipticity, the larger the area enclosed within the inner caustic, which is where a source needs to lie in order to produce four or more images.
Quad lenses tend to also have a slightly larger halo mass at fixed stellar mass, for $\theta_{\mathrm{Ein},min}<1''$.

\subsubsection{Impact on $H_0$}

A measurement of the time delay between the images of a lensed quasar can be converted into an estimate of the expansion rate of the Universe, $H_0$. 
In general, though, the value of $H_0$ is degenerate with the mass distribution of the lens (and of matter along the line of sight), which can be difficult to constrain on an individual lens basis.
One possible strategy to break this degeneracy is to use information gathered from a larger set of lenses as a prior for the time-delay lenses.
This was the strategy adopted by \citet{Bir++20} in the most recent measurement of $H_0$ by the TDCOSMO collaboration: they combined a larger sample of extended source lenses with a smaller set of time-delay lenses.
In this section we want to quantify the possible bias on $H_0$ that can result from this approach.

The lens mass model used by \citet{Bir++20} consists in a power-law density profile modified by a mass-sheet transformation. The corresponding projected density profile is
\begin{equation}\label{eq:mstprofile}
\kappa(\theta) \propto \lmst\theta^{1-\gamma} + (1 - \lmst),
\end{equation}
where $\lmst$ is called the mass-sheet transformation parameter.
Strong lensing data can constrain  the power-law index $\gamma$ very well, but not at all $\lmst$.
Given the time delay, the inferred value of the Hubble constant $H_0$ scales linearly with $1/\lmst$.
\citet{Bir++20} measured the population distribution of $\lmst$ on the sample of extended source lenses, by combining strong lensing and stellar kinematics data, and then used it as a prior in the time-delay analysis.
Although the mass model used in our work is different from that of Eq. \ref{eq:mstprofile}, we can estimate the bias associated with this approach by computing the value of $\lmst$ for each lens and then comparing the strong lensing bias on $\lmst$ of extended source and quasar lenses.
Appendix~\ref{sect:appendixc} explains how we obtained $\lmst$ for each lens. The strong lensing bias on the median value of $\lmst$ is shown in the fifth panel of Fig. \ref{fig:quasarbias}\footnote{Since $\lmst$ is only defined for strong lenses, the median value of $\lmst$ of the general population of galaxies is not defined.}.

At fixed minimum Einstein radius, there is a small difference between the median $\lmst$ in extended source and quasar lenses, typically of order 1\%.
The associated bias on $H_0$ is, therefore, equally small.
The value of $\lmst$, however, increases with increasing minimum $\tein$. This implies that if the two samples probe different ranges in Einstein radius, the bias on $H_0$ can be larger than that.
The bias can also be larger if the properties of the general population of galaxies (as opposed to the strong lenses) are used to inform the structure of time-delay lenses \citep[see][]{C+C16}. This is simply because quasar lenses are on average more similar to galaxy-galaxy lenses than they are to the underlying galaxy population (the blue and red curves in Fig. \ref{fig:quasarbias} are closer to each other than they are to the dashed horizontal lines).


\section{Discussion}\label{sect:discuss}

\subsection{Key results}

In order for a galaxy-source pair to be included as a strong lens in a survey, three conditions must be met.
First of all, at least part of the source must be multiply imaged.
Second, the multiple images must be detectable.
Third, the lens must be recognised as such.
Each one of these conditions introduces a bias with respect to the parent population of foreground galaxies and background sources. Together, they define the lens selection probability term $\psel$ of Eq. \ref{eq:one}.
The first two points are intrinsic to a strong lensing survey and constitute an unavoidable source of bias. 
The best case scenario occurs when the efficiency of including a detected strong lens in a survey is always one, and the sample is $100\%$ complete. In this case, the third condition does not introduce any further selection and the strong lensing bias is minimised.
We explore this scenario in Sect. \ref{sect:results} when setting the minimum Einstein radius to zero.
If only lenses with Einstein radii larger than a given threshold are selected, however, the bias generally increases.

The strong lensing bias affects all quantities that are related to the mass distribution of the lens, as well as the redshifts of lens and source and the source surface brightness parameters.
Some of these quantities, such as the observed stellar mass and half-light radius of the lenses, can be directly measured, and it is straightforward to quantify their lensing bias.
Other quantities, however, such as the stellar population synthesis mismatch parameter or the dark matter content, are difficult to obtain via traditional, non-lensing, observations. Our simulations are particularly useful to quantify the bias on these properties.

In the fiducial scatter scenario, the one consistent with the observed scatter around the fundamental plane, the bias on $\asps$ varies from $0.03$~dex, when no restrictions on the lens Einstein radius are applied, to $0.09$~dex, corresponding to the extreme case in which only lenses with $\tein > 2.0''$ are selected (see Fig. \ref{fig:lensmassbias}).
A reasonable value for the minimum Einstein radius in a space-based survey like \textit{Euclid} or the Chinese Space Station Telescope is $\theta_{\mathrm{Ein,min}}=0.5''$. In this case, the bias on $\asps$ is slightly smaller than $0.04$~dex.
The current systematic uncertainty on $\asps$ is $0.2-0.3$~dex: this is roughly the difference in measurements of the stellar mass of a galaxy obtained with a Chabrier or a Salpeter IMF.
Compared with this uncertainty, the amplitude of the strong lensing bias on $\asps$ is small: strong lensing observations can be used directly to discriminate between these two alternative choices of IMF, without the need to correct for selection effects. 
Such a goal could be reached with a sample size of a thousand lenses and a statistical study of the kind proposed by \citet{S+C21}.

Strong lenses are also biased towards larger halo masses.
Nevertheless, when focusing on the dark matter content in the inner regions, the amplitude of the strong lensing bias is relatively small, especially when controlling for the stellar distribution of the lens galaxies.
For instance, at fixed observed stellar mass and half-light radius, the bias on the projected dark matter mass enclosed within $5$~kpc, $\mdmfive$, is only a few percent in the fiducial scatter scenario with $\theta_{\mathrm{Ein,min}} < 1.0''$.
This means that strong lenses can indeed be used to understand the inner dark matter distribution of galaxies, as long as the dependence of the dark matter distribution on the properties of the baryonic component is accurately modelled \citep[for example, by following the approach of][]{S+C21}.

We also looked at the bias on source-related parameters.
Strong lensing causes the luminosity function of background sources to be broadened, compared to the distribution of detectable sources.
On the one hand, it preferentially selects brighter sources, because the lensing cross-section increases with increasing source brightness.
On the other hand, it allows for the detection of sources that are intrinsically fainter than the detection limit in the absence of lensing.
Interestingly, we did not find any significant bias on the source size, at fixed magnitude.
This result follows from the fact that our lens detection criterion relies on a surface brightness threshold, and surface brightness is preserved by lensing.
However, it appears to be in contradiction with the work of \citet{O+A17}, who argued that their strong lens sample selected preferentially compact sources.
The origin of this discrepancy probably lies in the differences between the criteria used to define a strong lens in the two studies.
The \citet{O+A17} sample was selected primarily via spectroscopy, by looking for signatures of two galaxies at different redshifts in the Sloan Digital Sky Survey \citep[SDSS;][]{Yor++00} data.
SDSS spectra were taken in fibres with a $1.5''$ radius.
Lensed galaxies that are comparable in size to this scale, or larger, are less likely to be detected, because part of their flux extends outside of the fibre. Compact galaxies with a large overall magnification, instead, are more likely to be detected \citep[see the discussion in Sect. 5.3 of][]{O+A17}.
We refer the reader to \citet{Arn++12} for a thorough study of selection effects associated with spectroscopy.

There is also an indirect way in which strong lensing could preferentially probe more compact sources than possible in the field.
Sources that are smaller than the size of the PSF can be easily confused as stars in our own Galaxy.
Since these stars cannot be strongly lensed into multiple images, the detection of a strong lens automatically confirms the extragalactic nature of a lensed source.
This type of selection effect could explain, for example, the detection of an apparent outlier in the magnitude-size relation of Lyman-break galaxy by means of strong lensing, by \citet{Jae++20}.
It is, however, improper to refer to this effect as strong lensing bias, since the star-galaxy separation is a bias that primarily affects field observations.
Our analysis showed that the tendency to select compact sources is not a general feature of strong lens samples.

\subsection{Combining lensed quasars with lensed galaxies}

Our experiments were also useful for understanding the possible biases that might incur when combining information from samples of galaxy-galaxy lenses with samples of galaxy-quasar lenses.
On the one hand the biases in the mass-related quantities ($\asps$ and $\mdmfive$) are very similar in simulations with the same foreground galaxy population and different background sources (see Fig. \ref{fig:quasarbias}).
On the other hand, these biases are a function of the completeness limit. In order to use a sample of strong lenses as a prior for another sample, then, it is important to make sure that 1) the parent population of foreground galaxies among which lenses are searched for is the same in both surveys, and 2) the two surveys can probe the same distribution in Einstein radius.

The above argument applies to samples of quasars selected regardless of the number of multiple images.
When dealing with quad lenses, however, the situation is more complicated.
First of all, because our experiment revealed differences in the bias on the dark matter distribution between quads and the entire sample of lensed quasars.
Second, because quad lenses tend to have a preferentially higher ellipticity.
The ellipticity is important in the context of stellar dynamical analyses,
which are often used in combination with strong lensing to constrain lens mass parameters \citep[see e.g.][]{Yil++20}.
In order to correctly interpret stellar dynamics data, assumptions on the three-dimensional structure of a lens must be made.
This is directly related to the projected ellipticity: galaxies with an axis ratio close to one tend to be preferentially elongated in the line-of-sight direction, and vice versa.
When using stellar dynamics-based mass measurements to inform the properties of a sample of quad lenses, it is therefore important to take into account possible biases due to the different three-dimensional structure of the two samples.

\subsection{The importance of the source population properties}

Strictly speaking, the results shown in Sect. \ref{sect:results} apply only to samples of lenses and background sources with the same properties as our simulations.
Real lens samples can have more complex source distributions. For example, many existing lens-finding algorithms are tuned to identify sources with different colours from the lenses, introducing a selection that can modify the source distribution in redshift-magnitude space.
However, the comparison of Sect. \ref{ssec:quasarbias} shows that, at fixed foreground galaxy population, the strong lensing bias for the sample of extended sources is indistinguishable from that of the lensed quasar population, at least for values of $\theta_{\mathrm{Ein,min}} < 1.0''$.
This result suggests that the dependence of the strong lensing bias on the details of the background source population is very weak, as long as no additional selection on the image configuration is applied (e.g. by selecting only quad lenses).
This is not a coincidence, but is a consequence of the fact that the dependence of the strong lensing cross-section on the lens parameters is a weak function of source magnitude\footnote{The cross-section itself is a strong function of source magnitude, but the trends between $\crosssect$ and the lens parameters are not.}, especially for magnitudes close to the detection limit (see Sect. \ref{ssec:axisymmpoint}).
We can then conclude that the details of the properties of the background source population play a secondary role in determining the strong lensing bias.
A selection aimed at specific kinds of sources, such as blue star-forming galaxies or quasars, results primarily in different numbers of lenses being discovered, but does not change how biased the lenses are with respect to the general galaxy population.

\subsection{Mitigation strategies}

As we discussed throughout this paper, the amplitude of the strong lensing bias depends on the intrinsic scatter in the mass parameters of the foreground galaxy population.
One way to minimise the bias, therefore, is to identify scaling relations between observable quantities and mass-related properties that can account for part of the scatter.
Describing the inner dark matter distribution as a function of stellar mass and half-light radius, as we did in Sect. \ref{ssec:galbias}, is the first step in this direction: the bias in $\mdmfive$ at fixed $\mstar$ and $\reff$ is smaller than the overall shift in the median $\mdmfive$ of the lens population (albeit by a marginal amount, as we pointed out earlier).
This description could be extended by including the central velocity dispersion as an additional control parameter. 
The velocity dispersion is directly related to the mass distribution of the lens galaxy, which means that, for example, the distribution in $\asps$ at fixed velocity dispersion should be narrower than its global distribution marginalised over the whole population. 
The central velocity dispersion, however, is very sensitive to the orbital anisotropy, to the three-dimensional structure, and to gradients in stellar mass-to-light ratio, which are not well known. Therefore, it is difficult to quantitatively estimate the benefit of including it in the description of the lensing bias.

If one wishes to directly account for the strong lensing bias, the formally correct procedure is to explicitly model all of the selection steps in a Bayesian hierarchical formalism, as explained by \citet{Son22}.
Although this can be computationally challenging, machine learning can offer an efficient alternative \citep{Leg++23}.
In order for either of these approaches to work, however, it is essential that the lens selection procedure can be simulated.
This, in turn, requires having an objective definition of a strong lensing event. A peak-based definition such as that introduced in Sect. \ref{ssec:lensdefext} could be used for this purpose.

Nevertheless, it can still be difficult to fully forward model strong lensing selection if visual inspection by humans is involved in the definition of the lens sample.
In that case, a possible alternative is to approximate the lens finding probability $\pfind$. 
For example, we can make the assumption that $\pfind$ depends purely on the Einstein radius.
This is a reasonable assumption, as long as the lens finding procedure does not selectively pick lenses with different image configurations depending on their Einstein radius.
The dependence of $\pfind$ on $\tein$ could then be described empirically and inferred during the analysis.
This is essentially the approach adopted by \citet{Son++19} in the analysis of strong and weak lensing data from the Hyper Suprime-Cam Survey.

\subsection{Limitations of our analysis}\label{ssec:limitations}

The simulations on which our analysis is based are as complex as required by the goal of the analysis itself, which is to estimate the amplitude of the strong lensing bias in a few key quantities.
We did, however, make some simplifying assumptions.
One such assumption consisted in neglecting the contribution from line-of-sight structure and the environment to the lensing signal, which typically introduce an external shear and convergence.
The effect of including external shear in a lens model is similar to that of changing the ellipticity of the lens. 
External convergence mimics the effect of adding or removing a constant sheet of invisible mass, producing an effect similar to varying the distribution of dark matter.
Typical values of the external shear and external convergence are $|\gamma|<0.1$ and $|\kappa_{\mathrm{ext}}|<0.1$ \citep{Mil++20}: these values are small compared to the typical ellipticities and dark matter fractions of our simulated lenses.
Therefore, while including them might modify slightly the amplitude of the strong lensing bias in the axis ratio and dark matter distribution parameters, the conclusions of our analysis would not be affected.

We also assumed that the dark matter density profile of the galaxies is completely determined by the halo mass and by the stellar mass distribution (see Sect. \ref{ssub:dmprofile}). In reality we expect there to be a range of density profiles, for example as a result of halos having a non-zero scatter in their initial (i.e. before baryonic infall) concentration parameter.
We could in principle carry out an experiment with such an additional source of variation in the dark matter profile.
Adding a scatter in concentration while keeping the halo mass scatter parameter $\sigma_h$ fixed results in a larger spread in the inner dark matter distribution, which would increase the amplitude of the lensing bias. However, the value of $\sigma_h$ that we chose for our fiducial experiment was tuned on the basis of the predicted scatter in velocity dispersion around the fundamental plane. That scatter would also be increased by adding flexibility to the dark matter density profile. Then, in order to be consistent with the analysis carried out in this paper, the value of $\sigma_h$ would need to be lowered accordingly. This, in turn, would reduce the amplitude of the strong lensing bias and introduce a bias on the halo concentration.
Since lensing and dynamics are sensitive to the mass distribution at comparable scales (the half-light radius and Einstein radius are similar for most of the lenses), we expect these two effects to cancel out to first approximation. The lensing bias on quantities directly related to the inner mass distribution, such as $\asps$ and $\mdmfive$, would be left roughly unchanged. 
The main effect of adding scatter to the concentration is to reduce the correlation between the total halo mass and the dark matter mass enclosed within the Einstein radius. This would then reduce the amplitude of the strong lensing bias on the halo mass.

Another simplifying assumption that we made was that of adopting a constant mass-to-light ratio for the stellar component. Massive galaxies are known to have colour gradients, the main effect of which is to cause the stellar half-mass radius to be smaller than the half-light radius \citep{Szo++13,Sue++19}. Since, as shown in Sect. \ref{sect:results}, strong lensing bias tends to select galaxies with a more compact stellar distribution, it would also preferentially select galaxies with a steeper mass-to-light ratio gradient.
However, testing for the amplitude of the strong lensing bias on the mass-to-light ratio gradient is beyond the scope of this paper.

\subsection{Lens versus source selection}

The simulations described in Sect. \ref{sect:lenspop}, on which our experiment is based, are examples of so-called lens-selected surveys:
the lens samples are built by first defining a population of possible lens galaxies and then searching for strong lenses among them. Photometric lens searches are typically carried out in this way \citep[see e.g.][]{Gav++12,Son++18a,Pet++19}.
An alternative approach is to build a source-selected sample, by first searching for signatures of strong lensing within a population of possible background sources and then obtaining information on the lens galaxies \citep[see e.g.][]{Hez++13}.
In principle, these two approaches lead to different strong lensing biases. 
For example, from Fig. \ref{fig:lenspars} we can see that the stellar mass distribution of the lenses extends below the minimum mass cut that we applied to the population of foreground galaxies: a source-based search that is able detect those lenses would result in a different sample and a different associated strong lensing bias.
However, when considering only lenses with $\tein > 1''$, lens-selection and source-selection yield the same sample (see the red histogram in Fig. \ref{fig:lenspars}): in the limit of large Einstein radius, the two operations (identifying suitable lens galaxies and detecting a strongly lensed source) commute.

In general, there is no such thing as a purely lens- or source-selected sample. 
A lens-based search is always also source-based, because a detection of the lensed source is required to identify a lens.
A source-based search is also lens-based, because the lens properties determine important observational features of the lensed source, such as the number of images, the image separation and the total flux.

\subsection{Alternative lens definitions}

\begin{figure}
\includegraphics[width=\columnwidth]{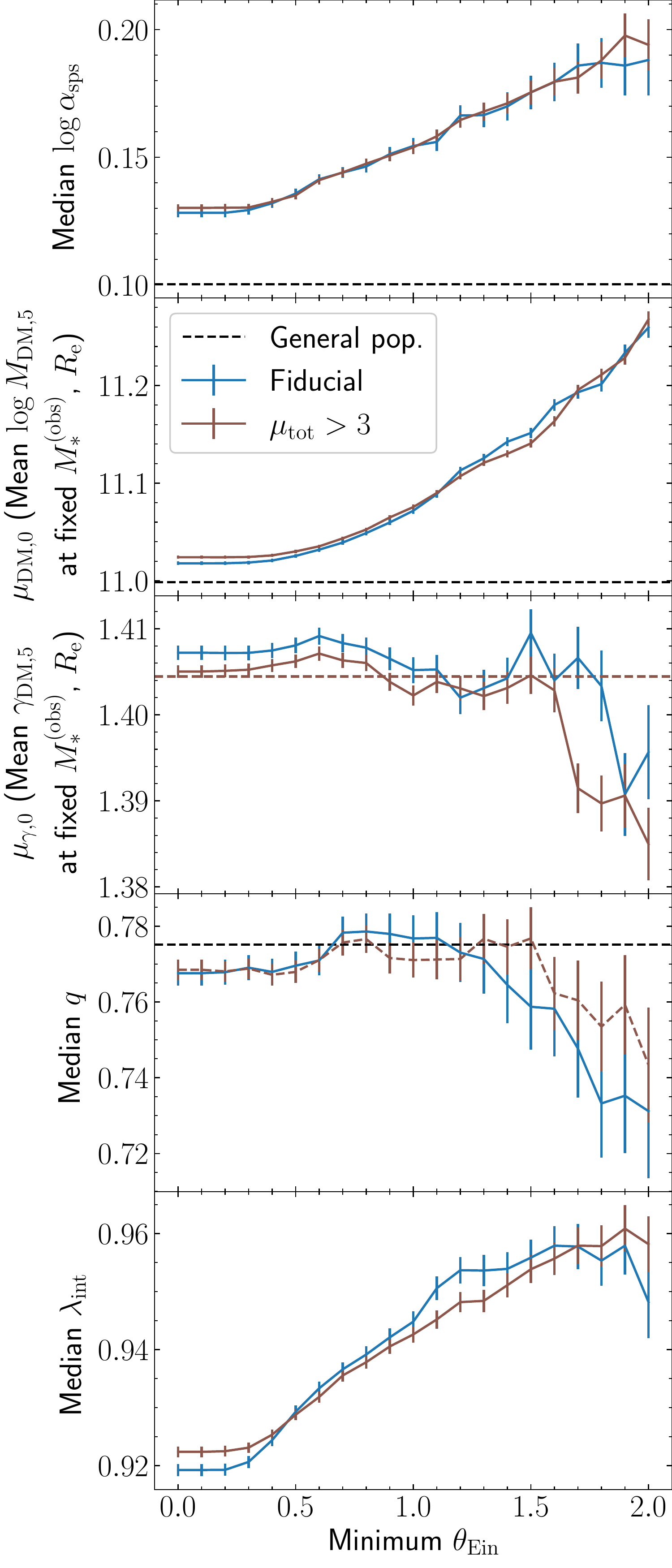}
\caption{
Lens strong lensing bias in the case of a magnification-based lens definition (brown curves), compared to that of the fiducial analysis (blue curves).
First panel: Median $\log{\asps}$.
Second panel: Median $\mdmfive$ at $\log{\mstar}=11.5$ and $\log{\reff}=1.2$.
Third panel: Median $\gammadm$ at $\log{\mobs}=11.4$ and $\log{\reff}=1.2$.
Fourth panel: Median axis ratio.
Fifth panel: Median mass-sheet transformation parameter, defined in Appendix~\ref{sect:appendixc}.
In each panel, the dashed line indicates the value of the parent population.
}
\label{fig:minmulbias}
\end{figure}

\begin{figure}
\includegraphics[width=\columnwidth]{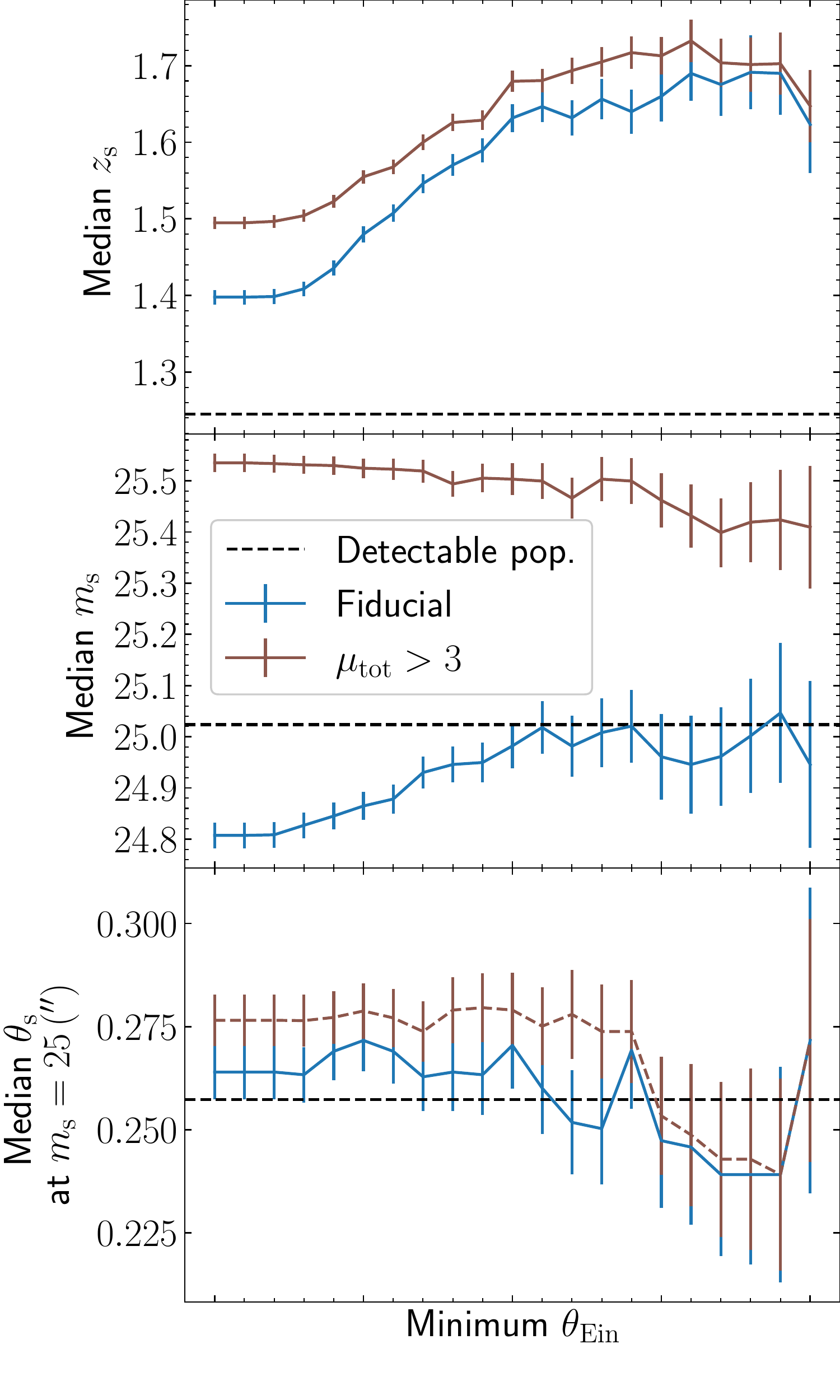}
\caption{
Source strong lensing bias in the case of a magnification-based lens definition (brown curves), compared to that of the fiducial analysis (blue curves). 
First panel: Median source redshift.
Second panel: Median source magnitude.
Third panel: Median half-light radius in the magnitude bin $24.8 < \msource < 25.2$.
The dashed line indicates the median value of the population of sources detectable in the absence of lensing.
}
\label{fig:minmusbias}
\end{figure}

Our extended source analysis was based on a peak detection criterion to define strong lensing events, as explained in Sect. \ref{ssec:lensdefext}.
However, many previous studies aimed at simulating populations of lenses adopted a magnification-based definition, in which a lens-source system is classified as a strong lens if the magnification of the source is larger than a minimum value \citep[see e.g.][]{Col15, Robertson++20}.
In order to check whether our results are robust to the choice of definition of a strong lensing event, we re-analysed the fiducial sample with this alternative definition. 
In particular, we defined as strong lenses all systems with a detected background source and $\mu_{\mathrm{tot}} > 3$, where $\mu_{\mathrm{tot}}$ is the total magnification (obtained by integrating the flux of all of the lensed images and dividing it by the intrinsic source flux).
The total number of strong lenses increased by $34\%$ with this new definition, mostly due to the inclusion of fainter sources with no detected multiple images.
Figures \ref{fig:minmulbias} and  \ref{fig:minmusbias} show the strong lensing bias on a series of lens- and source-related properties.
Besides the change in the source redshift and magnitude (first two panels in Fig. \ref{fig:minmusbias}), we see very little differences in the bias with respect to the peak-based definition of the fiducial analysis. We then conclude that the exact definition of a strong lensing event does not have a big impact on the strong lensing bias of lens-related properties.
We stress, however, that magnification is not a directly observable quantity, and therefore we advise against applying magnification-based lens definitions to define lens samples, for the reasons discussed in Sect. \ref{ssec:lensdefext}.

\section{Conclusions}\label{sect:concl}

Strong lensing is a very active line of research, but a comprehensive understanding of the selection effects associated with it has so far been lacking.
This work takes a major step towards filling that knowledge gap.
After a thorough investigation, we have learned several lessons regarding the strong lensing bias in photometrically selected lens samples. The following are the most important ones.
\begin{enumerate}
\item The strong lensing cross-section increases primarily with increasing lens mass, with decreasing half-mass radius, and with increasing source brightness. 
At fixed stellar distribution and fixed dark matter mass enclosed within a given aperture, varying the inner dark matter slope has little impact on the strong lensing cross-section (as long as the aperture within which the dark matter mass is normalised is comparable to the Einstein radius).
\item The strong lensing cross-section has little dependence on the size of the source if the former is smaller than the Einstein radius and is detectable in the absence of lensing.
Sources with a surface brightness that is too low to be detected are still undetected when strongly lensed. 
\item Lens galaxies tend to be more massive and more compact than their non-lens counterparts. Their redshift distribution also differs from that of the general galaxy population.
At fixed observed stellar mass (i.e. inferred by means of stellar population synthesis), lens galaxies have a higher intrinsic stellar mass (i.e. a larger stellar population synthesis mismatch parameter, $\asps$) and a higher dark matter halo mass.
At fixed stellar mass and size, lens galaxies are still biased towards a larger dark matter content.
\item The amplitude of the strong lensing bias depends on how broad the distribution of the parameters describing the lens population is. 
Important quantities in this regard are the intrinsic scatter in the stellar population synthesis mismatch parameter, $\sigma_{\mathrm{sps}}$, and in the dark matter halo mass, $\sigma_{\mathrm{h}}$.
Increasing the values of $\sigma_{\mathrm{sps}}$ and $\sigma_\mathrm{h}$ results in a stronger bias on $\asps$ and on the dark matter mass, and a weaker bias on the observed stellar mass and half-light radius.
This implies that, in principle, we could constrain the intrinsic scatter parameters, which are currently poorly known, by measuring the amplitude of the strong lensing bias on $\mobs$ and $\reff$, which is easily observable.
\item The strong lensing bias varies depending on the completeness of the lens sample as a function of the Einstein radius. Surveys that can discover lenses with a smaller Einstein radius have a smaller associated strong lensing bias in all quantities.
\item 
Under reasonable assumptions regarding the intrinsic scatter parameters, for a \textit{Euclid}-like survey that is complete down to $\tein=0.5''$ the bias on $\asps$ is smaller than $0.04$~dex (10\%). This bias is much smaller than the current systematic uncertainty on the stellar population synthesis-based stellar masses. Therefore, strong lensing measurements could be used directly to calibrate stellar mass measurements of massive galaxies to $10\%$ accuracy, without the need to correct for selection effects.
Under the same assumptions, the strong lensing bias on the average halo mass at fixed stellar mass is $0.07$~dex, while that on the inner dark matter distribution at fixed stellar mass and size is $0.02$~dex.
\item Strong lensing selection broadens the magnitude distribution of background sources, compared to the population of objects that are detectable without lensing.
At the same time, we did not find any evidence for a bias in the size distribution of background sources at fixed magnitude.

\item Simulations with lensed quasars in place of extended sources showed that the amplitude of the strong lensing bias in the lens-related parameters is not very sensitive to the details of the source population.
This result has positive implications for time-delay lensing studies: it means that information from a sample of galaxy-galaxy lenses can be used as a prior on the properties of a set of galaxy-quasar lenses, as long as the two samples probe the same range in Einstein radius and lens observable properties.
The associated bias on the inference of $H_0$ is of the order of $1\%$.
\item Samples of quad lenses are biased towards galaxies with larger ellipticities, which implies that their three-dimensional structure is also biased. This means that particular care must be taken when stellar dynamics measurements obtained on galaxy-galaxy lenses are used to inform the properties of quads.
\end{enumerate}

In conclusion, strong lensing selection introduces unavoidable biases in the properties of the lens galaxy and background source populations.
Biases that affect observable properties, such as the redshift and the light distribution of the lens, can be easily quantified.
Biases on mass-related quantities, such as the stellar mass-to-light ratio or the dark matter distribution, are more difficult to measure directly and must be modelled taking selection effects into account. 
Designing strong lensing surveys with clearly defined and easily modellable selection criteria would help greatly in this task.

\begin{acknowledgements}

The collaboration leading to this work was initiated at the 2022 Lorentz Center workshop “Bridging gaps between dynamical probes of galaxies”.
AS and SSL are supported by NOVA, the Netherlands Research School for Astronomy.
Support for this work was provided by NASA through the NASA Hubble Fellowship grant HST-HF2-51492 awarded to AJS by the Space Telescope Science Institute, which is operated by the Association of Universities for Research in Astronomy, Inc., for NASA, under contract NAS5-26555.

\end{acknowledgements}

\bibliographystyle{aa}
\bibliography{references}

\begin{thebibliography}{72}
\expandafter\ifx\csname natexlab\endcsname\relax\def\natexlab#1{#1}\fi

\bibitem[{{Ahumada} {et~al.}(2020){Ahumada}, {Prieto}, {Almeida}, {Anders},
  {Anderson}, {Andrews}, {Anguiano}, {Arcodia}, {Armengaud}, {Aubert}, {Avila},
  {Avila-Reese}, {Badenes}, {Balland}, {Barger}, {Barrera-Ballesteros}, {Basu},
  {Bautista}, {Beaton}, {Beers}, {Benavides}, {Bender}, {Bernardi}, {Bershady},
  {Beutler}, {Bidin}, {Bird}, {Bizyaev}, {Blanc}, {Blanton}, {Boquien},
  {Borissova}, {Bovy}, {Brandt}, {Brinkmann}, {Brownstein}, {Bundy}, {Bureau},
  {Burgasser}, {Burtin}, {Cano-D{\'\i}az}, {Capasso}, {Cappellari}, {Carrera},
  {Chabanier}, {Chaplin}, {Chapman}, {Cherinka}, {Chiappini}, {Doohyun Choi},
  {Chojnowski}, {Chung}, {Clerc}, {Coffey}, {Comerford}, {Comparat}, {da
  Costa}, {Cousinou}, {Covey}, {Crane}, {Cunha}, {Ilha}, {Dai}, {Damsted},
  {Darling}, {Davidson}, {Davies}, {Dawson}, {De}, {de la Macorra}, {De Lee},
  {Queiroz}, {Deconto Machado}, {de la Torre}, {Dell'Agli}, {du Mas des
  Bourboux}, {Diamond-Stanic}, {Dillon}, {Donor}, {Drory}, {Duckworth},
  {Dwelly}, {Ebelke}, {Eftekharzadeh}, {Davis Eigenbrot}, {Elsworth},
  {Eracleous}, {Erfanianfar}, {Escoffier}, {Fan}, {Farr},
  {Fern{\'a}ndez-Trincado}, {Feuillet}, {Finoguenov}, {Fofie},
  {Fraser-McKelvie}, {Frinchaboy}, {Fromenteau}, {Fu}, {Galbany}, {Garcia},
  {Garc{\'\i}a-Hern{\'a}ndez}, {Oehmichen}, {Ge}, {Maia}, {Geisler}, {Gelfand},
  {Goddy}, {Gonzalez-Perez}, {Grabowski}, {Green}, {Grier}, {Guo}, {Guy},
  {Harding}, {Hasselquist}, {Hawken}, {Hayes}, {Hearty}, {Hekker}, {Hogg},
  {Holtzman}, {Horta}, {Hou}, {Hsieh}, {Huber}, {Hunt}, {Chitham}, {Imig},
  {Jaber}, {Angel}, {Johnson}, {Jones}, {J{\"o}nsson}, {Jullo}, {Kim},
  {Kinemuchi}, {Kirkpatrick}, {Kite}, {Klaene}, {Kneib}, {Kollmeier}, {Kong},
  {Kounkel}, {Krishnarao}, {Lacerna}, {Lan}, {Lane}, {Law}, {Le Goff}, {Leung},
  {Lewis}, {Li}, {Lian}, {Lin}, {Long}, {Longa-Pe{\~n}a}, {Lundgren}, {Lyke},
  {Ted Mackereth}, {MacLeod}, {Majewski}, {Manchado}, {Maraston}, {Martini},
  {Masseron}, {Masters}, {Mathur}, {McDermid}, {Merloni}, {Merrifield},
  {M{\'e}sz{\'a}ros}, {Miglio}, {Minniti}, {Minsley}, {Miyaji}, {Mohammad},
  {Mosser}, {Mueller}, {Muna}, {Mu{\~n}oz-Guti{\'e}rrez}, {Myers}, {Nadathur},
  {Nair}, {Nandra}, {do Nascimento}, {Nevin}, {Newman}, {Nidever}, {Nitschelm},
  {Noterdaeme}, {O'Connell}, {Olmstead}, {Oravetz}, {Oravetz}, {Osorio},
  {Pace}, {Padilla}, {Palanque-Delabrouille}, {Palicio}, {Pan}, {Pan},
  {Parker}, {Paviot}, {Peirani}, {Ram{\'r}ez}, {Penny}, {Percival},
  {Perez-Fournon}, {P{\'e}rez-R{\`a}fols}, {Petitjean}, {Pieri},
  {Pinsonneault}, {Poovelil}, {Povick}, {Prakash}, {Price-Whelan}, {Raddick},
  {Raichoor}, {Ray}, {Rembold}, {Rezaie}, {Riffel}, {Riffel}, {Rix}, {Robin},
  {Roman-Lopes}, {Rom{\'a}n-Z{\'u}{\~n}iga}, {Rose}, {Ross}, {Rossi},
  {Rowlands}, {Rubin}, {Salvato}, {S{\'a}nchez}, {S{\'a}nchez-Menguiano},
  {S{\'a}nchez-Gallego}, {Sayres}, {Schaefer}, {Schiavon}, {Schimoia},
  {Schlafly}, {Schlegel}, {Schneider}, {Schultheis}, {Schwope}, {Seo},
  {Serenelli}, {Shafieloo}, {Shamsi}, {Shao}, {Shen}, {Shetrone}, {Shirley},
  {Aguirre}, {Simon}, {Skrutskie}, {Slosar}, {Smethurst}, {Sobeck}, {Sodi},
  {Souto}, {Stark}, {Stassun}, {Steinmetz}, {Stello}, {Stermer},
  {Storchi-Bergmann}, {Streblyanska}, {Stringfellow}, {Stutz}, {Su{\'a}rez},
  {Sun}, {Taghizadeh-Popp}, {Talbot}, {Tayar}, {Thakar}, {Theriault}, {Thomas},
  {Thomas}, {Tinker}, {Tojeiro}, {Toledo}, {Tremonti}, {Troup}, {Tuttle},
  {Unda-Sanzana}, {Valentini}, {Vargas-Gonz{\'a}lez}, {Vargas-Maga{\~n}a},
  {V{\'a}zquez-Mata}, {Vivek}, {Wake}, {Wang}, {Weaver}, {Weijmans}, {Wild},
  {Wilson}, {Wilson}, {Wolthuis}, {Wood-Vasey}, {Yan}, {Yang}, {Y{\`e}che},
  {Zamora}, {Zarrouk}, {Zasowski}, {Zhang}, {Zhao}, {Zhao}, {Zheng}, {Zheng},
  {Zhu}, \& {Zou}}]{Ahu++20}
{Ahumada}, R., {Prieto}, C.~A., {Almeida}, A., {et~al.} 2020, \apjs, 249, 3

\bibitem[{{Arneson} {et~al.}(2012){Arneson}, {Brownstein}, \&
  {Bolton}}]{Arn++12}
{Arneson}, R.~A., {Brownstein}, J.~R., \& {Bolton}, A.~S. 2012, \apj, 753, 4

\bibitem[{{Auger} {et~al.}(2010){Auger}, {Treu}, {Bolton}, {Gavazzi},
  {Koopmans}, {Marshall}, {Moustakas}, \& {Burles}}]{Aug++10}
{Auger}, M.~W., {Treu}, T., {Bolton}, A.~S., {et~al.} 2010, \apj, 724, 511

\bibitem[{{Binney} \& {Tremaine}(1987)}]{B+T87}
{Binney}, J. \& {Tremaine}, S. 1987, {Galactic dynamics}

\bibitem[{{Birrer} {et~al.}(2020){Birrer}, {Shajib}, {Galan}, {Millon}, {Treu},
  {Agnello}, {Auger}, {Chen}, {Christensen}, {Collett}, {Courbin}, {Fassnacht},
  {Koopmans}, {Marshall}, {Park}, {Rusu}, {Sluse}, {Spiniello}, {Suyu},
  {Wagner-Carena}, {Wong}, {Barnab{\`e}}, {Bolton}, {Czoske}, {Ding},
  {Frieman}, \& {Van de Vyvere}}]{Bir++20}
{Birrer}, S., {Shajib}, A.~J., {Galan}, A., {et~al.} 2020, \aap, 643, A165

\bibitem[{{Birrer} \& {Treu}(2021)}]{B+T21}
{Birrer}, S. \& {Treu}, T. 2021, \aap, 649, A61

\bibitem[{{Bolton} {et~al.}(2006){Bolton}, {Burles}, {Koopmans}, {Treu}, \&
  {Moustakas}}]{Bol++06}
{Bolton}, A.~S., {Burles}, S., {Koopmans}, L. V.~E., {Treu}, T., \&
  {Moustakas}, L.~A. 2006, \apj, 638, 703

\bibitem[{{Burak Dogruel} {et~al.}(2023){Burak Dogruel}, {Taylor}, {Cluver},
  {D'Eugenio}, {de Graaff}, {Colless}, \& {Sonnenfeld}}]{Dog++23}
{Burak Dogruel}, M., {Taylor}, E.~N., {Cluver}, M., {et~al.} 2023, arXiv
  e-prints, arXiv:2306.10693

\bibitem[{{Cappellari} {et~al.}(2013){Cappellari}, {Scott}, {Alatalo}, {Blitz},
  {Bois}, {Bournaud}, {Bureau}, {Crocker}, {Davies}, {Davis}, {de Zeeuw},
  {Duc}, {Emsellem}, {Khochfar}, {Krajnovi{\'c}}, {Kuntschner}, {McDermid},
  {Morganti}, {Naab}, {Oosterloo}, {Sarzi}, {Serra}, {Weijmans}, \&
  {Young}}]{Cap++13}
{Cappellari}, M., {Scott}, N., {Alatalo}, K., {et~al.} 2013, \mnras, 432, 1709

\bibitem[{{Cautun} {et~al.}(2020){Cautun}, {Ben{\'\i}tez-Llambay}, {Deason},
  {Frenk}, {Fattahi}, {G{\'o}mez}, {Grand}, {Oman}, {Navarro}, \&
  {Simpson}}]{Cau++20}
{Cautun}, M., {Ben{\'\i}tez-Llambay}, A., {Deason}, A.~J., {et~al.} 2020,
  \mnras, 494, 4291

\bibitem[{{Chabrier}(2003)}]{Cha03}
{Chabrier}, G. 2003, \pasp, 115, 763

\bibitem[{{Ciotti} \& {Bertin}(1999)}]{C+B99}
{Ciotti}, L. \& {Bertin}, G. 1999, \aap, 352, 447

\bibitem[{{Collett}(2015)}]{Col15}
{Collett}, T.~E. 2015, \apj, 811, 20

\bibitem[{{Collett} \& {Cunnington}(2016)}]{C+C16}
{Collett}, T.~E. \& {Cunnington}, S.~D. 2016, \mnras, 462, 3255

\bibitem[{{Conroy} \& {van Dokkum}(2012)}]{CvD12}
{Conroy}, C. \& {van Dokkum}, P.~G. 2012, \apj, 760, 71

\bibitem[{{de Graaff} {et~al.}(2021){de Graaff}, {Bezanson}, {Franx}, {van der
  Wel}, {Holden}, {van de Sande}, {Bell}, {D'Eugenio}, {Maseda}, {Muzzin},
  {Sobral}, {Straatman}, \& {Wu}}]{deG++21}
{de Graaff}, A., {Bezanson}, R., {Franx}, M., {et~al.} 2021, \apj, 913, 103

\bibitem[{{Elahi} {et~al.}(2018){Elahi}, {Welker}, {Power}, {Lagos},
  {Robotham}, {Ca{\~n}as}, \& {Poulton}}]{Ela++18}
{Elahi}, P.~J., {Welker}, C., {Power}, C., {et~al.} 2018, \mnras, 475, 5338

\bibitem[{{Euclid Collaboration} {et~al.}(2022){Euclid Collaboration},
  {Scaramella}, {Amiaux}, {Mellier}, {Burigana}, {Carvalho}, {Cuillandre}, {Da
  Silva}, {Derosa}, {Dinis}, {Maiorano}, {Maris}, {Tereno}, {Laureijs},
  {Boenke}, {Buenadicha}, {Dupac}, {Gaspar Venancio}, {G{\'o}mez-{\'A}lvarez},
  {Hoar}, {Lorenzo Alvarez}, {Racca}, {Saavedra-Criado}, {Schwartz}, {Vavrek},
  {Schirmer}, {Aussel}, {Azzollini}, {Cardone}, {Cropper}, {Ealet}, {Garilli},
  {Gillard}, {Granett}, {Guzzo}, {Hoekstra}, {Jahnke}, {Kitching}, {Maciaszek},
  {Meneghetti}, {Miller}, {Nakajima}, {Niemi}, {Pasian}, {Percival},
  {Pottinger}, {Sauvage}, {Scodeggio}, {Wachter}, {Zacchei}, {Aghanim},
  {Amara}, {Auphan}, {Auricchio}, {Awan}, {Balestra}, {Bender}, {Bodendorf},
  {Bonino}, {Branchini}, {Brau-Nogue}, {Brescia}, {Candini}, {Capobianco},
  {Carbone}, {Carlberg}, {Carretero}, {Casas}, {Castander}, {Castellano},
  {Cavuoti}, {Cimatti}, {Cledassou}, {Congedo}, {Conselice}, {Conversi},
  {Copin}, {Corcione}, {Costille}, {Courbin}, {Degaudenzi}, {Douspis},
  {Dubath}, {Duncan}, {Dusini}, {Farrens}, {Ferriol}, {Fosalba}, {Fourmanoit},
  {Frailis}, {Franceschi}, {Franzetti}, {Fumana}, {Gillis}, {Giocoli},
  {Grazian}, {Grupp}, {Haugan}, {Holmes}, {Hormuth}, {Hudelot}, {Kermiche},
  {Kiessling}, {Kilbinger}, {Kohley}, {Kubik}, {K{\"u}mmel}, {Kunz},
  {Kurki-Suonio}, {Lahav}, {Ligori}, {Lilje}, {Lloro}, {Mansutti}, {Marggraf},
  {Markovic}, {Marulli}, {Massey}, {Maurogordato}, {Melchior}, {Merlin},
  {Meylan}, {Mohr}, {Moresco}, {Morin}, {Moscardini}, {Munari}, {Nichol},
  {Padilla}, {Paltani}, {Peacock}, {Pedersen}, {Pettorino}, {Pires}, {Poncet},
  {Popa}, {Pozzetti}, {Raison}, {Rebolo}, {Rhodes}, {Rix}, {Roncarelli},
  {Rossetti}, {Saglia}, {Schneider}, {Schrabback}, {Secroun}, {Seidel},
  {Serrano}, {Sirignano}, {Sirri}, {Skottfelt}, {Stanco}, {Starck},
  {Tallada-Cresp{\'\i}}, {Tavagnacco}, {Taylor}, {Teplitz}, {Toledo-Moreo},
  {Torradeflot}, {Trifoglio}, {Valentijn}, {Valenziano}, {Verdoes Kleijn},
  {Wang}, {Welikala}, {Weller}, {Wetzstein}, {Zamorani}, {Zoubian}, {Andreon},
  {Baldi}, {Bardelli}, {Boucaud}, {Camera}, {Di Ferdinando}, {Fabbian},
  {Farinelli}, {Galeotta}, {Graci{\'a}-Carpio}, {Maino}, {Medinaceli}, {Mei},
  {Neissner}, {Polenta}, {Renzi}, {Romelli}, {Rosset}, {Sureau}, {Tenti},
  {Vassallo}, {Zucca}, {Baccigalupi}, {Balaguera-Antol{\'\i}nez}, {Battaglia},
  {Biviano}, {Borgani}, {Bozzo}, {Cabanac}, {Cappi}, {Casas}, {Castignani},
  {Colodro-Conde}, {Coupon}, {Courtois}, {Cuby}, {de la Torre}, {Desai},
  {Dole}, {Fabricius}, {Farina}, {Ferreira}, {Finelli}, {Flose-Reimberg},
  {Fotopoulou}, {Ganga}, {Gozaliasl}, {Hook}, {Keihanen}, {Kirkpatrick},
  {Liebing}, {Lindholm}, {Mainetti}, {Martinelli}, {Martinet}, {Maturi},
  {McCracken}, {Metcalf}, {Morgante}, {Nightingale}, {Nucita}, {Patrizii},
  {Potter}, {Riccio}, {S{\'a}nchez}, {Sapone}, {Schewtschenko}, {Schultheis},
  {Scottez}, {Teyssier}, {Tutusaus}, {Valiviita}, {Viel}, {Vriend}, \&
  {Whittaker}}]{Sca++22}
{Euclid Collaboration}, {Scaramella}, R., {Amiaux}, J., {et~al.} 2022, \aap,
  662, A112

\bibitem[{{Falco} {et~al.}(1985){Falco}, {Gorenstein}, \& {Shapiro}}]{FGS85}
{Falco}, E.~E., {Gorenstein}, M.~V., \& {Shapiro}, I.~I. 1985, \apjl, 289, L1

\bibitem[{{Garvin} {et~al.}(2022){Garvin}, {Kruk}, {Cornen}, {Bhatawdekar},
  {Ca{\~n}ameras}, \& {Mer{\'\i}n}}]{Gar++22}
{Garvin}, E.~O., {Kruk}, S., {Cornen}, C., {et~al.} 2022, \aap, 667, A141

\bibitem[{{Gavazzi} {et~al.}(2012){Gavazzi}, {Treu}, {Marshall}, {Brault}, \&
  {Ruff}}]{Gav++12}
{Gavazzi}, R., {Treu}, T., {Marshall}, P.~J., {Brault}, F., \& {Ruff}, A. 2012,
  \apj, 761, 170

\bibitem[{{Glikman} {et~al.}(2006){Glikman}, {Helfand}, \& {White}}]{Gli++06}
{Glikman}, E., {Helfand}, D.~J., \& {White}, R.~L. 2006, \apj, 640, 579

\bibitem[{{Griffith} {et~al.}(2012){Griffith}, {Cooper}, {Newman}, {Moustakas},
  {Stern}, {Comerford}, {Davis}, {Lotz}, {Barden}, {Conselice}, {Capak},
  {Faber}, {Kirkpatrick}, {Koekemoer}, {Koo}, {Noeske}, {Scoville}, {Sheth},
  {Shopbell}, {Willmer}, \& {Weiner}}]{Gri++12}
{Griffith}, R.~L., {Cooper}, M.~C., {Newman}, J.~A., {et~al.} 2012, \apjs, 200,
  9

\bibitem[{{Hezaveh} {et~al.}(2016){Hezaveh}, {Dalal}, {Marrone}, {Mao},
  {Morningstar}, {Wen}, {Blandford}, {Carlstrom}, {Fassnacht}, {Holder},
  {Kemball}, {Marshall}, {Murray}, {Perreault Levasseur}, {Vieira}, \&
  {Wechsler}}]{Hez++16}
{Hezaveh}, Y.~D., {Dalal}, N., {Marrone}, D.~P., {et~al.} 2016, \apj, 823, 37

\bibitem[{{Hezaveh} {et~al.}(2013){Hezaveh}, {Marrone}, {Fassnacht}, {Spilker},
  {Vieira}, {Aguirre}, {Aird}, {Aravena}, {Ashby}, {Bayliss}, {Benson},
  {Bleem}, {Bothwell}, {Brodwin}, {Carlstrom}, {Chang}, {Chapman}, {Crawford},
  {Crites}, {De Breuck}, {de Haan}, {Dobbs}, {Fomalont}, {George}, {Gladders},
  {Gonzalez}, {Greve}, {Halverson}, {High}, {Holder}, {Holzapfel}, {Hoover},
  {Hrubes}, {Husband}, {Hunter}, {Keisler}, {Lee}, {Leitch}, {Lueker},
  {Luong-Van}, {Malkan}, {McIntyre}, {McMahon}, {Mehl}, {Menten}, {Meyer},
  {Mocanu}, {Murphy}, {Natoli}, {Padin}, {Plagge}, {Reichardt}, {Rest}, {Ruel},
  {Ruhl}, {Sharon}, {Schaffer}, {Shaw}, {Shirokoff}, {Stalder}, {Staniszewski},
  {Stark}, {Story}, {Vanderlinde}, {Wei{\ss}}, {Welikala}, \&
  {Williamson}}]{Hez++13}
{Hezaveh}, Y.~D., {Marrone}, D.~P., {Fassnacht}, C.~D., {et~al.} 2013, \apj,
  767, 132

\bibitem[{{Hilbert} {et~al.}(2007){Hilbert}, {White}, {Hartlap}, \&
  {Schneider}}]{Hil++07}
{Hilbert}, S., {White}, S. D.~M., {Hartlap}, J., \& {Schneider}, P. 2007,
  \mnras, 382, 121

\bibitem[{{Hyde} \& {Bernardi}(2009)}]{H+B09}
{Hyde}, J.~B. \& {Bernardi}, M. 2009, \mnras, 396, 1171

\bibitem[{{Jaelani} {et~al.}(2020){Jaelani}, {More}, {Sonnenfeld}, {Oguri},
  {Rusu}, {Wong}, {Chan}, {Suyu}, {Kayo}, {Lee}, \& {Inoue}}]{Jae++20}
{Jaelani}, A.~T., {More}, A., {Sonnenfeld}, A., {et~al.} 2020, \mnras, 494,
  3156

\bibitem[{{Jones} {et~al.}(2013){Jones}, {Ellis}, {Schenker}, \&
  {Stark}}]{Jon++13}
{Jones}, T.~A., {Ellis}, R.~S., {Schenker}, M.~A., \& {Stark}, D.~P. 2013,
  \apj, 779, 52

\bibitem[{{Keeton} {et~al.}(1997){Keeton}, {Kochanek}, \& {Seljak}}]{KKS97}
{Keeton}, C.~R., {Kochanek}, C.~S., \& {Seljak}, U. 1997, \apj, 482, 604

\bibitem[{{Kroupa}(2001)}]{Kro01}
{Kroupa}, P. 2001, \mnras, 322, 231

\bibitem[{{Lagos} {et~al.}(2018){Lagos}, {Tobar}, {Robotham}, {Obreschkow},
  {Mitchell}, {Power}, \& {Elahi}}]{Lag++18}
{Lagos}, C. d.~P., {Tobar}, R.~J., {Robotham}, A. S.~G., {et~al.} 2018, \mnras,
  481, 3573

\bibitem[{{Laigle} {et~al.}(2016){Laigle}, {McCracken}, {Ilbert}, {Hsieh},
  {Davidzon}, {Capak}, {Hasinger}, {Silverman}, {Pichon}, {Coupon}, {Aussel},
  {Le Borgne}, {Caputi}, {Cassata}, {Chang}, {Civano}, {Dunlop}, {Fynbo},
  {Kartaltepe}, {Koekemoer}, {Le F{\`e}vre}, {Le Floc'h}, {Leauthaud}, {Lilly},
  {Lin}, {Marchesi}, {Milvang-Jensen}, {Salvato}, {Sanders}, {Scoville},
  {Smolcic}, {Stockmann}, {Taniguchi}, {Tasca}, {Toft}, {Vaccari}, \&
  {Zabl}}]{Lai++16}
{Laigle}, C., {McCracken}, H.~J., {Ilbert}, O., {et~al.} 2016, \apjs, 224, 24

\bibitem[{{Langeroodi} {et~al.}(2023){Langeroodi}, {Sonnenfeld}, {Hoekstra}, \&
  {Agnello}}]{Lan++23}
{Langeroodi}, D., {Sonnenfeld}, A., {Hoekstra}, H., \& {Agnello}, A. 2023,
  \aap, 669, A154

\bibitem[{{Legin} {et~al.}(2023){Legin}, {Hezaveh}, {Perreault-Levasseur}, \&
  {Wandelt}}]{Leg++23}
{Legin}, R., {Hezaveh}, Y., {Perreault-Levasseur}, L., \& {Wandelt}, B. 2023,
  \apj, 943, 4

\bibitem[{{Li} {et~al.}(2023){Li}, {Kuijken}, {Hoekstra}, {Miller}, {Heymans},
  {Hildebrandt}, {van den Busch}, {Wright}, {Yoon}, {Bilicki}, {Bravo}, \&
  {Lagos}}]{Li++23}
{Li}, S.-S., {Kuijken}, K., {Hoekstra}, H., {et~al.} 2023, \aap, 670, A100

\bibitem[{{Mandelbaum} {et~al.}(2009){Mandelbaum}, {van de Ven}, \&
  {Keeton}}]{MVK09}
{Mandelbaum}, R., {van de Ven}, G., \& {Keeton}, C.~R. 2009, \mnras, 398, 635

\bibitem[{{Manti} {et~al.}(2017){Manti}, {Gallerani}, {Ferrara}, {Greig}, \&
  {Feruglio}}]{Man++17}
{Manti}, S., {Gallerani}, S., {Ferrara}, A., {Greig}, B., \& {Feruglio}, C.
  2017, \mnras, 466, 1160

\bibitem[{{Meert} {et~al.}(2015){Meert}, {Vikram}, \& {Bernardi}}]{Mee++15}
{Meert}, A., {Vikram}, V., \& {Bernardi}, M. 2015, \mnras, 446, 3943

\bibitem[{{Mendel} {et~al.}(2014){Mendel}, {Simard}, {Palmer}, {Ellison}, \&
  {Patton}}]{Men++14}
{Mendel}, J.~T., {Simard}, L., {Palmer}, M., {Ellison}, S.~L., \& {Patton},
  D.~R. 2014, \apjs, 210, 3

\bibitem[{{Millon} {et~al.}(2020){Millon}, {Galan}, {Courbin}, {Treu}, {Suyu},
  {Ding}, {Birrer}, {Chen}, {Shajib}, {Sluse}, {Wong}, {Agnello}, {Auger},
  {Buckley-Geer}, {Chan}, {Collett}, {Fassnacht}, {Hilbert}, {Koopmans},
  {Motta}, {Mukherjee}, {Rusu}, {Sonnenfeld}, {Spiniello}, \& {Van de
  Vyvere}}]{Mil++20}
{Millon}, M., {Galan}, A., {Courbin}, F., {et~al.} 2020, \aap, 639, A101

\bibitem[{{Muzzin} {et~al.}(2013){Muzzin}, {Marchesini}, {Stefanon}, {Franx},
  {McCracken}, {Milvang-Jensen}, {Dunlop}, {Fynbo}, {Brammer}, {Labb{\'e}}, \&
  {van Dokkum}}]{Muz++13}
{Muzzin}, A., {Marchesini}, D., {Stefanon}, M., {et~al.} 2013, \apj, 777, 18

\bibitem[{{Nierenberg} {et~al.}(2020){Nierenberg}, {Gilman}, {Treu}, {Brammer},
  {Birrer}, {Moustakas}, {Agnello}, {Anguita}, {Fassnacht}, {Motta}, {Peter},
  \& {Sluse}}]{Nie++20}
{Nierenberg}, A.~M., {Gilman}, D., {Treu}, T., {et~al.} 2020, \mnras, 492, 5314

\bibitem[{{Oguri}(2021)}]{Ogu21}
{Oguri}, M. 2021, \pasp, 133, 074504

\bibitem[{{Oguri} \& {Marshall}(2010)}]{O+M10}
{Oguri}, M. \& {Marshall}, P.~J. 2010, \mnras, 405, 2579

\bibitem[{{Oguri} {et~al.}(2014){Oguri}, {Rusu}, \& {Falco}}]{ORF14}
{Oguri}, M., {Rusu}, C.~E., \& {Falco}, E.~E. 2014, \mnras, 439, 2494

\bibitem[{{Oldham} {et~al.}(2017){Oldham}, {Auger}, {Fassnacht}, {Treu},
  {Brewer}, {Koopmans}, {Lagattuta}, {Marshall}, {McKean}, \&
  {Vegetti}}]{O+A17}
{Oldham}, L., {Auger}, M.~W., {Fassnacht}, C.~D., {et~al.} 2017, \mnras, 465,
  3185

\bibitem[{{Petrillo} {et~al.}(2019){Petrillo}, {Tortora}, {Vernardos},
  {Koopmans}, {Verdoes Kleijn}, {Bilicki}, {Napolitano}, {Chatterjee},
  {Covone}, {Dvornik}, {Erben}, {Getman}, {Giblin}, {Heymans}, {de Jong},
  {Kuijken}, {Schneider}, {Shan}, {Spiniello}, \& {Wright}}]{Pet++19}
{Petrillo}, C.~E., {Tortora}, C., {Vernardos}, G., {et~al.} 2019, \mnras, 484,
  3879

\bibitem[{{Robertson} {et~al.}(2020){Robertson}, {Smith}, {Massey}, {Eke},
  {Jauzac}, {Bianconi}, \& {Ryczanowski}}]{Robertson++20}
{Robertson}, A., {Smith}, G.~P., {Massey}, R., {et~al.} 2020, \mnras, 495, 3727

\bibitem[{{Robotham} {et~al.}(2020){Robotham}, {Bellstedt}, {Lagos}, {Thorne},
  {Davies}, {Driver}, \& {Bravo}}]{Rob++20}
{Robotham}, A.~S.~G., {Bellstedt}, S., {Lagos}, C. d.~P., {et~al.} 2020,
  \mnras, 495, 905

\bibitem[{{Salpeter}(1955)}]{Sal55}
{Salpeter}, E.~E. 1955, \apj, 121, 161

\bibitem[{{Savary} {et~al.}(2022){Savary}, {Rojas}, {Maus}, {Cl{\'e}ment},
  {Courbin}, {Gavazzi}, {Chan}, {Lemon}, {Vernardos}, {Ca{\~n}ameras},
  {Schuldt}, {Suyu}, {Cuillandre}, {Fabbro}, {Gwyn}, {Hudson}, {Kilbinger},
  {Scott}, \& {Stone}}]{Sav++22}
{Savary}, E., {Rojas}, K., {Maus}, M., {et~al.} 2022, \aap, 666, A1

\bibitem[{{Schneider} {et~al.}(1992){Schneider}, {Ehlers}, \& {Falco}}]{SEF92}
{Schneider}, P., {Ehlers}, J., \& {Falco}, E.~E. 1992, {Gravitational Lenses}

\bibitem[{{Shajib} {et~al.}(2021){Shajib}, {Treu}, {Birrer}, \&
  {Sonnenfeld}}]{Sha++21}
{Shajib}, A.~J., {Treu}, T., {Birrer}, S., \& {Sonnenfeld}, A. 2021, \mnras,
  503, 2380

\bibitem[{{Smith} {et~al.}(2015){Smith}, {Lucey}, \& {Conroy}}]{SLC15}
{Smith}, R.~J., {Lucey}, J.~R., \& {Conroy}, C. 2015, \mnras, 449, 3441

\bibitem[{{Sonnenfeld}(2018)}]{Son18}
{Sonnenfeld}, A. 2018, \mnras, 474, 4648

\bibitem[{{Sonnenfeld}(2020)}]{Son20}
{Sonnenfeld}, A. 2020, \aap, 641, A143

\bibitem[{{Sonnenfeld}(2022)}]{Son22}
{Sonnenfeld}, A. 2022, \aap, 659, A132

\bibitem[{{Sonnenfeld} \& {Cautun}(2021)}]{S+C21}
{Sonnenfeld}, A. \& {Cautun}, M. 2021, \aap, 651, A18

\bibitem[{{Sonnenfeld} {et~al.}(2018){Sonnenfeld}, {Chan}, {Shu}, {More},
  {Oguri}, {Suyu}, {Wong}, {Lee}, {Coupon}, {Yonehara}, {Bolton}, {Jaelani},
  {Tanaka}, {Miyazaki}, \& {Komiyama}}]{Son++18a}
{Sonnenfeld}, A., {Chan}, J. H.~H., {Shu}, Y., {et~al.} 2018, \pasj, 70, S29

\bibitem[{{Sonnenfeld} {et~al.}(2019){Sonnenfeld}, {Jaelani}, {Chan}, {More},
  {Suyu}, {Wong}, {Oguri}, \& {Lee}}]{Son++19}
{Sonnenfeld}, A., {Jaelani}, A.~T., {Chan}, J., {et~al.} 2019, \aap, 630, A71

\bibitem[{{Sonnenfeld} {et~al.}(2022){Sonnenfeld}, {Tortora}, {Hoekstra},
  {Asgari}, {Bilicki}, {Heymans}, {Hildebrandt}, {Kuijken}, {Napolitano},
  {Roy}, {Valentijn}, \& {Wright}}]{Son++22}
{Sonnenfeld}, A., {Tortora}, C., {Hoekstra}, H., {et~al.} 2022, \aap, 662, A55

\bibitem[{{Sonnenfeld} {et~al.}(2015){Sonnenfeld}, {Treu}, {Marshall}, {Suyu},
  {Gavazzi}, {Auger}, \& {Nipoti}}]{Son++15}
{Sonnenfeld}, A., {Treu}, T., {Marshall}, P.~J., {et~al.} 2015, \apj, 800, 94

\bibitem[{{Suess} {et~al.}(2019){Suess}, {Kriek}, {Price}, \&
  {Barro}}]{Sue++19}
{Suess}, K.~A., {Kriek}, M., {Price}, S.~H., \& {Barro}, G. 2019, \apjl, 885,
  L22

\bibitem[{{Szomoru} {et~al.}(2013){Szomoru}, {Franx}, {van Dokkum}, {Trenti},
  {Illingworth}, {Labb{\'e}}, \& {Oesch}}]{Szo++13}
{Szomoru}, D., {Franx}, M., {van Dokkum}, P.~G., {et~al.} 2013, \apj, 763, 73

\bibitem[{{Treu} \& {Marshall}(2016)}]{T+M16}
{Treu}, T. \& {Marshall}, P.~J. 2016, \aapr, 24, 11

\bibitem[{{van de Ven} {et~al.}(2009){van de Ven}, {Mandelbaum}, \&
  {Keeton}}]{VMK09}
{van de Ven}, G., {Mandelbaum}, R., \& {Keeton}, C.~R. 2009, \mnras, 398, 607

\bibitem[{{Vanden Berk} {et~al.}(2001){Vanden Berk}, {Richards}, {Bauer},
  {Strauss}, {Schneider}, {Heckman}, {York}, {Hall}, {Fan}, {Knapp},
  {Anderson}, {Annis}, {Bahcall}, {Bernardi}, {Briggs}, {Brinkmann}, {Brunner},
  {Burles}, {Carey}, {Castander}, {Connolly}, {Crocker}, {Csabai}, {Doi},
  {Finkbeiner}, {Friedman}, {Frieman}, {Fukugita}, {Gunn}, {Hennessy},
  {Ivezi{\'c}}, {Kent}, {Kunszt}, {Lamb}, {Leger}, {Long}, {Loveday}, {Lupton},
  {Meiksin}, {Merelli}, {Munn}, {Newberg}, {Newcomb}, {Nichol}, {Owen}, {Pier},
  {Pope}, {Rockosi}, {Schlegel}, {Siegmund}, {Smee}, {Snir}, {Stoughton},
  {Stubbs}, {SubbaRao}, {Szalay}, {Szokoly}, {Tremonti}, {Uomoto}, {Waddell},
  {Yanny}, \& {Zheng}}]{vdB++01}
{Vanden Berk}, D.~E., {Richards}, G.~T., {Bauer}, A., {et~al.} 2001, \aj, 122,
  549

\bibitem[{{Vegetti} {et~al.}(2012){Vegetti}, {Lagattuta}, {McKean}, {Auger},
  {Fassnacht}, \& {Koopmans}}]{Veg++12}
{Vegetti}, S., {Lagattuta}, D.~J., {McKean}, J.~P., {et~al.} 2012, \nat, 481,
  341

\bibitem[{{Wyithe} {et~al.}(2001){Wyithe}, {Turner}, \& {Spergel}}]{WTS01}
{Wyithe}, J.~S.~B., {Turner}, E.~L., \& {Spergel}, D.~N. 2001, \apj, 555, 504

\bibitem[{{Y{\i}ld{\i}r{\i}m} {et~al.}(2020){Y{\i}ld{\i}r{\i}m}, {Suyu}, \&
  {Halkola}}]{Yil++20}
{Y{\i}ld{\i}r{\i}m}, A., {Suyu}, S.~H., \& {Halkola}, A. 2020, \mnras, 493,
  4783

\bibitem[{{York} {et~al.}(2000){York}, {Adelman}, {Anderson}, {Anderson},
  {Annis}, {Bahcall}, {Bakken}, {Barkhouser}, {Bastian}, {Berman}, {Boroski},
  {Bracker}, {Briegel}, {Briggs}, {Brinkmann}, {Brunner}, {Burles}, {Carey},
  {Carr}, {Castander}, {Chen}, {Colestock}, {Connolly}, {Crocker}, {Csabai},
  {Czarapata}, {Davis}, {Doi}, {Dombeck}, {Eisenstein}, {Ellman}, {Elms},
  {Evans}, {Fan}, {Federwitz}, {Fiscelli}, {Friedman}, {Frieman}, {Fukugita},
  {Gillespie}, {Gunn}, {Gurbani}, {de Haas}, {Haldeman}, {Harris}, {Hayes},
  {Heckman}, {Hennessy}, {Hindsley}, {Holm}, {Holmgren}, {Huang}, {Hull},
  {Husby}, {Ichikawa}, {Ichikawa}, {Ivezi{\'c}}, {Kent}, {Kim}, {Kinney},
  {Klaene}, {Kleinman}, {Kleinman}, {Knapp}, {Korienek}, {Kron}, {Kunszt},
  {Lamb}, {Lee}, {Leger}, {Limmongkol}, {Lindenmeyer}, {Long}, {Loomis},
  {Loveday}, {Lucinio}, {Lupton}, {MacKinnon}, {Mannery}, {Mantsch}, {Margon},
  {McGehee}, {McKay}, {Meiksin}, {Merelli}, {Monet}, {Munn}, {Narayanan},
  {Nash}, {Neilsen}, {Neswold}, {Newberg}, {Nichol}, {Nicinski}, {Nonino},
  {Okada}, {Okamura}, {Ostriker}, {Owen}, {Pauls}, {Peoples}, {Peterson},
  {Petravick}, {Pier}, {Pope}, {Pordes}, {Prosapio}, {Rechenmacher}, {Quinn},
  {Richards}, {Richmond}, {Rivetta}, {Rockosi}, {Ruthmansdorfer}, {Sandford},
  {Schlegel}, {Schneider}, {Sekiguchi}, {Sergey}, {Shimasaku}, {Siegmund},
  {Smee}, {Smith}, {Snedden}, {Stone}, {Stoughton}, {Strauss}, {Stubbs},
  {SubbaRao}, {Szalay}, {Szapudi}, {Szokoly}, {Thakar}, {Tremonti}, {Tucker},
  {Uomoto}, {Vanden Berk}, {Vogeley}, {Waddell}, {Wang}, {Watanabe},
  {Weinberg}, {Yanny}, {Yasuda}, \& {SDSS Collaboration}}]{Yor++00}
{York}, D.~G., {Adelman}, J., {Anderson}, John~E., J., {et~al.} 2000, \aj, 120,
  1579

\end{thebibliography}

\appendix
\section{Lens galaxy surface brightness distribution}\label{sect:appendixa}

To assign half-light radii and ellipticities to the simulated lenses we relied on observations of a sample of early-type galaxies selected from the SDSS \citep[][]{Yor++00}.
This sample was selected as follows. Starting from the SDSS spectroscopic sample, we defined a narrow redshift slice around $z=0.2$. Then we applied a selection in colour, by choosing objects with $g-r>1.2$, and on S\'{e}rsic index, by selecting only galaxies with $n>2.5$. We used the S\'{e}rsic fit measurements by \citet{Mee++15} for this purpose. These cuts produced a sample of $8078$ galaxies.
We then focused on the $r-$band de Vaucouleurs model-based photometric measurements of \citet{Mee++15}.
Using the $r-$band total flux from the de Vaucouleurs model and the stellar mass-to-light ratio estimates of \citet{Men++14}, we obtained measurements of $\mobs$. Finally, we fitted the stellar mass-size relation with the following model:
\begin{equation}
\log{\reff} \sim \mathcal{N}(\mu_{R,0} + \beta_R(\log{\mobs} - 11.4), \sigma_R^2).
\end{equation}
We obtained $\mu_{R,0}=1.20$, $\beta_R=0.63$ and $\sigma_R=0.14$.

\begin{figure}
\includegraphics[width=\columnwidth]{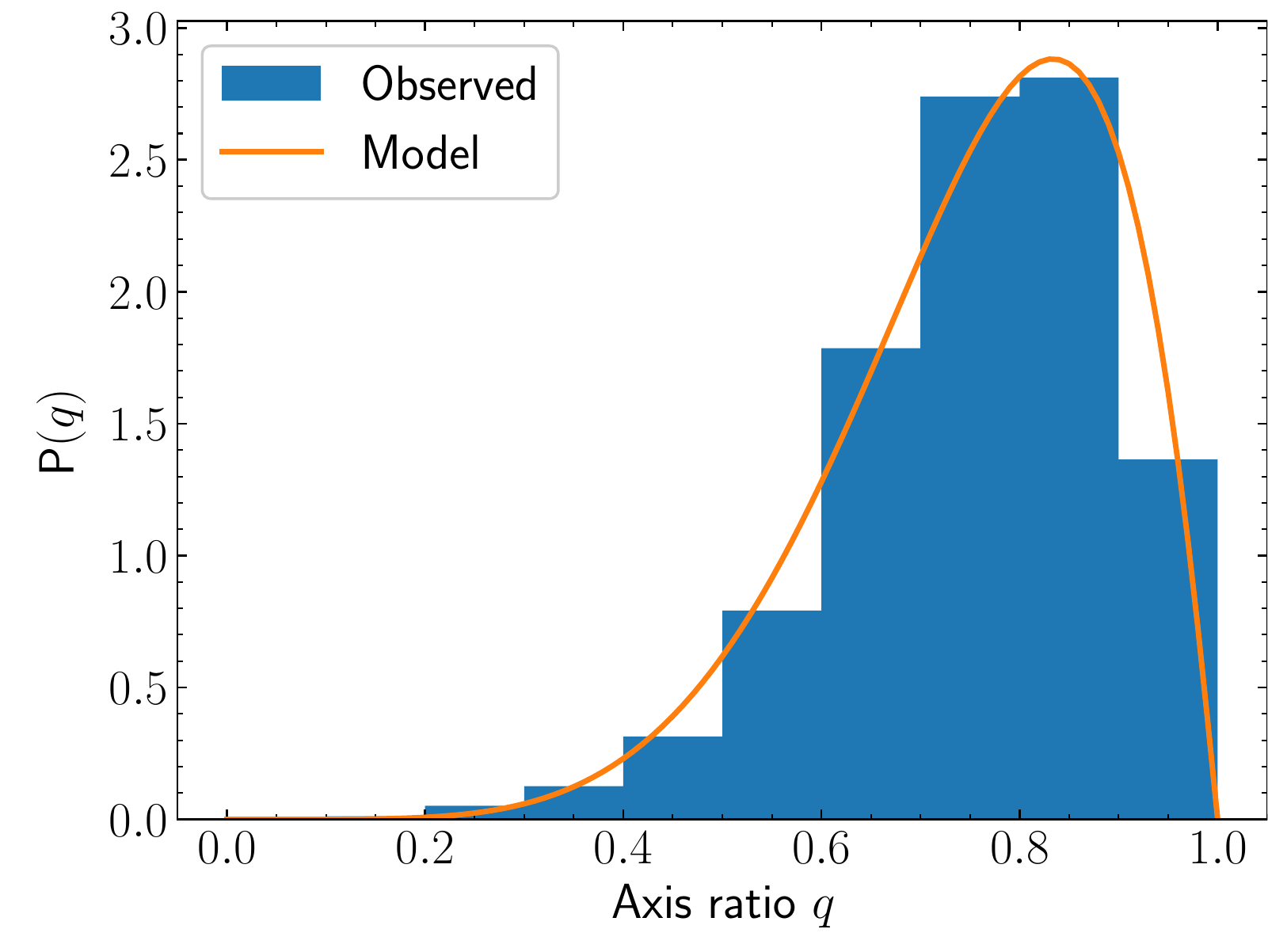}
\caption{
Distribution in axis ratio of a sample of early-type galaxies.
The sample consists of $8078$ galaxies from the SDSS, selected by means of cuts in redshift, colour, and S\'{e}rsic index, as explained in the text.
Measurements of the axis ratio are taken from the de Vaucouleurs model fits of \citet{Mee++15}.
The model curve is a beta distribution (Eq. \ref{eq:betaappendix}) with $\alpha=6.28$ and $\beta=2.05$.
\label{fig:qhist}
}
\end{figure}

We then proceeded to fit for the axis ratio distribution of the same sample of galaxies.
Figure \ref{fig:qhist} shows a histogram of the observed distribution.
We fitted this with a beta distribution:
\begin{equation}\label{eq:betaappendix}
\pr(q) \propto q^{\alpha-1}(1 - q)^{\beta-1}.
\end{equation}
We obtained $\alpha=6.28$ and $\beta=2.05$.

\section{Upper limits on the intrinsic scatter parameters}\label{sect:appendixb}

We used the fundamental plane of early-type galaxies to set an upper limit on the intrinsic scatter parameters of the simulation: $\sigma_{\mathrm{sps}}$, $\sigma_h$ and $\sigma_\gamma$.
First, we measured the fundamental plane of the sample of SDSS early-type galaxies introduced in Appendix~\ref{sect:appendixa}.
We took measurements of the line-of-sight stellar velocity dispersion within the SDSS spectroscopic aperture, $\sigma_{\mathrm{ap}}$, from the SDSS data release 16 catalogue \citep{Ahu++20}.
Then, we fitted the following model to the distribution of $\sigma_{\mathrm{ap}}$ of the sample:
\begin{equation}\label{eq:fpfull}
\pr(\sigma_{\mathrm{ap}}) \sim \mathcal{N}(\mu_{\sigma} + \beta(\log{\mobs} - 11.4) + \xi(\log{\reff} - 1.2), \sigma_{\sigma}).
\end{equation}
We accounted for observational uncertainties on $\sigma_{\mathrm{ap}}$ when doing the fit; therefore, the parameter $\sigma_{\sigma}$ describes the intrinsic scatter in the logarithm of the velocity dispersion, deconvolved from the observational scatter.
In principle we should also account for observational uncertainties on the stellar mass measurement, as they too contribute to the inferred scatter in $\sigma_{\mathrm{ap}}$. In practice, however, it is difficult to estimate observational uncertainties on stellar population synthesis measurements \citep[but see][]{Dog++23}.
For this reason we chose not to propagate uncertainties on $\mobs$. As a result, the inferred scatter parameter $\sigma_{\sigma}$ is slightly overestimated.
We obtained $\mu_{\sigma}=2.36$, $\beta_{\sigma}=0.33$, $\xi_\sigma=-0.17$, and $\sigma_{\sigma}=0.035$.

We then generated samples of $z=0.2$ galaxies from the model of Sect. \ref{ssec:lenses} and used the spherical Jeans equation to predict their central velocity dispersion.
The spherical Jeans equation is \citep{B+T87}
\begin{equation}\label{eq:jeans}
\frac{d(\rho_*\sigma_r^2)}{dr} + \frac{\beta(r)}{r}\rho_*\sigma_r^2 = -\rho_*(r)\frac{GM(r)}{r^2},
\end{equation}
where $\rho_*(r)$ is the three-dimensional distribution of dynamical tracers (i.e. the stars), $\sigma_r$ the radial component of the velocity dispersion, $\beta(r)$ the orbital anisotropy parameter, and $M(r)$ the mass enclosed within a spherical shell of radius $r$. 
We assumed isotropic orbits ($\beta=0$), then integrated Eq. \ref{eq:jeans} to obtain the seeing-convolved, surface brightness-weighted line-of-sight velocity dispersion within the SDSS spectroscopic aperture\footnote{We used the Python code available at \url{https://github.com/astrosonnen/spherical_jeans} for this purpose.}.
Finally, we fitted the fundamental plane relation of Eq. \ref{eq:fpfull} to the observed stellar mass, half-light radius and central velocity dispersion of the mock sample.
We repeated this procedure for different values of the intrinsic scatter parameters $\sigma_{\mathrm{sps}}$, $\sigma_\mathrm{h}$ and $\sigma_\gamma$, then settled on the three scenarios listed in \Tref{tab:scatter}.

\section{Mass-sheet transformation parameter}\label{sect:appendixc}

In the time-delay study of \citet{Bir++20}, lenses are described with a mass density profile given by Eq. \ref{eq:mstprofile}.
\citet{Bir++20} defined the profile by first fitting a pure power-law mass model to the strong lensing data, thus constraining the Einstein radius and power-law index $\gamma$, and then applying a mass-sheet transformation of the following kind:
\begin{equation}\label{eq:mst}
\kappa(\boldsymbol\theta) \rightarrow \lmst\kappa(\boldsymbol\theta) + 1 - \lmst.
\end{equation}
The mass-sheet transformation parameter $\lmst$ can be interpreted as a quantity describing a departure from a pure power-law profile.
One possible way of estimating $\lmst$ for our lenses is to fit a power-law to the simulated images and then optimise for the value of $\lmst$ that best matches the true density profile. That, however, would be too time-consuming. Instead, we emulate this process as follows.

\citet{Son18} showed that, when the main arc and the counter-image of an extended source are well resolved, their relative widths constrain the following combination of derivatives of the lens potential at the Einstein radius:
\begin{equation}\label{eq:invariant}
\frac{\psi'''}{1 - \psi''}.
\end{equation}
In other words, the above quantity can be measured robustly (i.e. in a model-independent way).
When fitting a pure power-law lens model to lensing data, then, the value of $\gamma$ that we obtain is the one that reproduces the true $\psi'''/(1-\psi'')$\footnote{Strictly speaking, this is only true in the limit of well-resolved arcs and nearly axisymmetric lenses}. This is given by
\begin{equation}
\gamma = 2 + \tein \frac{\psi'''}{1-\psi''}
.\end{equation}
Hence, given a lens in our sample, we first define a pure power-law model with the same Einstein radius as the true one and with $\gamma$ given by the equation above. 
Then, given a pure power-law model, we find $\lmst$ such that the first three derivatives of the lens potential match the true values.
In particular, since a mass-sheet transformation changes the second derivative of the potential as
\begin{equation}
\psi'' \rightarrow \lmst\psi'' + 1 - \lmst,
\end{equation}
we set
\begin{equation}
\lmst = \frac{1 - \psi''_{\mathrm{PL}}}{1 - \psi''},
\end{equation}
where $\psi''_{\mathrm{PL}}$ is the value of $\psi''$ of the pure power-law lens model.

\end{document}